\documentclass[10pt,a4paper,usenames,dvipsnames]{article}
\RequirePackage{pdf14}
\usepackage{jheppub}
\usepackage[utf8]{inputenc}
\usepackage[parfill]{parskip}
\usepackage{comment} 
\usepackage{ytableau}
\usepackage{adjustbox}
\usepackage{tikz}
\usepackage{tikz-cd}
\usepackage{graphicx}
\usepackage{amsfonts,amsmath,amssymb,amscd}
\usepackage{array,setspace,mathrsfs,yfonts,dsfont,bbm,colonequals,mathtools}
\usepackage{relsize,suffix,cancel,bbm,capt-of,upgreek,verbatim,xcolor}
\usepackage{dsfont}
\usepackage{multirow,colortbl,arydshln}
\usepackage{adjustbox}
\usepackage{rotating}
\usepackage{caption}
\usepackage{subcaption}
\usepackage{footnote}
\usepackage{tablefootnote}
\usepackage{stmaryrd}
\usepackage{makecell}

\usetikzlibrary{calc,shadings,patterns,tikzmark}

\tikzset{gauge1/.style={draw=none,minimum size=0.6cm,fill=white,circle, draw}}
\tikzset{gauge3/.style={draw=none,minimum size=0.4cm,fill=white,circle, draw}}
\tikzset{crosses/.style={cross out, draw=black, minimum size=0.3cm, inner sep=0pt, outer sep=0pt},
cross/.default={1pt}}
\tikzset{blank/.style={draw=none,minimum size=0.4cm,fill=none,circle, draw}}
\tikzset{flavour2/.style={draw=none,minimum size=0.5cm,fill=white,regular polygon sides=4,draw}}
\tikzset{flavourBlue/.style={draw=none,minimum size=0.4cm,fill=blue,regular polygon sides=4,draw}}
\tikzset{flavourRed/.style={draw=none,minimum size=0.4cm,fill=red,regular polygon sides=4,draw}}
\tikzset{none/.style={draw=none}}
\tikzset{redgauge/.style={draw=none,minimum size=0.4cm,fill=red,circle, draw}}
\tikzset{miniU/.style={draw=none,minimum size=0.1cm,fill=red,circle, draw}}
\tikzset{smallgauge1/.style={draw=none,minimum size=0.1cm,fill=white,circle, draw}}
\tikzset{miniBlue/.style={draw=none,minimum size=0.1cm,fill=blue,circle, draw}}
\tikzset{gauge2/.style={draw=none,minimum size=0.35mm,fill=red,circle, draw}}
\tikzset{bluegauge/.style={draw=none,minimum size=0.4cm,fill=blue,circle, draw}}
\tikzset{flavour1/.style={draw=none,minimum size=0.35mm,fill=blue, regular polygon,regular polygon sides=4,draw}}
\tikzset{flavour0/.style={draw=none,minimum size=0.35mm,fill=white, regular polygon,regular polygon sides=4,draw}}
\tikzset{smalldot/.style={draw=none,minimum size=0.1mm,fill=black, circle,draw}}
\tikzset{dotsize/.style={circle,fill,inner sep=1.5pt,draw}}
\tikzset{doubleguys/.style={double, double distance = 3pt}}
\tikzset{tripleguys/.style={triple}}
\tikzset{new edge style 1/.style={dashed}}
\tikzset{thickline/.style={line width=0.06cm}}
\tikzset{darke/.style={line width=0.3mm,black}}
\tikzset{brace/.style={decorate,decoration={brace,amplitude=10pt}}}
\pgfdeclarelayer{edgelayer}
\pgfdeclarelayer{nodelayer}
\usepackage[noabbrev,capitalise]{cleveref}
\usetikzlibrary{decorations.pathreplacing}
\pgfsetlayers{edgelayer,nodelayer,main}

\tikzset{hasse/.style={circle, fill,inner sep=2pt}}
\tikzset{gauge/.style={inner sep=1mm,draw=none,fill=white,minimum size=2mm,circle, draw}}
\tikzset{flavour/.style={draw=none,minimum size=0.3mm,fill=white, regular polygon,regular polygon sides=4,draw}}
\tikzset{bd/.style={circle, draw=black, inner sep=0pt, fill=black, minimum size=2mm}}
\tikzset{gd/.style={circle, draw=green, inner sep=0pt, fill=green, minimum size=2mm}}
\tikzset{gauge3/.style={draw=none,minimum size=0.35mm,fill=white,circle, draw}}
\tikzset{none/.style={draw=none}}
\tikzset{flavor1/.style={draw=none,minimum size=0.35mm,fill=white, regular polygon,regular polygon sides=4,draw}}
\tikzset{flavour2/.style={draw=none,minimum size=0.35mm,fill=white, regular polygon,regular polygon sides=4,draw}}
\tikzset{blankflavor/.style={draw=none,minimum size=0.5mm,fill=none, regular polygon,regular polygon sides=4,draw}}
\pgfdeclarelayer{edgelayer}
\pgfdeclarelayer{nodelayer}
\pgfsetlayers{edgelayer,nodelayer,main}
\tikzset{brace/.style={decorate,decoration={brace,amplitude=10pt}}}

\def\quatdim{d}
\def\quatdimtwo{d}
\def\gammaclass{\upgamma}
\def\betaclass{\upbeta}
\def\unbrSYMM{\mathfrak{g}_{\text{UV}}^{\natural}}

\title{Free field realizations for rank-one SCFTs}

\author[a]{Christopher Beem,\!}
\author[b]{Anirudh Deb,\!}
\author[c]{Mario Martone,\!}
\author[d]{Carlo Meneghelli,\!}
\author[b]{Leonardo Rastelli}

\affiliation[a]{Mathematical Institute, University of Oxford, Woodstock Road, Oxford, OX2 6GG, UK}
\affiliation[b]{C.~N.~Yang Institute for Theoretical Physics, Stony Brook University, Stony Brook, NY 11794, USA}
\affiliation[c]{Department of Mathematics, King’s College London, The Strand, London WC2R 2LS, UK}
\affiliation[d]{Dipartimento SMFI, Universit`a di Parma, Viale G.P. Usberti 7/A, 43100, Parma, Italy and INFN Gruppo Collegato di Parma}

\usepackage{amsmath}
\usepackage{ytableau}
\usepackage{booktabs}

\numberwithin{equation}{section}

\def\bea{\begin{eqnarray}}
\def\eea{\end{eqnarray}}

\def\g{\gamma}
\def\G{\Gamma}

\def\ksu2{k_{2d}^{\suf(2)}}

\DeclarePairedDelimiterX\MeijerM[3]{\lparen}{\rparen}%
{\begin{smallmatrix}#1 \\ #2\end{smallmatrix}\delimsize\vert\,#3}

\newcommand\MeijerG[8][]{%
  G^{\,#2,#3}_{#4,#5}\MeijerM[#1]{#6}{#7}{#8}}

\WithSuffix\newcommand\MeijerG*[7]{%
  G^{\,#1,#2}_{#3,#4}\MeijerM*{#5}{#6}{#7}}

\def\SU{\mathrm{SU}}

\def\cH{\mathcal{H}}

\def\cN{\mathcal{N}}
\def\cV{\mathbb{V}}

\def \beg#1{\begin{#1}} 

\def \bea{\beg{eqnarray}}
\def \eea{\end{eqnarray}}
\def \ee{\end{equation}}

\def \af{\mf{a}}
\def \df{\mf{d}}
\def \ef{\mf{e}}

\def \uf{\mf{u}}

\def \suf{\mf{su}}

\def \spf{\mf{sp}}

\def \ef{\mf{e}}

\def \restr#1#2{{\left.\kern-\nulldelimiterspace#1\vphantom{\big|}\right|_{#2}}}

\def \nn{\nonumber}

\def \cf{{\it cf.}}
\def \Tr{{\rm Tr}}

\def \NN{\mathcal{N}}

\newcommand{\IL}{{\cal U}}

\newcommand{\CB}{{\text{CB}}}
\newcommand{\HB}{{\text{HB}}}
\newcommand{\ECB}{{\text{ECB}}}

\newcommand\HatchedCell[4][0pt]{%
  \begin{tikzpicture}[overlay,remember picture]%
    \fill[#4] ( $ (pic cs:#2) + (0,1.9ex) $ ) rectangle ( $ (pic cs:#3) + (0pt,-#1*\baselineskip-.8ex) $ );
  \end{tikzpicture}%
}%
\newcommand*{\hatch}[2]{\multicolumn{#2}{!{\hspace*{-0.4pt}\tikzmark{start#1}}c!{\tikzmark{end#1}}}{}}
\allowdisplaybreaks[2]  
\usepackage{pifont}

\def\blue#1{{\color{blue}{#1}}}
\def\green#1{{\color{black!25!green}{#1}}}

\def\red#1{{\color{red}{#1}}}

\def\rcy{\rowcolor{black!25!yellow!10}}

\def\blue#1{{\color{blue}{#1}}}
\def\green#1{{\color{black!25!green}{#1}}}

\def\red#1{{\color{red}{#1}}}

\def\rcy{\rowcolor{black!25!yellow!10}}

\def\red#1{{\color{red}{#1}}}

\def\bar{\overline}

\def\hat{\widehat}

\def\^{\wedge}

\def\Tr{\mathop{\rm Tr}}

\def\U{{\rm U}}
\def\SU{{\rm SU}}
\def\O{\mathop{\rm o}}

\def\H{\mathbb{H}}
 
\def\O{\mathbb{O}}

\def\V{\mathbb{V}} 
\def\Z{\mathbb{Z}} 
\def\af{{\mathfrak a}}
\def\cf{{\mathfrak c}}
\def\df{{\mathfrak d}}
\def\ef{{\mathfrak e}}
\def\ff{{\mathfrak f}}

\def\uf{{\mathfrak u}}
\def\spf{\mathfrak{sp}}

\def\cH{{\mathcal H}}

\def\cN{{\mathcal N}}

\def\cQ{{\mathcal Q}}

\def\cS{{\mathcal S}}
\def\cT{{\mathcal T}}
\def\cV{{\mathcal V}}

\def\g{{\gamma}}

\def\G{{\Gamma}}

\def\af{\mathfrak{a}}
\def\uf{\mathfrak{u}}

\def\beq{\begin{equation}}
\def\eeq{\end{equation}}

\def\af{{\mathfrak a}}
\def\cf{{\mathfrak c}}
\def\df{{\mathfrak d}}
\def\ef{{\mathfrak e}}
\def\ff{{\mathfrak f}}

\def\spf{\mathfrak{sp}}
\def\suf{\mathfrak{su}}

\def\uf{\mathfrak{u}}

\abstract{In this paper, we construct  the associated vertex operator algebras for all $\mathcal{N}=2$ superconformal field theories of rank one. We give a uniform presentation through free-field realizations, which turns out to be a particularly suitable framework for this task. The elementary building blocks of the construction are dictated by the low energy degrees of freedom on the Higgs branch, which are well understood for rank-one theories. We further analyze the interplay between Higgs and Coulomb data on the moduli space of vacua, which tightly constrain the overall structure of the free field realizations. Our results suggest a plausible bottom-up classification scheme for low-rank SCFTs incorporating vertex algebra techniques.}

\begin{document}
\maketitle

\section{\label{sec:intro}Introduction}

To any four-dimensional $\cN=2$ superconformal field theory (SCFT), one can canonically associate a two-dimensional vertex operator algebra (VOA) via the construction of \cite{Beem:2013sza},
\begin{equation}\label{eqn:correspondence}
    \mathbb{V}: \;\; {\rm 4d \;\; \cN=2 \; \; SCFT ~\longrightarrow~ VOA}~.
\end{equation}
The VOA arises as a cohomological reduction of the full local OPE algebra of a four-dimensional theory $\cT$ with respect to a certain nilpotent supercharge.\footnote{The supercharge takes the schematic form $\cQ + \widetilde{\cS}$, where $\cQ$ and $\widetilde{\cS}$ denote certain Poincar\'e and special conformal supercharges, respectively.} This correspondence has been investigated extensively in recent years and there are many indications that $\mathbb{V}[\cT]$ is deeply connected with the physics of the Higgs branch of vacua \HB$[\cT]$. An important observation---conjectured to be universally true \cite{Beem:2017ooy}---is that the Higgs Branch can be recovered directly from $\mathbb{V}[\cT]$ by applying a certain canonical map~\cite{Arakawa:2010ni} that extracts from any VOA $\cV$ a Poisson variety $X_\cV$ known as the \emph{associated variety}. In short, the Higgs Branch Conjecture of~\cite{Beem:2017ooy} reads as
\begin{equation}\label{eqn:M=X}
    \HB[\cT]= X_{\mathbb{V} [\cT]}~.
\end{equation}

Starting with the work of \cite{Beem:2019tfp,Beem_2020rank2inst}, the connection between VOA and Higgs branch physics has been seen to go deeper, with the latter being in some cases sufficient to determine the full VOA through geometrically motivated free field realizations. Though the principles underlying these free field realizations have yet to be completely elucidated, the intuitive picture that arises from the examples studied thus far is that geometric data of the Higgs branch as a holomorphic symplectic variety, supplemented by knowledge of the low-energy degrees of freedom in a generic Higgs branch vacuum, largely determine the free field realization. In particular, at low energies at a given point on the Higgs branch the theory will generally flow to another interacting SCFT (with associated VOA ${\cal V}_{\rm IR}$) plus a collection of free hypermultiplets and/or free vector multiplets. The details of the low energy physics dictate the number of (lattice) chiral bosons, symplectic bosons, and symplectic fermions that should arise in the construction. The sought after VOA is identified as a subVOA of ${\cal V}_{\rm IR}$ tensored with the requisite free fields.

More formally, we expect that the vertex algebras associated to SCFTs possess some good (micro)localization properties on their associated varieties, and so can be interpreted as algebras of sections of appropriately defined sheaves of vertex algebras on $X_{\mathbb{V}[\cT]}$.\footnote{More accurately, they should be something like sections of sheaves of asymptotic algebras of chiral differential operators as in \cite{Arakawa:2011vpo}; see \cite{Arakawa:2023cki,Coman:2023xcq} for related recent work connected to supersymmetric field theories.} Absent a general computational handle on this local structure, in practice the examples in works past and present rely on identifying a Zariski-open patch on the Higgs branch whose symplectic geometry takes a particularly simple form, so that the assignment of our hypothetical sheafified vertex algebra to this set is determined straightforwardly from the low energy physics on that patch. We can then construct by hand certain elements of the (global section) vertex algebra such as the stress tensor operator and geometrically meaningful operators (\emph{e.g.}, those corresponding to Higgs branch chiral ring generators), the details of which are constrained by physical considerations and consistency conditions for the vertex algebra. Additional strong generators of the VOA may arise in the singular terms in the operator product expansion (OPE) of those geometric operators, which are \emph{a priori} fixed given the free field constructions. It must be acknowledged, however, that all SCFTs for which free field realizations have been obtained so far, including our new set, are still somewhat special due to the existence of an appropriate open chart. 


In this work we apply these free-field methods to produce uniform constructions of the VOAs associated to all rank-one SCFTs---the simplest set of four dimensional $\cN=2$ SCFTs---which have been classified on the basis of consistency conditions for Coulomb branch geometries in \cite{Argyres:2015ffa,Argyres:2015gha,Argyres:2016xua,Argyres:2016xmc}. The results of this classification are summarized in Table~\ref{tab:summary}. The VOAs of certain rank-one SCFTs----those in the first series in Table~\ref{tab:summary}---have been known since the original work of \cite{Beem:2013sza}. Here we construct the VOAs for the remaining entries. This result fills a conspicuous gap in the systematic study of VOAs for four dimensional SCFTs. The common theme of the theories analyzed is the presence of free hypermultiplets at a generic point of their \emph{Coulomb} branch, \emph{i.e.}, these theories have \emph{Enhanced Coulomb Branches}. This fact plays an important role in dictating the structure of our free field constructions, linking Higgs and Coulomb branch data with the latter providing additional guidance regarding the VOA building blocks.

A central conjecture about Higgs branch free-field realizations is that they realize the \emph{simple quotient} of the relevant VOA, \emph{i.e.}, all null vectors vanish identically when expressed in terms of the free fields. This is a delicate property, as the VOAs in question occur at specific values of the central charge and other parameters such that there generally exists an array of non-trivial singular vectors, whose nullity is essential for the VOA to exist (as associativity of the OPE would fail if these null vectors were not set to zero). While a general argument remains elusive, this property has been observed in all geometrically inspired free-field realizations to date. We provide strong evidence for the continued validity of this property in our new examples as well.

The interplay among Coulomb, Higgs, and VOA data appears to be so delicate and constrained that a more ambitious program naturally suggests itself: a \emph{bottom-up} approach to classifying SCFTs. The aspiration would be to identify a set of consistency conditions on the allowed moduli space geometry and on the free-field construction of the VOA such that Table~\ref{tab:summary} might be \emph{derived} rather than assumed. The continued (nontrivial) success of these geometrically motivated free field methods in the larger class of models studied in this paper is a welcome affirmation of the general program. A further encouraging sign in this direction is that when we lift some of the assumptions that follow from our focus on rank-one theories, a larger set of allowed geometries arise, all of which are indeed realized in known higher-rank $\NN=2$ SCFTs. Such a classification program would share a kindred spirit with that of~\cite{Kaidi:2022sng}, which also aimed to combine Higgs, Coulomb, and VOA data. Here, though, we envision the full VOA being realized as a natural output. We leave this promising line of work for the future.

This paper is organized as follows. In Section \ref{sec:ECBHiggs} we recall in brief the structure of the moduli space of vacua of rank-one SCFTs, with an emphasis on Higgs and Enhanced Coulomb branches (ECB). We explain how the structure of the ECB serves to constrain the allowed IR theories that can remain after Higgsing the UV SCFT. The VOAs of these IR theories act as building blocks in the construction of VOAs for the more elaborate rank-one theories under consideration. In Section \ref{Delignereview} we review the structure of Higgs branch free field realizations in the case of rank-one and rank-two Deligne SCFTs to illustrate the intuition and general idea behind these constructions. We give an overview of our strategy to obtain free fields realizations and discuss the $C_2U_1$ and $A_1U_1$ theory in detail in Section \ref{OverviewofStrategyandC2U1}.  In Section \ref{FFallrank1} we discuss the generalization of the construction to the remaining rank-one theories with Enhanced Coulomb branches. Appendix~\ref{appB} discusses anomaly matching constraints on the Higgs branch.


\begin{table}[t!]
\centering
\resizebox{1.145\textwidth}{!}{
$\begin{array}{|c@{}c@{}c@{}@{}c@{}@{}c@{}@{}c@{}|@{}c@{}@{}c@{}@{}c@{}|c@{}|cc@{}c@{}c@{}|}
\hline
\multicolumn{6}{c}{\text{Theory data}} &
\multicolumn{3}{c}{\text{Central charges}}&\multicolumn{1}{c}{\text{}}& \multicolumn{4}{c}{\text{VOA data}}
\\[1mm]
\hline
&{\rm Name}&\Delta(u) &\mathfrak{g}_{UV}&\ d_{\text{HB}}\
&\ \; \; \; \; \; n_h^\text{mix} 
\ \
&\ \ \ \ k_\ff\ \ \ \  &\ \ 24a\ \ &\ \ 12c\ \ &\ \ \ \ \ &\multicolumn{2}{c}{\cH}&\multicolumn{2}{c|}{\chi(\cT)}
\\[1mm]
\hline\hline
&&&&&&&&&&&&\\[-4mm]
&[II^*,E_8]& 6
& E_8 & 29 & 0
& 12 & 95 & 62
&\hatch{1}{1} &\multicolumn{2}{c}{\O_{\rm min}(\ef_8)}&\multicolumn{2}{c|}{V_{-6}(\ef_8) }\\
&[III^*,E_7]&4
& E_7 & 17 & 0
& 8 & 59 & 38
&\hatch{2}{1} &\multicolumn{2}{c}{\O_{\rm min}(\ef_7)}&\multicolumn{2}{c|}{V_{-4}(\ef_7)} \\
&[IV^*,E_6]& 3
& E_6 & 11 & 0
& 6 & 41 & 26
&\hatch{3}{1} &\multicolumn{2}{c}{\O_{\rm min}(\ef_6)}&\multicolumn{2}{c|}{V_{-3}(\ef_6)} \\
\rcy &[I_0^*,D_4]& 2
& D_4& 5 & 0
& 4 & 23 & 14
&\hatch{4}{1} &\multicolumn{2}{c}{\O_{\rm min}(\df_4)}&\multicolumn{2}{c|}{V_{-2}(\df_4)} \\
&[IV,A_2]&  3/2
& A_2& 2 & 0
& 3 & 14 & 8
&\hatch{5}{1} &\multicolumn{2}{c}{\O_{\rm min}(\af_2)}&\multicolumn{2}{c|}{V_{-\frac{3}{2}}(\af_2)} \\
&[III,A_1]& 4/3
& A_1 & 1 & 0
& 8/3 & 11 & 6
&\hatch{6}{1}&\multicolumn{2}{c}{\O_{\rm min}(\af_1)}&\multicolumn{2}{c|}{V_{-\frac{4}{3}}(\af_1)}  \\
&\red{[II,\varnothing]}& 6/5
& \varnothing & 0 & 0
& - & 43/5 & 22/5
&\hatch{7}{1}&\multicolumn{2}{c}{-}&\multicolumn{2}{c|}{{\rm Vir}_{2,5}  } 
\\
\rcy &\red{[I_0,\varnothing]}& 1
& \varnothing & 0 & 0
& - & 5 & 2
&\hatch{8}{1}&\multicolumn{2}{c}{-}&\multicolumn{2}{c|}{\mathbb{V}_\eta  } 

\\[.5mm]
\hline\hline\\[-3.5mm]
&&\multicolumn{8}{c}{}&\cH_{\rm mixed}&\unbrSYMM &\qquad\chi(\cT_{\rm IR})\qquad&\quad\qquad\cV_{\rm free}\quad\qquad  \\[1mm]
\hline\hline
 &[II^*,C_5]&6
& C_5& 16 & 5
& 7 & 82 & 49
&\hatch{9}{1}&\O_{\rm min}(\cf_5)&C_4&\qquad V_{-3}(\ef_6) &\V_{\xi}^{\spf(8)}\otimes \Pi_{\frac12}\\
&[III^*,C_3A_1] & 4
& C_3A_1& 8 & 3
& (5,8) & 50 & 29
&\hatch{10}{1}&\O_{\rm min}(\cf_3)&C_2A_1&\qquad V_{-2}(\df_4)&\V_{\xi}^{\spf(4)}\otimes \Pi_{\frac12}\\
&[IV^*,C_2] & 3
 & C_2U_1& 4 & 2
& (4,\bullet) & 34 & 19
&\hatch{11}{1}&\O_{\rm min}(\cf_2)&C_1U_1&\qquad V_{-\frac32}(\af_2)&\V_{\xi}^{\spf(2)}\otimes \Pi_{\frac12}\\
\rcy &\blue{[I_0^*,C_1]} &2
 & C_1& 1 & 1
& 3 & 18 & 9
&\hatch{12}{1}&\O_{\rm min}(\cf_1)&\varnothing&\qquad\V_{\eta}&\Pi_{\frac12}\\[.5mm]
\hline\hline
&&&&&&&&&&&&\\[-4mm]
&[II^*,A_3] & 6
& A_3{\rtimes}\Z_2& 9 & 4
& 14 & 75 & 42
&\hatch{13}{1}&\H^4/\Z_3&A_2&\qquad V_{-2}(\df_4)&(\mathbb{V}_{\beta\gamma})^{\otimes 3}\otimes\Pi_{\frac{1}{3}}\\
&[III^*,A_1] & 4
& A_1U_1{\rtimes}\Z_2& 3 & 2
& (10,\bullet)& 45 & 24
&\hatch{14}{1}&\H^2/\Z_3&U_1^2&\qquad V_{-\frac43}(\af_1)&\mathbb{V}_{\beta\gamma}\otimes\Pi_{\frac{1}{3}}\\
&\green{[IV^*,U_1]} &3
& U_1& 1 & 1
& 5 & 30 & 15
&\hatch{15}{1}&\H/\Z_3&U_1&\qquad\V_{\eta}&  \Pi_{\frac{1}{3}}\\[.5mm]
\hline\hline
&&&&&&&&&&&&\\[-3.5mm]
&[II^*,A_2]& 6
& A_2{\rtimes}\Z_2& 5 & 3
& 14 & 71 & 38
&\hatch{16}{1}&\H^3/\Z_4&A_1&\qquad V_{-\frac32}(\af_2)&(\mathbb{V}_{\beta\gamma})^{\otimes 2}\otimes\Pi_{\frac{1}{4}}\\
&\green{[III^*,U_1]}&4
& U_1{\rtimes}\Z_2& 1 & 1
& 7 & 42 & 21
&\hatch{17}{1}&\H/\Z_4&U_1&\qquad\V_{\eta}&\Pi_{\frac{1}{4}}\\
&\red{[IV^*,\varnothing]}& 3
& \varnothing& 0 & 0
& - & 55/2 & 25/2
&\hatch{18}{1}&-&\varnothing&\qquad?&\\[1.5mm]
\hline\hline
&&&&&&&&&&&&\\[-3.5mm]
&\green{[II^*,U_1]}&6
& U_1{\rtimes}\Z_2& 1 & 1
& 11 & 66 & 33
&\hatch{19}{1}&\H/\Z_6&U_1&\qquad\V_{\eta}&\Pi_{\frac{1}{6}}\\
&\red{[III^*,\varnothing]}&4
& \varnothing& 0 & 0
& - & 39 & 18
&\hatch{20}{1}&-&\varnothing &\qquad?&\\
&\red{[IV^*,\varnothing]_{\sqrt{2}}}& 3
& \varnothing& 0 & 0
& - & 29 & 14
&\hatch{21}{1}&-&\varnothing&\qquad?&\\[1.5mm]
\hline\hline
\end{array}$
\HatchedCell{start1}{end1}{%
 pattern color=black!70,pattern=north east lines}
\HatchedCell{start2}{end2}{%
 pattern color=black!70,pattern=north east lines}
\HatchedCell{start3}{end3}{%
 pattern color=black!70,pattern=north east lines}
\HatchedCell{start4}{end4}{%
 pattern color=black!70,pattern=north east lines}
\HatchedCell{start5}{end5}{%
 pattern color=black!70,pattern=north east lines}
\HatchedCell{start6}{end6}{%
 pattern color=black!70,pattern=north east lines}
\HatchedCell{start7}{end7}{%
 pattern color=black!70,pattern=north east lines}
 \HatchedCell{start8}{end8}{%
 pattern color=black!70,pattern=north east lines}
 \HatchedCell{start9}{end9}{%
 pattern color=black!70,pattern=north east lines}
 \HatchedCell{start10}{end10}{%
 pattern color=black!70,pattern=north east lines}
 \HatchedCell{start11}{end11}{%
 pattern color=black!70,pattern=north east lines}
 \HatchedCell{start12}{end12}{%
 pattern color=black!70,pattern=north east lines}
 \HatchedCell{start13}{end13}{%
 pattern color=black!70,pattern=north east lines}
 \HatchedCell{start14}{end14}{%
 pattern color=black!70,pattern=north east lines}
 \HatchedCell{start15}{end15}{%
 pattern color=black!70,pattern=north east lines}
 \HatchedCell{start16}{end16}{%
 pattern color=black!70,pattern=north east lines}
 \HatchedCell{start17}{end17}{%
 pattern color=black!70,pattern=north east lines}
 \HatchedCell{start18}{end18}{%
 pattern color=black!70,pattern=north east lines}
 \HatchedCell{start19}{end19}{%
 pattern color=black!70,pattern=north east lines}
 \HatchedCell{start20}{end20}{%
 pattern color=black!70,pattern=north east lines}
 \HatchedCell{start21}{end21}{%
 pattern color=black!70,pattern=north east lines}
 }
\caption{\label{tab:summary} Select details of VOA constructions for the rank-one SCFTs. In the above $\V_{\xi}^{\spf(2n)}:=(V_{\beta\gamma})^{\otimes(2n)}$ and $\Pi_{\ell}:=\bigoplus^{\infty}_{n=-\infty}\big(V_{\partial \varphi}\otimes V_{\partial \delta}\big)e^{\ell(\delta+\varphi)}$, where the $(\beta,\gamma)$ are a pair of symplectic bosons while $(\delta,\varphi)$ are chiral bosons. The notation is explained in greater detail in the main text. }
\end{table}

\section{\label{sec:ECBHiggs}Moduli spaces of rank-one SCFTs}

The free field realizations studied below are motivated by the structure of the moduli space of vacua of the corresponding SCFTs. In this section, we provided a targeted review of salient facts about these moduli spaces, with a focus on certain special features of the rank-one case.

We begin by recalling terminology. Different branches of the moduli space of vacua are distinguished by their patterns of spontaneous symmetry breaking of superconformal $R$-symmetries. Within this framework, the \emph{Coulomb branch} is the locus where the $SU(2)_R$ symmetry is completely unbroken, while $U(1)_r$ is broken. Alternatively, the \emph{Higgs branch} is the locus where $SU(2)_R$ is broken, while $U(1)_r$ remains unbroken.\footnote{In theories with enhanced supersymmetry, this division of branches is somewhat artificial, as the $SU(2)_R$ and $U(1)_r$ symmetries are unified into a larger $R$-symmetry group. Nevertheless, for the study of $\mathcal{N}=2$-based structures such as the associated VOA this division remains pertinent.} In general, there can also be vacua where $SU(2)_R$ and $U(1)_r$ are both spontaneously broken, and these are designated as \emph{mixed branches}. In a Lagrangian theory, the Coulomb branch comprises the vacua where complex scalars in the vector multiplets have acquire vacuum expectation values, while on the Higgs branch it is the complex scalars in hypermultiplets. For a given SCFT $\mathcal{T}$ we denote the Coulomb branch by $\CB[\mathcal{T}]$ and its Higgs branch by $\HB[\mathcal{T}]$.

The \emph{rank} of an $\mathcal{N}=2$ SCFT is the complex dimension of its Coulomb branch. A pervasive conjecture that underlies much of the classification-oriented work on $\mathcal{N}=2$ SCFTs---and which we will take this for granted whenever it is relevant in this work---is that the only SCFTs with rank zero (\emph{i.e.}, without a Coulomb branch) are theories of free hypermultiplets and their discrete gaugings, making rank-one theories the ``simplest'' (from a Coulomb branch perspective) interacting examples of $\mathcal{N}=2$ SCFTs. By contrast, there are many examples of interacting SCFTs with no Higgs branch, and indeed these are of some particular interest in the context of the SCFT/VOA relation.

The Coulomb branch $\CB[\mathcal{T}]$ is a complex affine variety, further endowed with a \emph{special K\"ahler} structure, which formalizes the physical notion of Seiberg--Witten geometry.%
\footnote{See \cite{Argyres:2020nrr,Martone:2020hvy} for modern reviews on this subject.}
There is a rich story to the analysis of the constraints of special Ka\"hler geometry, and in a series of papers \cite{Argyres:2015ffa,Argyres:2015gha,Argyres:2016xua,Argyres:2016xmc,Martone:2020nsy,Argyres:2020wmq,Martone:2021ixp,Argyres:2022lah,Argyres:2022fwy,Argyres:2022puv} a Coulomb-branch-based classification program for $\mathcal{N}=2$ SCFTs has been pursued. This program has been completed for rank one, leading to the list of theories in Table~\ref{tab:summary}. As the physics of the Coulomb branch is largely complementary to that of the Schur subsector captured by the associated VOA, the relation between this classification program and vertex algebraic considerations is not entirely transparent.

The Higgs branch $\HB[\mathcal{T}]$ is a hyperk\"ahler cone. From an algebraic-geometric viewpoint (fixing a complex structure), the Higgs branch is a complex affine variety endowed with a holomorphic symplectic two-form, and a $\mathbb{C}^*$ action associated to scaling symmetry (complexified by the Cartan of $SU(2)_R$) with respect to which the symplectic form has weight $-2$.%
\footnote{A more specific mathematical formalization of the Higgs branch geometry in a fixed complex structure is as a {\it symplectic singularity}, as defined by Beauville \cite{beauville2000symplectic}. The physical status of the technical conditions entering this definition is unclear, but for broad classes of examples, including all rank-one theories, they do appear to hold.}
A key feature of the Higgs branch is that it admits a finite stratification: it is partitioned into a finite union of symplectic leaves, with each leaf corresponding to a pattern of ``Higgsing'' \cite{brieskorn1970singular, slodowy1980simple, beauville2000symplectic, Strati}. The leaves are partially ordered by inclusion of closures. In the simplest scenario there are just two leaves: the origin (superconformal point) and the larger leaf covering the entire Higgs branch except for the origin. More generally there can be intermediate leaves corresponding to partial Higgsing.   

\subsection{Rank-one SCFTs and their moduli spaces}

Many details about the catalogue of rank-one SCFTs produced by the classification methods of \cite{Argyres:2015ffa,Argyres:2015gha,Argyres:2016xua,Argyres:2016xmc,Martone:2020nsy,Argyres:2020wmq,Martone:2021ixp,Argyres:2022lah,Argyres:2022fwy,Argyres:2022puv} can be found in Table~\ref{tab:summary}. Theories coded in red are those with trivial Higgs branch. According to the Higgs Branch Conjecture of \cite{Beem:2017ooy}, the corresponding VOAs should have trivial associated variety and so are (synonymously) \emph{lisse} or \emph{$C_2$-cofinite}. Lisse VOAs can enter our free field constructions as elementary building blocks, where they correspond to nontrivial IR degrees of freedom that remain in vacua on the generic stratum of the Higgs branch. Theories coded in blue (green) are those with enhanced $\mathcal{N}=4$ ($\mathcal{N}=3$) supersymmetry, respectively.

Table~\ref{tab:summary} is organised into five blocks. The theories within each block are connected by renormalisation group flows triggered by relevant deformations. The first block contains the most well-studied rank-one theories: those realized by a single D3 brane probing a Kodaira singularity in F-theory \cite{Banks:1996nj,Douglas:1996js,Dasgupta:1996ij,Sen:1996vd}. These theories have also been dubbed the \emph{Deligne--Cvitanovi\'c (DC) series}, and their associated VOAs are affine Kac-Moody vertex algebras associated to (a subset of) the DC exceptional series of simple Lie algebras \cite{Beem:2013sza,Beem:2019tfp}. Their moduli spaces are particularly simple. Indeed:
\begin{enumerate}
    \item[(i)] They have no mixed branches.
    \item[(ii)] The Higgs branches are minimal nilpotent orbit closures for the complexified flavour groups, and so have the simplest nontrivial stratification with just two leaves: the origin and the generic stratum.
\end{enumerate}
These theories have a single pattern of Higgsing. The IR theory at a generic point of the Higgs branch (all such points being related to one another by a flavor rotation) must have rank zero, and as such consist of a collection of free hypermultiplets which in this case are all Nambu--Goldstone bosons for the spontaneously broken flavour symmetry.%
\footnote{This is a special feature of these particularly simple examples. In general, Higgsing may yield massless hypermultiplets that are not Nambu--Goldstone bosons.}
This physical picture informs the free field constructions obtained in \cite{Beem:2019tfp,Beem_2020rank2inst}, where the associated VOAs are identified as subalgebras of a collection of symplectic bosons and a half-lattice vertex algebra, as we shall review in the next section.%
\footnote{The construction works in a uniform way for a slightly larger family of VOAs labeled by all the elements of the DC exceptional series of simple Lie algebras; this includes $\mathfrak{g}_2$ and $\mathfrak{f}_4$ in addition to the Kodaira algebras $\{\mathfrak{a}_1, \mathfrak{a}_2, \mathfrak{d}_4, \mathfrak{e}_6, \mathfrak{e}_7, \mathfrak{e}_8\}$. There is no known four-dimensional interpretation of the $\mathfrak{g}_2$ and $\mathfrak{f}_4$ cases (\emph{cf}. \cite{Shimizu:2017kzs}).\label{g2f4}}

\begin{figure}[t]
	\centering
	\includegraphics[width=0.3\linewidth]{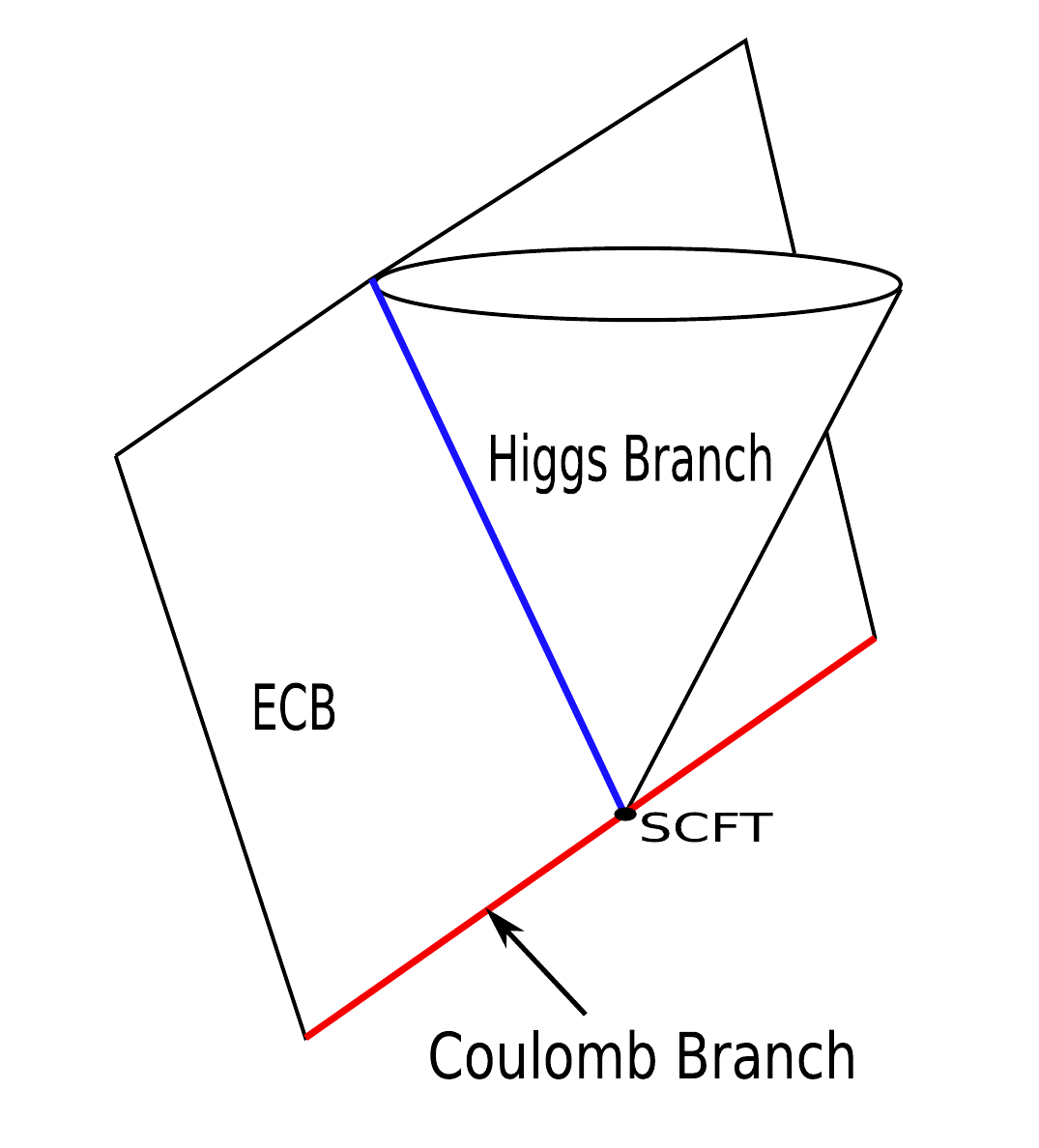}
	\caption{Sketch of different branches of Moduli Space. The SCFT sits at the tip of the cone. The red line represents the Coulomb branch which is a subvariety of the ECB. From the origin of the Coulomb branch emanates Higgs branch. The intersection of the Higgs branch and ECB is represented by the blue line.}
	\label{fig:branchesdiag}
\end{figure}

\begin{figure}[t]
\centering
\begin{tabular}{c|c|c||c|c||c||c} 
    $C_{5}\;\;$ &  $C_{3}A_1\;\;$&  $C_{2}U_1\;\;$&$A_3\rtimes\mathbb{Z}_2\;\;$& $A_1U_1\rtimes\mathbb{Z}_2\;\;$& $A_2\rtimes\mathbb{Z}_2\;\;$& $\mathcal{N}\geqslant 3$ theories\\   
{\begin{tikzpicture}
	\begin{pgfonlayer}{nodelayer}
	
		\node [style=bd] (0) at (0, 6) {};
		\node [style=bd] (1) at (0, 4) {};
		\node [style=bd] (2) at (0, 2) {};
		\node [style=none] (4) at (0.5, 5) { $e_6$};
		\node [style=none] (5) at (0.5, 3) {$c_5$};
		\end{pgfonlayer}
  
	\begin{pgfonlayer}{edgelayer}
		\draw (0) to (1);
		\draw (1) to (2);
	\end{pgfonlayer}
\end{tikzpicture}}&
{\begin{tikzpicture}
	\begin{pgfonlayer}{nodelayer}
		\node [style=bd] (0) at (0, 6) {};
		\node [style=bd] (1) at (0, 4) {};
		\node [style=bd] (2) at (0, 2) {};
		\node [style=none] (4) at (0.5, 5) { $d_4$};
		\node [style=none] (5) at (0.5, 3) {$c_3$};
		\end{pgfonlayer} 
	\begin{pgfonlayer}{edgelayer}
		\draw (0) to (1);
		\draw (1) to (2);
	\end{pgfonlayer}
\end{tikzpicture}}&
{\begin{tikzpicture}
	\begin{pgfonlayer}{nodelayer}
		\node [style=bd] (0) at (0, 6) {};
		\node [style=bd] (1) at (0, 4) {};
		\node [style=bd] (2) at (0, 2) {};
		\node [style=none] (4) at (0.5, 5) { $a_2$};
		\node [style=none] (5) at (0.5, 3) {$c_2$};
		\end{pgfonlayer}  
	\begin{pgfonlayer}{edgelayer}
		\draw (0) to (1);
		\draw (1) to (2);
	\end{pgfonlayer}
\end{tikzpicture}}&
{\begin{tikzpicture}
	\begin{pgfonlayer}{nodelayer}
		\node [style=bd] (0) at (0, 6) {};
		\node [style=bd] (1) at (0, 4) {};
		\node [style=bd] (2) at (0, 2) {};
		\node [style=none] (4) at (0.5, 5) { $d_4$};
		\node [style=none] (5) at (0.5, 3) {$h_{4,3}$};
		\end{pgfonlayer}  
	\begin{pgfonlayer}{edgelayer}
		\draw (0) to (1);
		\draw (1) to (2);
	\end{pgfonlayer}
\end{tikzpicture}}&
{\begin{tikzpicture}
	\begin{pgfonlayer}{nodelayer}
		\node [style=bd] (0) at (0, 6) {};
		\node [style=bd] (1) at (0, 4) {};
		\node [style=bd] (2) at (0, 2) {};
		\node [style=none] (4) at (0.5, 5) { $a_1$};
		\node [style=none] (5) at (0.5, 3) {$h_{2,3}$};
		\end{pgfonlayer}  
	\begin{pgfonlayer}{edgelayer}
		\draw (0) to (1);
		\draw (1) to (2);
	\end{pgfonlayer}
 \end{tikzpicture}}&
{\begin{tikzpicture}
	\begin{pgfonlayer}{nodelayer}
		\node [style=bd] (0) at (0, 6) {};
		\node [style=bd] (1) at (0, 4) {};
		\node [style=bd] (2) at (0, 2) {};
		\node [style=none] (4) at (0.5, 5) { $a_2$};
		\node [style=none] (5) at (0.5, 3) {$h_{3,4}$};
		\end{pgfonlayer}  
	\begin{pgfonlayer}{edgelayer}
		\draw (0) to (1);
		\draw (1) to (2);
	\end{pgfonlayer}
\end{tikzpicture}}&
{\begin{tikzpicture}
	\begin{pgfonlayer}{nodelayer}
		\node [style=bd] (0) at (0, 10) {};
		\node [style=bd] (1) at (0, 6) {};
		\node [style=none] (3) at (1, 8) {$h_{1,k={2,3,4,6}}$};	
		\end{pgfonlayer}  
	\begin{pgfonlayer}{edgelayer}
		\draw (0) to (1);
	\end{pgfonlayer}
\end{tikzpicture}}
\end{tabular}
\caption{Hasse diagrams for rank-one theories. The Lie algebra $\mathfrak{g}$ denotes its minimal nilpotent orbit $\mathfrak{g}\to\overline{\mathbb{O}_{\text{min}}(\mathfrak{g})}$ and $h_{n,m}=\mathbb{H}^n/\mathbb{Z}_m$. The last column is for theories with enhanced supersymmetry having Higgs branches of the form $\mathbb{H}/\mathbb{Z}_k$.}
\label{tab:Hasserank1}
\end{figure}
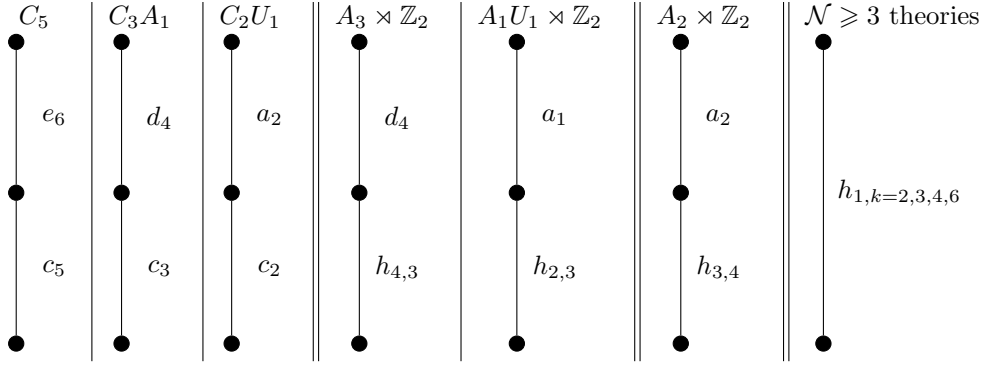

The moduli spaces of the theories in the lower four blocks of Table~\ref{tab:summary} are more involved, but still admit uniform descriptions. The relative novelty in these cases is the existence of a mixed branch. Following \cite{Argyres:2016xmc}, we will refer to the mixed branch in these cases as an \emph{extended Coulomb branch} ($\ECB$). The name captures the fact that the closure of the mixed branch contains the entire Coulomb branch $\overline{\ECB}\supset\CB$ as a smoothly embedded submanifold. For rank-one theories, the low energy physics of the $\ECB$ is that of a single massless vector multiplet and some number $n_h^\text{mix}$ of massless hypermultiplets; the total complex dimension of $\ECB$ is $2 n_h^\text{mix}+1$. One can think of the $\ECB$ as ``extending'' the Coulomb branch into $n_h^\text{mix}$ quaternionic directions. A fact related to the presence of an $\ECB$ is that the Higgs branch of these theories has a three-step stratification, as described by the Hasse diagrams in Figure \ref{tab:Hasserank1}. There is a single intermediate leaf $\IL$ between the origin and the generic locus, corresponding to the subvariety where the the $\ECB$ intersects the Higgs branch,\footnote{See figure \ref{fig:branchesdiag} for a schematic rendering of the various branches. The diagram is correct at the level of inclusion of different branches as sets, but fails to capture the subtleties of the global structure, which we discuss in more detail below.}
\begin{equation}\label{intermediateleaf}
    \overline{\ECB }\cap \HB = \IL~.
\end{equation}

The general structure of the $\ECB$ for the rank-one theories in the lower blocks of Table~\ref{tab:summary} can be anticipated reasonably well on general physical grounds. The IR theory at any smooth point of the Coulomb branch must consist of a free vector multiplet and $\quatdimtwo=n_h^\text{mix}$ decoupled massless hypermultiplets,\footnote{The hypermultiplets must be neutral with respect to the IR $\U(1)$ gauge field, otherwise giving them an expectation value would Higgs the gauge symmetry and lift the Coulomb branch, in contradiction with the existence of the $\ECB$.} possibly discretely gauged (though we will ignore this possibility as it does not appear to arise in practice at rank one). This follows from our assumption that free hypermultiplets and their discrete gaugings are the only rank-zero theories. In fact, the Higgs branch of the theory rooted at the given point in the Coulomb branch must be identified \emph{globally} (not just in an infinitesimal neighborhood) by the space $\mathbb{H}^\quatdimtwo$ as a holomorphic symplectic variety. This is because the complexified Cartan of the unbroken $\SU(2)_R$ symmetry gives an algebraic $\mathbb{C}^\ast$ action on the Higgs branch. Consequently the full \ECB\ (including the Coulomb branch itself) will have the structure of a local system over the Coulomb branch with fiber $\mathbb{H}^\quatdimtwo$. The global structure is then determined by the monodromy of the $\mathbb{H}^\quatdimtwo$ fiber around the origin of the \CB. 

There is a natural choice for the closure of an $\ECB$ described above which is a global quotient,
\begin{equation}\label{globalECB}
	\overline{\ECB} =\frac{\widetilde{\CB}\times \mathbb{H}^\quatdimtwo}{\Gamma}~,
\end{equation}
where $\G\cong\mathbb{Z}_\ell$ is the (cyclic) subgroup of $\mathrm{Sp}(\quatdimtwo)$ generated by the monodromy of the above local system, and $\widetilde{\CB}$ is a $\Gamma$-covering space of the $\CB$ \cite{Giacomelli:2020jel},
\begin{equation}
    \CB = \frac{\widetilde{\CB}}{\Gamma}~.
\end{equation}
The group $\G$ should act faithfully on $\widetilde{\CB}$ for the fiber of the local system to be $\mathbb{H}^\quatdimtwo$ rather than an orbifold thereof. We can additionally identify the intersection $\IL$ with the Higgs branch as the orbifold
\begin{equation}\label{intermediateleafOrbifold}
    \IL = \frac{\mathbb{H}^\quatdimtwo}{\Gamma} \subset \HB~.
\end{equation}
Upon partially Higgsing the UV SCFT at a point on $\IL$, the dimension of the Coulomb branch does not decrease. The residual IR theory will therefore consist of a (generally interacting) \emph{rank-one} SCFT$_{\rm IR}$ plus $\quatdimtwo$ decoupled hypermultiplets. The Higgs branch $\HB_{\rm IR}$ of SCFT$_{\rm IR}$---if it exists---which is given by the transverse slice of $\IL$ into $\HB_{\rm UV}$, must then have trivial stratification. Then the IR SCFT at a point on $\IL$ must be one of the Kodaira theories (first block of Table~\ref{tab:summary}) or the free vector multiplet SCFT (indicated as $[I_0,\varnothing]$ above).

From \eqref{globalECB} we deduce that the Coulomb branch of the IR SCFT at a point of $\IL$ should be identified with the covering space,
\begin{equation}
    \CB_{\rm IR} = \widetilde{\CB}~.
\end{equation}
This further implies a relation between the dimensions of the Coulomb branch operators of the UV and the IR theories, 
\begin{equation}
	\label{eq:cbdimuvir}
	\Delta_{\rm UV}=\ell \Delta_{\rm IR}~.
\end{equation}
Equation \ref{eq:cbdimuvir} restricts the allowed pairs of $\Delta_{\rm UV}$ and $\Delta_{\rm IR}$. Since $\Delta_{\rm IR}\in \left\{1,\frac{6}{5},\frac{4}{3},\frac{3}{2},2,3,4,6\right\}$ (with $\Delta_{\rm IR} =1 $ corresponding to the free vector multiplet) we can easily list the allowed pairs $(\Delta_{\rm IR},\Delta_{\rm UV})$. They are given in Table~\ref{tableoptions}. Remarkably, all but one of the naively allowed pairs is realized by one (and only one!) of the non-Kodaira rank-one SCFTs,\footnote{Excluding SCFTs with trivial Higgs branches, which are  outside the scope of this geometric discussion.} with the four distinct values of $\ell$ corresponding to the four non-Kodaira blocks. The unaccounted-for option (red in Tab~\ref{tableoptions} with $\ell=5$) can be ruled out if we further require that $\Gamma$ acts faithfully on the electromagnetic lattice, which restricts it to be a cyclic subgroup of $\mathrm{SL}(2,\mathbb{Z})$ allowing only $\mathbb{Z}_{\ell=2,3,4,6}$. Alternatively, we have applied the free-field techniques described later in this paper to attempt to construct a candidate vertex algebra associated to the choice with $\ell=5$ and any such construction fails; this is an interesting instance of the rigidity of our vertex algebraic methods apparently ruling out candidate SCFT data in a bottom-up construction. All in all, though it is entirely non-obvious that a one-to-one correspondence of this type should hold, we nevertheless arrive at a strikingly simple way to rationalize the complicated structure of Table~\ref{tab:summary}.

\begin{table}[t]
	\centering
	\begin{tabular}{ |c| c| }
		\hline
		$\ell$ & pairs of $(\Delta_{\rm IR},\Delta_{\rm UV})$ \\   
		\hline
		\rule{-0.7ex}{3ex} 
		2 & (1,2), ($\tfrac{3}{2}$,3), (2,4), (3,6)\\
		3 & (1,3), ($\tfrac{4}{3}$,4), (2,6)\\
		4 & (1,4), ($\tfrac{3}{2}$,6)\\
        \textcolor{red}{5} & \textcolor{red}{($\tfrac{6}{5}$,6)}\\
		6 & (1,6)\\
		\hline
	\end{tabular}
	\caption{\label{tableoptions}Table listing the allowed pairs for $\Delta_{IR}$ and $\Delta_{UV}$. }
\end{table}

\section{\label{Delignereview}Review of geometrically inspired free field realizations}

In this section we review the main conceptual and technical features of ``geometric'' free field realizations of vertex operator algebras associated to four dimensional $\mathcal{N}=2$ SCFTs. In the SCFT/VOA setting, this line of study was initiated in \cite{Beem:2019tfp} (see also \cite{Bonetti:2018fqz} for related ideas that have a somewhat different geometric flavour) and was further developed in \cite{Beem_2020rank2inst,Beem:2021jnm,Beem:2022vfz}. 

In physical terms, the general idea is to construct the VOA of a given SCFT using ``simpler'' building blocks corresponding to the low energy degrees of freedom on some stratum/symplectic leaf of the Higgs branch. In other words, the contention is that, at least in certain circumstances, it should be possible to ``invert'' Higgsing at the level of the associated VOA.

For the case of the generic stratum (``maximal Higgsing''), these degrees of freedom should roughly be encoded in $n=\text{dim}_{\mathbb{C}}(\mathcal{M}_{\text{Higgs}})$ chiral bosons associated to the massless half-hypermultiplets on a generic point of the \HB, along with $k$ symplectic fermions associated to massless vector multiplets if an abelian gauge group $U(1)^k$ remains unbroken at generic points of the \HB, and a further $C_2$ co-finite VOA in the case of a residual interacting SCFT with trivial \HB.

It is often convenient to consider partial Higgsing as well, in which case the VOA of the given theory is realized as a subVOA of the product of $\mathbb{V}(\mathcal{T}_\text{IR})\otimes \mathcal{V}_{\text{free}}$, where $\mathcal{V}_{\text{free}}$ includes all chiral bosons and symplectic fermions in the IR and $\mathbb{V}(\mathcal{T}_\text{IR})$ is the VOA of the interacting theory (or product of interacting theories) obtained in the IR. It may be possible to iterate this procedure and obtain a free field realization for $\mathbb{V}(\mathcal{T}_\text{IR})$. When such an iteration is possible the key operation to understand will be the case of \emph{minimal Higgsings}.

In practice, such free field realizations have been worked out mainly for a very special class of Higgsings in which the construction envisioned above can be implemented in a fairly uniform fashion. These are cases where the (partial) Higgsing in question can be realized by assigning a non-zero expectation value to a single Higgs branch generator\footnote{One of the simplest examples that does not belong to this class is the case of $\mathcal{N}=4$ SYM with gauge group $SU(3)$, see \emph{e.g.}, \cite{Bonetti:2018fqz}.}. We will denote the corresponding chiral ring generator by $\mathsf{e}$ with the warning that it is not necessarily an $\mathfrak{sl}(2)$ nilpotent generator\footnote{In these cases our construction could be regarded as an inverse Drinfel'd--Sokolov reduction.}. We can now consider the open patch $\mathcal{U}_{\mathsf{e}}:=\{\mathsf{e}\neq 0 \}\subset\mathcal{M}_{\text{Higgs}}$. This patch is invariant under the scaling $\mathbb{C}^*$ action on the Higgs Branch. An observation which is crucial for the geometric construction is that this big open patch can be identified with
\begin{equation}\label{Ueidentification}
    \mathcal{U}_{\mathsf{e}}\simeq\frac{ T^*(\mathbb{C}^*)\times \mathbb{C}^{2(\quatdim-1)} \times \mathcal{M}_H[\mathcal{T}_{IR}]}{\mathbb{Z}_{\ell}}~,
\end{equation}
where the $T^*(\mathbb{C}^*)$ factor\footnote{Here $T^*M$ denotes the cotangent bundle of $M$.} has coordinates $\mathsf{e}^{1/\ell}$ for the $\mathbb{C}^*$ direction and $\mathsf{h}$ for the fiber direction and we denote the coordinates of $\mathbb{C}^{2(\quatdim-1)}$ by $(\betaclass_a,\gammaclass^b)$ with $a,b=1,\dots,\quatdim-1$. The action of the group $\mathbb{Z}_{\ell}$ on the factors in the numerator is given by
\begin{equation}\label{Zellactiononfreefields}
    (\mathsf{e}^{1/\ell},\mathsf{h})\mapsto (\omega_{\ell}\,\mathsf{e}^{1/\ell},\mathsf{h})\,,\qquad
    (\betaclass,\gammaclass)\mapsto (\omega_{\ell}^{}\,\betaclass,\omega_{\ell}^{-1}\gammaclass)\,,\qquad
    \omega_{\ell}^{}=e^{2\pi i /\ell}~,
\end{equation}
on the coordinates and by a $\mathbb{Z}_{\ell}$ automorphism on $\mathcal{M}_H[\mathcal{T}_{IR}]$. Notice that when $\ell=2$ all the $\mathbb{C}^{2(\quatdim-1)}$ directions get the same factor. The relation \eqref{Ueidentification} provides an isomorphism of Poisson varieties, where we have the canonical Poisson brackets $\{\mathsf{h}, \mathsf{e}^{1/\ell}\} =\mathsf{e}^{1/\ell}$ and $\{\gammaclass^a,\betaclass_b\} =\delta^a_b$. The identification \eqref{Ueidentification} can be shown in the examples in which we know well the Higgs branch, like in the cases analyzed in \cite{Beem:2019tfp,Beem_2020rank2inst,Beem:2021jnm} reviewed below. Alternatively, we could take \eqref{Ueidentification} as an approximation of the Higgs branch we want to construct.

The intuitive picture behind \eqref{Ueidentification} is simple: from each point of the subset $T^*(\mathbb{C}^*)$ sprouts the product of $\mathbb{C}^{2(\quatdim-1)}$, associated to free hypermultiplets in the IR, with the Higgs branch of the residual IR effective theory $\mathcal{M}_H[\mathcal{T}_{IR}]$. Via the identification \eqref{Ueidentification} one can express the elements of the Higgs branch chiral ring in terms of the building blocks on the right hand side but in general it remains an open problem to determine which holomorphic function on \eqref{Ueidentification} has a counterpart on the whole HB. The simplest example being $\mathsf{e}^{-1}$ which is a function on $\mathcal{U}_{\mathsf{e}}$ with no counterpart in the HB chiral ring.

Generalized free field realizations of the associated VOAs should be regarded as a sort of ``chiralization'' or ``affinization'' of the geometric construction above. To proceed, we introduce the VOAs associated to the low energy degrees of freedom in the following way. The affinization of the $T^*(\mathbb{C}^*)$ coordinates $(\mathsf{e}^{1/\ell},\mathsf{h})$ is achieved by introducing two chiral bosons $\delta$ and $\varphi$ satisfying the OPEs
\begin{equation}\label{deltaphiOPE}
    \delta(z_1)\delta(z_2)\sim \langle \delta,\delta\rangle\,\log z_{12}~,\qquad
    \varphi(z_1)\varphi(z_2)\sim \langle \varphi,\varphi\rangle\,\log z_{12}~,\qquad
    \delta(z_1)\varphi(z_2)\sim0~,
\end{equation}
with $z_{12}=z_1-z_2$ and $\langle \varphi,\varphi\rangle+\langle \delta,\delta\rangle=0$. In terms of the latter we define the isotropic lattice vertex algebra%
\footnote{These vertex algebras are constructed after having identified an isotropic sublattice, i.e. a sublattice on which the non-degenerate bilinear form $\langle -,- \rangle$ vanishes. Roughly speaking, only chiral bosons associated to the isotropic directions can be put in exponent, see \cite{Berman2001RepresentationsOA,Adamovic:simple_affine} for more details.}
\begin{equation}\label{Pi1elldef}
    \Pi_{\frac{1}{\ell}}:=\bigoplus_{n=-\infty}^{\infty}\left(V_{\partial \varphi}\otimes V_{\partial \delta}\right)e^{\frac{n}{\ell}(\delta+\varphi)}~,
\end{equation}
where $V_j$ is the $\widehat{\mathfrak{gl}(1)}$ affine current algebra associated with the current $j$. The condition $\langle \varphi,\varphi\rangle+\langle \delta,\delta\rangle=0$ guarantees that the exponentials have regular OPEs with each other.

The next ingredients correspond to the affinization of the $\mathbb{C}^{2(\quatdim-1)}$ factor in \eqref{Ueidentification}. These are symplectic bosons $\xi_m$, with $m=1,\dots,2(\quatdim-1)$, with OPEs
\begin{equation}
    \xi_m(z_1)\xi_n(z_2)\sim\frac{\Omega_{mn}}{z_{12}}~,
\end{equation}
where $\Omega_{mn}$ is a non degenerate symplectic form. We will sometimes split the symplectic bosons as $\xi_m=(\beta_a,\gamma^a)$ where $a,b=1,\dots,(\quatdim-1)$ and OPEs
\begin{equation}
    \beta_a(z_1)\gamma^b(z_2)\sim\frac{-\delta_a^b}{z_{12}}\,,\qquad
    \beta_a(z_1)\beta_b(z_2)\sim\gamma^a(z_1)\gamma^b(z_2)\sim 0~.
\end{equation}
The other free field ingredient is given by symplectic fermions\footnote{They appear in the cases of $\mathcal{N}>2$ theories discussed in Section \ref{sec:moresusy} and have featured in the case studied in \cite{Beem:2021jnm}.} $\eta_I$, $I=1,\dots, 2r$ with OPEs
\begin{equation}\label{symplecticFermions}
    \eta_I(z_1)\eta_J(z_2)\sim\frac{\omega_{IJ}}{z_{12}^2}~,
\end{equation}
where $\omega_{IJ}$ is a non degenerate symplectic form. To summarize, the free field realization, in the cases in which the identification \eqref{Ueidentification} holds, consists in realizing $\mathbb{V}[\mathcal{T}]$ as a subVOA, namely\footnote{In all the examples treated here the four factors will never all be simultaneously non-trivial.}
\begin{equation}\label{SUBVOAgeneral}
    \mathbb{V}[\mathcal{T}]\subset \left(\Pi_{\frac{1}{\ell}} \otimes \mathbb{V}_{\xi}\otimes \mathbb{V}_{\eta} \otimes \mathbb{V}[\mathcal{T}_{IR}]\right)^{\mathbb{Z}_{\ell}}~,
\end{equation}
where $(\dots)^{\Gamma}$ denotes the $\Gamma$-invariant subVOA. The aim of the game is to identify the appropriate subVOA $\mathbb{V}[\mathcal{T}]$. The strategy that we adopt to achieve this is the following. We construct a set of generators that we either know exist from a partial knowledge of $\mathbb{V}[\mathcal{T}]$ (or of the Higgs branch) or we assume they exist, and generate the whole algebra starting from these. It should be noticed that in general the complete list of strong generators will be bigger than the starting set. The explicit construction of the generator will be guided by the knowledge of quantum numbers related to the flavor symmetry and the conformal weight.

An important feature which is expected to hold for these generalized free field realizations is that they realize \emph{simple} vertex algebras, so any null states naively present in a strong-generators-and-OPEs presentation should vanish identically in the free field realization.%
\footnote{We note that typically, the VOAs associated to four dimensional SCFTs are not obviously simple quotients of some universal VOA as is the case for, \emph{e.g.}, current algebras. Rather, they tend be rigid VOAs which, when presented in terms of strong generators and OPEs, are only well-defined when certain null operators are set to zero.}
The free field realization also allows to propose a canonical prescription to recover the four-dimensional R-filtration of these VOAs \cite{Beem:2017ooy,Beem:2019tfp} which works as follows. We assume that we know the R-filtration of the IR VOA $\mathbb{V}[\mathcal{T}_{IR}]$ appearing in \eqref{SUBVOAgeneral}. We then supplement it with an R-grading for the free field ingredients that we summarize in Table \ref{tab:Rdegree}.
\begingroup
\begin{table}[t]
\centering
\renewcommand{\arraystretch}{1.4}
\begin{tabular}{| c||c | c |c | c|c |c |c|c  | }
\hline
& $\xi=(\beta,\gamma)$ & $\partial(\delta-\varphi)$ & $e^{\frac{n}{\ell}(\delta+\varphi)}$ & $\partial(\delta+\varphi)$ & $\eta_1$ & $\eta_2$  & $\partial$ \\
\hline \hline
$R$ & $\frac 12$ &$1$ & $\frac n2$ & $0$& $\frac 12$ & $\frac 12$ & $0$ \\
\hline
$h-R$ & $0$ & $0$ & $0$& $1$ & $\frac 12$ & $\frac 12$ & $1$ \\
\hline
$r$ & $0$ & $0$ & $0$ & $0$ & $+\frac 12$ & $-\frac 12$ & $0$ \\
\hline
\end{tabular}
 \caption{$R$ degrees of the free field ingredients.}
  \label{tab:Rdegree}
\end{table}
\endgroup

In the following we will highlight how this strategy works in the examples of rank-one Deligne theories and their rank-two generalizations. We conclude this section emphasizing some important features of these examples, most of which will continue to hold for the remaining rank-one theories with some interesting twists.

\subsection{Example 1: rank-one Kodaira/Deligne--Cvitanovi\'c series}

The VOAs associated to the Kodaira rank-one theories are affine Kac--Moody VOAs for the corresponding simple Lie algebras at level $k=-\frac{h^\vee}{6}-1$, where $h^\vee$ is the dual Coxeter number of $\mathfrak{g}$. These VOAs naturally sit within a slightly larger family corresponding to the Deligne--Cvitanovi\'c exceptional series of simple Lie algebras, 
\begin{equation}
    \mathfrak{a}_0\subset\mathfrak{a}_1\subset\mathfrak{a}_2\subset\mathfrak{g}_2\subset\mathfrak{d}_4\subset\mathfrak{f}_4\subset\mathfrak{e}_6\subset\mathfrak{e}_7\subset\mathfrak{e}_8~,
\end{equation}
where for $\mathfrak{a}_0$ the associated VOA is the Virasoro vertex algebra at central charge $c=-22/5$. The Higgs branches (associated varieties) for these SCFTs (VOAs) are the closures of the minimal nilpotent orbits of the relevant Lie algebra, namely,
\begin{equation}
    \mathcal{M}_H=\overline{\mathbb{O}_{\text{min}}(\mathfrak{g})}~.
\end{equation}
We recall that the minimal nilpotent orbit is the orbit of the nilpotent element $e_{\theta}\in\mathfrak{g}$ associated to the highest root $\theta$ of $\mathfrak{g}$. One has the following decomposition for $\mathfrak{g}$,
\begin{equation}\label{gdecompositionDeligne}
    \mathfrak{g}=\mathfrak{g}^{\natural}\oplus \mathfrak{sl}(2)_{\theta} \oplus (\mathfrak{R},2)~,
\end{equation}
where $\mathfrak{sl}(2)_{\theta}=\langle e_\theta,f_\theta,h_\theta\rangle$, $\mathfrak{g}^{\natural}$ is the commutant of  $\mathfrak{sl}(2)_{\theta}$ in $\mathfrak{g}$ and $\mathfrak{R}$ is a specific quaternionic representation of $\mathfrak{g}^{\natural}$. These nilpotent orbit closures are smooth everywhere apart from the origin; correspondingly there is only one pattern of non-trivial Higgsing for these theories, and in the IR the theory becomes that of free hypermultiplets. At a generic point of the Higgs branch the symmetry is broken spontaneously to $\mathfrak{g}^{\natural}$ with one full hyper transforming as a singlet under this symmetry and the rest transforming in the representation $\mathfrak{R}$ introduced in \eqref{gdecompositionDeligne}.%
\footnote{This is the unbroken symmetry as a holomorphic symplectic variety. As a hyperk\"ahler manifold, the unbroken symmetry contains an $\mathfrak{sl}(2)_{\bar{R}}=\text{diag}(\mathfrak{sl}(2)_{R},\mathfrak{sl}(2)_{\theta})$ which matches the $SU(2)_R$ symmetry of the IR fixed point.}
Indeed, one can verify the relation $\text{dim}_{\mathbb{C}}(\mathbb{O}_{\text{min}}(\mathfrak{g}))=2+\text{dim}\,\mathfrak{R}$. 

These nilpotent orbits possess open charts of the form \eqref{Ueidentification} where the factor $\mathcal{M}_H(\mathcal{T}_{IR})$ is absent and $\ell=2$. According to the general scheme advertised in equation \eqref{SUBVOAgeneral}, these current algebras are realized as subalgebras of free field vertex algebras that chiralize the coordinate rings of these open charts,
\begin{equation}\label{VsubVOADeligne}
    \mathcal{V}^{(1)}_{\mathfrak{g}}=V_{-\frac{h^\vee}{6}-1}(\mathfrak{g})\subset \left(\Pi_{\frac{1}{2}} \otimes \mathbb{V}_{\xi}\right)^{\mathbb{Z}_2}~,
\end{equation}
where $\mathbb{V}_{\xi}$ is the VOA of $\text{dim}\,\mathfrak{R}$ symplectic bosons. Certain generators of the current algebras take a simple form in terms of free fields. This is explained by the fact that part of the UV symmetry, namely the $U(1)$ generated by $h_{\theta}$ and the unbroken symmetry $\mathfrak{g}^{\natural}$, is realized linearly and in a known manner in the IR. Consequently the $U(1)$ charge assignments and $\mathfrak{g}^{\natural}$ transformation properties of both the free field ingredients on the left hand side of \eqref{VsubVOADeligne} and the current algebra generators are fixed. Additionally, scaling dimensions on both sides of \eqref{VsubVOADeligne} are understood, leaving unique candidates
\footnote{Up to a field redefinition of $\delta$ and $\varphi$ discussed in \cite[footnote 16]{Beem:2019tfp}.}
for the $\mathfrak{sl}(2)_{\theta}$ highest weight generators (\emph{i.e.}, elements that are annihilated by the action of the generator $e$) in \eqref{gdecompositionDeligne},
\begin{equation}\label{eandeDeligne}
    e_{\theta}(z)=e^{\delta+\varphi}~,\qquad 
    e_{A}(z)=\xi_A\,\,e^{\frac{\delta+\varphi}{2}}~,\qquad
    J^{\natural}_{\alpha}(z)=T_{\alpha}^{AB}\,\xi_A\xi_B~,
\end{equation}
where the tensor $T_{\alpha}^{AB}$ is determined by the embedding of $\mathfrak{g}^{\natural}\subset \mathfrak{sp}(\mathfrak{R})$, which in turn is specified by the fact that $\mathfrak{R}$ is an irreducible representation of $\mathfrak{g}^{\natural}$. The generator corresponding to $h_\theta$ takes the form%
\footnote{Though the value of the level $k$ is fixed as in \eqref{VsubVOADeligne}, we keep it as an indeterminate here as similar expressions appear in other examples.}
\begin{equation}\label{hgeneratorrank1}
    h(z)=k\partial\varphi~.
\end{equation}
Completing the construction requires expressions for the remaining generators in \eqref{gdecompositionDeligne}, namely the $\mathfrak{sl}(2)_{\theta}$ descendants denoted as $f_{\theta}(z)$ and $f_{A}(z)$. There are several ways to determine these. One is to make the most general Ansatz in terms of free fields compatible with the charge assignment and fix coefficients by requiring that the appropriate OPEs are satisfied. In this case it is more convenient to realize $f_\theta(z)$ using a general construction due to \cite{Semikhatov:1993pr,Adamovic:2004zi} and extract $f_A(z)$ from the OPEs of $f_\theta(z)$ with $e_A(z)$. In this scheme, $f_\theta(z)$ has the form\footnote{Here and in the following we adopt the conventions for normal ordering used in \cite{Beem:2021jnm} which differ from the one employed in \cite{Beem:2019tfp}.}
\begin{equation}\label{fthetaDELIGNE}
    f_{\theta}(z)= \Big(S^{\natural}-\left((\tfrac{k}{2}\,\partial\delta )^2-\tfrac{k(k+3)}{2}\partial^2 \delta\right)\Big)e^{-(\delta+\varphi)}~.
\end{equation}
The operator $S^\natural$, which does not involve the free fields $\delta$ and $\varphi$, needs to satisfy certain properties. Firstly, in order for the $\mathfrak{sl}(2)_{\theta}$ OPEs to come out right one requires
\begin{equation}\label{SnaturalSnaturalOPES}
    S^{\natural}(z_1)S^{\natural}(z_2) \sim (k+2)\left(\frac{(k+2)\tfrac{c^{\natural}}{2}}{z_{12}^4}+\frac{2S^{\natural}(z_2) }{z_{12}^2}+\frac{\partial S^{\natural}(z_2) }{z_{12}}\right)\,,\qquad
    c^{\natural}=1-6\tfrac{(k+1)^2}{k+2}~.
\end{equation}
Additionally, for the other OPEs to work we need $S^{\natural}$ to have regular OPEs with the $J^\natural$ currents. The resulting explicit expressions for $S^{\natural}$ and $f_A$ can be found in \cite{Beem:2019tfp}. It is interesting to look at what happens if one tries to apply the free field construction for a generic pair $(\mathfrak{g},k)$. One finds that the construction works only for the cases such that $\mathcal{W}_k(\mathfrak{g},f_{\theta})\simeq \mathbb{C}$ which are listed in \cite[Theorem 7.2]{Arakawa:2015jya}. Apart from the DC exceptional series at the appropriate level, the list includes only $\mathfrak{c}_n$ at level $k=-\frac{1}{2}$ (which corresponds to a $\mathbb{Z}_2$ quotient of symplectic bosons) and $\mathfrak{a}_1$ at the critical level $k=-2$.

To conclude this example, let us recall how the level $k$ of the flavor symmetries and the central charge $c$ are quickly obtained from the free field realization. The level of the $\mathfrak{g}^{\natural}$ is obtained by recalling that the symmetry is a subalgebra of the $\mathfrak{sp}(\mathfrak{R})$ transformations of the symplectic bosons, so that 
\begin{equation}\label{Jnaturalrank1Deligne}
    J^{\natural}=J^{\natural}_{\xi}\,,\qquad
    \Longrightarrow\qquad
    k^{\natural}=-\tfrac{1}{2}\,I_{\mathfrak{g}^{\natural}\hookrightarrow \mathfrak{sp}(\mathfrak{R})}~,
\end{equation}
where $I$ denotes the embedding index.%
\footnote{We recall that the embedding index is defined as
\begin{equation}
    I_{G\hookrightarrow H}=\frac{\sum_i T(\bf{r}_i)}{T(\bf{r})}~,
\end{equation}
where $T(\bf{r})$ is the quadratic index of representation $\bf{r}$. While the definition above employs a specific representation and its branching $\bf{r} \rightarrow \sum_i \bf{r}_i$, the index is independent of the choice of $\bf{r}$. The levels of the embedded current algebra is given by $k_{G}=I_{G\hookrightarrow H}k_{H}$. (See, \emph{e.g.}, \cite{Argyres:2007cn} for more details.)}
The stress tensor, which in these examples happens to coincide with the Sugawara stress tensor, takes the form
\begin{equation}
    T=T_{\delta,\varphi}+T_{\xi}~.
\end{equation}
Here $T_{\xi}=\partial\xi\Omega^{-1}\xi$ is such that $\xi_A$ are Virasoro primaries of dimension $\frac{1}{2}$ and its contribution to the central charge is given by $c_{\xi}=-\frac{1}{2}\text{dim}\,\mathfrak{R}=2-h^{\vee}$. The contribution of the chiral bosons $(\delta,\varphi)$ is more interesting. The associated stress tensor takes the form
\begin{equation}\label{stresstensordeltaphi}
    T_{\delta,\varphi}= T_\delta+T_{\varphi}\,,\qquad\,
    T_\delta=\frac{1}{2\langle\delta,\delta\rangle}((\partial\delta)^2-\alpha\,\partial^2\delta)\,,\qquad 
    T_\varphi=\frac{1}{2\langle\varphi,\varphi\rangle}(\partial\varphi)^2~.
\end{equation}
It follows that $c_{\delta}=1-\frac{3 \alpha^2}{\langle \delta,\delta \rangle}$, $c_{\varphi}=1$. We have assumed that the chiral boson $\varphi$ has zero background charge, which is related to the fact that $\partial \varphi$ is the generator of a physical $U(1)$ current algebra. From the requirement that $e_{\theta}(z)=e^{(\delta+\varphi)(z)}$ has conformal dimension $1$, and recalling that $\langle\delta,\delta\rangle=-\langle\varphi,\varphi\rangle=-\frac{2}{k}$, we find $\alpha=2$ and we conclude that
\begin{equation}\label{cdeltaphiSnatural}
    c_{\delta,\varphi}=c_{\delta}+c_{\varphi}=2+6k~.
\end{equation}
Putting everything together we reproduce the correct value of the central charge for the DC exceptional series
\begin{equation}
    c=c_\xi+c_{\delta,\varphi}=(2-h^\vee)+2+6(-\tfrac{1}{6}h^\vee-1)=-2-2h^\vee~.
\end{equation}

\subsection{Example 2: rank-two Deligne series}

We will now briefly recall the free field realizations of the rank-two Deligne series \cite{Beem_2020rank2inst}. These cases are particularly interesting since they are some of the simplest examples for which the moduli space of vacua contains an ECB. Accordingly, there is a non-trivial rank preserving Higgsing, so that the theory obtained in the IR is the product of two rank-one Deligne theories. 

The \HB\ is the (centered) two-instanton moduli space for the corresponding group. These enjoy $\mathfrak{sl}(2) \oplus \mathfrak{g}$ symmetry and the singular locus $\IL$ is identified with the subspace where the $\mathfrak{g}$ symmetry is unbroken, which is isomorphic to $\mathbb{C}^2/\mathbb{Z}_2$. The relevant open chart \eqref{Ueidentification} associated to partial Higgsing takes the form
\begin{equation}\label{UeidentificationRank2Deligne}
    \mathcal{U}_{\mathsf{e}}\simeq\frac{ T^*(\mathbb{C}^*)\times \widetilde{\mathcal{M}}_{\mathfrak{g}}^{(1)}\times \widetilde{\mathcal{M}}_{\mathfrak{g}}^{(1)}}{\mathbb{Z}_{2}}~,
\end{equation}
where the $\mathbb{Z}_2$ act on $T^*(\mathbb{C}^*)$ as in \eqref{Zellactiononfreefields} and exchanges the two copies of the instanton moduli spaces $\widetilde{\mathcal{M}}_{\mathfrak{g}}^{(1)}$. The symmetry preserved along the partial Higgsing flow is just $\mathfrak{g}$, and the distinguished $U(1)$ that acts on $\mathsf{e}$ is generated by the Cartan of the $\mathfrak{sl}(2)$ factor of the flavor symmetry. According to the general scheme, the VOA associated to the rank-two Deligne theories as will be realized as%
\footnote{As for the rank-one DC theories, the case of $\mathfrak{a}_0$ is a little bit different. In this case the two copies of the current algebras should be replaced with two copies of Virasoro at level $c=-22/5$. See \cite{Beem_2020rank2inst} for more details.}
\begin{equation}\label{VsubVOADelignerank2}
    \mathcal{V}^{(2)}_{\mathfrak{g}}\subset\left(\Pi_{\frac{1}{2}} \otimes V_{-\frac{h^\vee}{6}-1}(\mathfrak{g})\otimes V_{-\frac{h^\vee}{6}-1}(\mathfrak{g})\right)^{\mathbb{Z}_2}~.
\end{equation}
We denote by $\mathcal{J}^{\text{IR}}_1$ and $\mathcal{J}^{\text{IR}}_2$ the generators of the current algebra factors in \eqref{VsubVOADelignerank2}. Again some of the generators of the VOA are very easy to construct. The reason is again that we know the $\mathfrak{g}$ transformation properties, $U(1)$ assignment, and conformal grading of the free field ingredients and we have to select $\mathbb{Z}_2$ invariant elements. With these observations in mind, the free field realization of the $\mathfrak{sl}(2)$ primaries is easy to determine:
\begin{equation}\label{eandwDelignerank2}
    e(z)=e^{\delta+\varphi}~,\qquad \mathcal{W}_+(z)=\,\left(\mathcal{J}^{\text{IR}}_1-\mathcal{J}^{\text{IR}}_2\right)\,e^{\frac{\delta+\varphi}{2}}~,\qquad \mathcal{J}_{\mathfrak{g}}(z)=\mathcal{J}^{\text{IR}}_1+\mathcal{J}^{\text{IR}}_2~,
\end{equation}
where we leave ${\rm Adj}(\mathfrak{g})$ indices implicit. The $\mathfrak{sl}(2)$ Cartan generator takes the same form as in \eqref{hgeneratorrank1}, $h(z)=k\partial\varphi$. To complete the free field realization it is sufficient to construct the lowering $\mathfrak{sl}(2)$ generator $f(z)$. This is done again with the formula \eqref{fthetaDELIGNE} for an appropriate choice of $S^{\natural}$. Now the condition that $S^{\natural}$ has regular OPE with $\mathcal{J}_{\mathfrak{g}}(z)$ immediately singles out a unique element proportional to the stress tensor for the diagonal coset VOA,
\begin{equation}\label{SnaturalRank2}
    S^{\natural}(z)=(k_{\mathfrak{sl}(2)}+2)(T_1^{\text{Sug}}+T_2^{\text{Sug}}-T_{12}^{\text{Sug}})~,
\end{equation}
where $T_1^{\text{Sug}}$, $T_2^{\text{Sug}}$ and $T_{12}^{\text{Sug}}$ are Sugawara stress tensors built using $\mathcal{J}^{\text{IR}}_1$, $\mathcal{J}^{\text{IR}}_2$ and $\mathcal{J}^{\text{IR}}_1+\mathcal{J}^{\text{IR}}_2$ respectively.%
\footnote{Notice that the normalization in \eqref{SnaturalRank2} has been modified from that of \eqref{SnaturalSnaturalOPES}. See \cite{Beem_2020rank2inst} for more details.}

It is instructive to see how the levels of the current algebras and the central charge $c$ are uniquely fixed from the free field construction. The level of the $\mathfrak{g}$ symmetry is immediately obtained from the expression of $\mathcal{J}_{\mathfrak{g}}$ given in \eqref{eandwDelignerank2} to be the sum of the levels of the rank-one theories. The central charge is the sum of the contribution from the chiral bosons $(\varphi,\delta)$, which follows the general formula \eqref{cdeltaphiSnatural}, and from the two copies of the rank-one theories in the IR. The way in which the level $k_{\mathfrak{sl}(2)}$ is fixed by the free field construction is a little more subtle: it comes from the requirement that the most singular term in the OPE of the lowering operator $f(z)$ given in \eqref{fthetaDELIGNE} with $\mathcal{W}_+(w)$ in \eqref{eandwDelignerank2} is a simple pole, \emph{i.e.}, that $\mathcal{W}_+(w)$ be an AKM primary. This can also be interpreted as the requirement that there be no states of conformal weight $1/2$, which would necessarily correspond to free fields in four dimensions.

An alternative, intrinsically four dimensional, derivation of central charges proceeds as follows. As reviews in Appendix \ref{appB}, since $U(1)_r$ is unbroken on the HB, the anomaly $\Tr(r^3)$ can be matched on different strata. This implies a relation between the central charges of the UV theory and the one of the IR theory obtained after higgsing which takes the form
\begin{equation}\label{cminusaformual}
    24(c-a)_{4d}^{\text{UV}}=
    24(c-a)_{4d}^{\text{IR}}
    +\text{dim}_{\mathbb{H}}(\mathcal{M}_{H}(\mathcal{T}_{\text{UV}}))
    -\text{dim}_{\mathbb{H}}(\mathcal{M}_{H}(\mathcal{T}_{\text{IR}}))
    -n_{\text{v}}~\,,
\end{equation}
where $n_{\text{v}}$ denotes the number of vector multiplets supported on the stratum associated to the choice of Higgsing.
Additionally, the Shapere--Tachikawa formula \cite{Shapere:2008zf} relates a second combination of conformal anomaly coefficients to the set of scaling dimensions  $\{\Delta(u_i)\}$ of the generators  of the CB in the following way
\begin{equation}\label{STformula}
    4(2a-c)_{4d}=\sum_{i=1}^{\text{rank}} (2\Delta(u_i)-1)~.
\end{equation}
According to the discussion of the previous section (see \cite{Giacomelli:2020jel} for more details), in the case of minimal Higgsing in the presence of an $\ECB$, the set of scaling weights of the Coulomb branch generators in the UV are simply related to those in the IR. In the case of rank-two Deligne theories, this gives $(\Delta(u_1),\Delta(u_2))=(\Delta,2\Delta)$ where $\Delta$ is the scaling dimension of the Coulomb branch generator of the rank-one theory. From this, together with the absence of additional vector multiplets, one immediately derives the value of $a$ and $c$ from \eqref{cminusaformual} and \eqref{STformula}.

\subsection{General remarks}

Let us make a few general remarks and observations concerning these examples that will help guide us through the free field constructions for the remaining rank-one theories. 

\begin{itemize}
    \item At a point on the singular locus associated to the choice of Higgsing, the flavor symmetry is spontaneously broken.\footnote{In all examples considered the flavor symmetry is non trivial.} We denote the semi-simple part of the unbroken subalgebra of $\mathfrak{g}_{\text{UV}}$ by $\unbrSYMM$. This symmetry is visible all the way from the UV to the IR and it is consequently very useful for the free field construction. The generators of the corresponding affine currents take the simple form
\begin{equation}\label{JnaturalasSUMgeneral}
    \mathcal{J}_{\text{UV}}=\mathcal{J}_{\xi}+\mathcal{J}_{\text{IR}}~.
\end{equation}
In this formula $\mathcal{J}_{\xi}$ generate a $\unbrSYMM$ subalgebra of $\mathfrak{sp}(V)$ determined by the representation (possibly trivial) of the symplectic bosons/free hypermultiplets under $\unbrSYMM$. $\mathcal{J}_{\text{IR}}$ generate a $\unbrSYMM$ current subalgebra of VOAs associated to the interacting theory (or theories) in the IR. In the examples of rank-one and rank-two Deligne theories one of the two factors in \eqref{JnaturalasSUMgeneral} always vanishes (\emph{cf.} \eqref{Jnaturalrank1Deligne} and the last equation in \eqref{eandwDelignerank2}). The level of the current algebra generated by \eqref{JnaturalasSUMgeneral} then follows directly from the IR levels and the action of the unbroken symmetry on the free hypermultiplets. 

\item There is a distinguished $\mathfrak{u}(1)$ factor in the UV flavor symmetry that commutes with the unbroken symmetry \eqref{JnaturalasSUMgeneral} and under which the generator $e$ is charged  and we choose to normalize this current so that $e$ has charge and conformal dimension equal to $\ell$ (as appearing in the $\mathbb{Z}_{\ell}$ orbifold describing the $\overline{\ECB}$). This $\mathfrak{u}(1)$ combines with the Cartan generator of the $SU(2)_{R}$ symmetry of the UV theory to provide Cartan generator of the infrared $SU(2)_{R}$ symmetry. We elaborate further on this point in Appendix \ref{appB}.

\item The stress tensor takes the general form
\begin{equation}\label{TUVgeneral}
    T_{UV}=T_{\delta,\varphi}+T_{\xi}+T_{\eta}+T_{IR}~.
\end{equation}
The central charge of the UV theory is then determined once the contribution of the chiral bosons $(\delta,\varphi)$ is known. When the chiral boson $\varphi$ has zero background charge---which is related to the fact that $\partial \varphi$ will be part of the distinguished $\mathfrak{u}_1$ current---this is determined by $c_{\delta}=1-\frac{3 \alpha^2}{\langle \delta,\delta \rangle}$, $c_{\varphi}=1$. The value of the parameter $\alpha$ is fixed to be $2\omega_e=\ell$ from the condition that $e(z):=e^{\delta+\varphi}$ has conformal weight $\omega_e$. We conclude that 
\begin{equation}\label{cdeltaphiSectionreview}
    c_{\delta,\varphi}=2+\,\tfrac{3\,\ell^2}{\langle\varphi,\varphi\rangle}~.
\end{equation}
In the cases in which $e$ is associated with a current we have that $\ell=2$ and $\langle\varphi,\varphi\rangle=2/k$ and this equation reduces to \eqref{cdeltaphiSnatural}. As we will see, the value of $\langle\varphi, \varphi\rangle$ in the remaining cases is fixed by the requirement that the algebra of the affine currents, realized in terms of free fields, closes.
\end{itemize}
\section{\label{OverviewofStrategyandC2U1}Overview of the Strategy and
\texorpdfstring{$C_2U_1$}{C2U1}, 
 \texorpdfstring{$A_1U_1$}{A1U1}
 in detail}

We now move on to discuss two examples of new rank-one free field realizations in detail. The first example is the $C_2 U_1$ theory which has $\ell=2$. The VOA for this theory was previously discussed in \cite{Beem:2020pry} without using free field techniques. The second example is the $A_1 U_1$ theory which has $\ell=3$. This case exemplifies new technical features that are absent from the $\ell=2$ series. We discuss the general structure and its extension to the remaining rank-one theories in Section \ref{FFallrank1}.

In each of our two cases we start from an analysis of the singular locus of the Higgs branch, namely $\mathbb{H}^{\quatdim}/\mathbb{Z}_\ell$ with $\ell=2$ and $\ell=3$. For $\ell=2$ this is the minimal nilpotent orbit closure $\overline{\mathbb{O}_{\text{min}}(\mathfrak{c}_\quatdim)}$. We parametrize these by coordinates $(X_1,...,X_{\quatdim})$ and $(Y^1,...,Y^{\quatdim})$ on which $\mathbb{Z}_\ell$ act as $(X,Y)\mapsto (\omega X,\omega^{-1} Y)$ with $\omega$ and $\ell$'th primitive root of unity. The ring of invariants under this action describes the singular locus. Then we explain how to fiber the Deligne theory over this singular locus and perform an affine uplift to a free field realization. This requires specifying a $\mathbb{Z}_{\ell}$ action on the rank-one Deligne VOA, see \eqref{Ueidentification}. The $\mathbb{Z}_{\ell}$ invariant combinations of the ingredients are arranged appropriately to form the strong generators of the VOA. Once we finish the construction of generators, we also discuss how to use the free fields to compute various limits of the superconformal index. Since the free field realization that we discuss is a simple quotient, all the nulls are automatically zero. This makes the computation of different indices easier. For the case of the $C_2U_1$ theory, we reproduce the indices obtained using class $\mathcal{S}$ techniques (see \cite{Beem:2020pry}) but the computation of indices for the $A_1U_1$ case, to our knowledge, is new. Thus, the free fields also facilitate the computation of indices for theories for which class $\mathcal{S}$e or other known methods are not available.

\subsection{\texorpdfstring{$C_2U_1$}{C2U1}}
\label{sec:C2U1indetails}

The example of the $C_2U_1$ theory corresponds to the UV-IR pair $(\Delta_{IR},\Delta_{UV})=(\frac{3}{2},3)$ where the infrared theory is the  Deligne $\mathfrak{a}_2$ theory. The singular locus is $\mathbb{H}^2/\mathbb{Z}_2=\overline{\mathbb{O}_{\text{min}}(\mathfrak{c}_2)}$. Let us denote the coordinates of $\mathbb{H}^2=\mathbb{C}^2\times\mathbb{C}^2$ by 
$(X_1,X_2)$ and $(Y^1,Y^2)$. The $\mathbb{Z}_2$ acts on $(X_A,Y^A)$ as follows
\begin{equation}
	(X_A,Y^A)\mapsto (-X_A,-Y^A)\,.
\end{equation}
This action is compatible with the symplectic structure $\left\{X_A,Y^A\right\}=\delta^A_B$. The ring of invariants is generated by the following combinations
\begin{equation}
	\begin{split}
		b_{AB}&=X_A X_B,\;\; \overline{b}^{AB}=Y^A Y^B\\
		M^{B}_A&=X_AY^B-\frac{1}{2}\delta^A_B X_CY^C,\;\;m=X_C Y^C\,,
	\end{split}
\end{equation}
which combine together to form the $\mathfrak{c}_2$ moment map. These combinations satisfy the quadratic relations describing the minimial $\mathfrak{c}_2$ nilpotent orbit leading to the identification $\mathbb{H}^2/\mathbb{Z}_2=\overline{\mathbb{O}_{\text{min}}(\mathfrak{c}_2)}$.
Notice that we have split the $\mathfrak{c}_2$ generators to anticipate what will happen in the case $\ell\neq 2$.
In that case only the generator $M^B_A$ and $m$ are moment map generators.

The subset of $\mathbb{H}^2/\mathbb{Z}_2$ where $\mathsf{e}=b_{22}=(X_2)^2\neq 0$ is identified with  

\begin{equation}
	\frac{ T^*(\mathbb{C}^*)\times \mathbb{C}^{2}}{\mathbb{Z}_{2}}\,,
\end{equation}
where we have written

\begin{equation}
	\label{C2hmodZlgenerators}
		(X_1,X_2)\,=\, (\xi_1,\,\mathsf{e}^{1/2})\,,\qquad 
		(Y^1,Y^2)\,=\, (\xi_2,\mathsf{h}\mathsf{e}^{-1/2})\,.
	\end{equation}
Here $\mathsf{e}^{1/2},\mathsf{h}$ are coordinates of $T^*(\mathbb{C}^*)$
with $\{\mathsf{h},\mathsf{e}^{1/2}\}=\mathsf{e}^{1/2}$ and $\left\{\xi_1,\xi_2\right\}=1$ with $\mathbb{Z}_2$ acting as $(\mathsf{h},\mathsf{e}^{1/2},\xi)\mapsto (\mathsf{h},-\mathsf{e}^{1/2},-\xi)$.
In this patch, the generators of the $\mathfrak{c}_2$ flavour symmetry are rewritten as follows
\begin{equation}
	\begin{pmatrix}
		X_1 X_1 & X_1 Y^1 & X_1 X_2 & X_2Y^1 \\
		X_1 Y^1 & Y^1Y^1 & X_1 Y^2 & Y^2Y^1 \\
		X_1 X_2 & X_2 Y^1 & X_2 X_2 & X_2Y^2 \\
		X_2 Y^1 & Y^1 Y^2 & X_2 Y^2 & Y^2Y^2 \\
	\end{pmatrix}=
 \begin{pmatrix}
		\xi_1^2 & \xi_1 \xi_2 & \xi_1\mathsf{e}^{1/2} & \xi_2\mathsf{e}^{1/2} \\
		\xi_1\xi_2 & \xi_2^2 & \xi_1\mathsf{h}\mathsf{e}^{-1/2} & \xi_2\mathsf{h}\mathsf{e}^{-1/2} \\
		\xi_1\mathsf{e}^{1/2} & \xi_1\mathsf{h}\mathsf{e}^{-1/2} & \mathsf{e} & \mathsf{h} \\
		\xi_2\mathsf{e}^{1/2}  & \xi_2\mathsf{h}\mathsf{e}^{-1/2} & \mathsf{h} & \mathsf{h}^2\mathsf{e}^{-1} \\
	\end{pmatrix}\,.
\end{equation}
Having gained some understanding of the singular locus, let us try to fiber the Deligne $\mathfrak{a}_2$ theory over this. The infrared Deligne theory now introduces new ingredients $\mathcal{J}^{\text{IR}}_{\mathfrak{a}_2}$.
The next step is to specify a $\mathbb{Z}_2$ action on these currents.
This should act as a Lie algebra automorphism and, in this case, there are three inequivalent\footnote{The even subalgebras are respectively $\mathfrak{a}_2$, $\mathfrak{a}_1$ and  $\mathfrak{c}_1\oplus \mathfrak{u}_1$} such automorphisms. The choice that produces the correct flavor symmetry is  
\begin{equation}
	\mathcal{J}^{\text{IR}}_{\mathfrak{a}_2} 
	\mapsto 
	\mathcal{J}^{\text{IR}}_{\mathfrak{c}_1}\,\oplus \,
	\mathcal{J}^{\text{IR}}_{\mathfrak{u}_1}\,\oplus \,
	\mathcal{J}^{\text{IR}}_{(\mathbf{2},\pm)}\,,
\end{equation}
where the first two factors 
are $\mathbb{Z}_2$ even while the third is $\mathbb{Z}_2$ odd.
The latter transform under the $\mathbb{Z}_2$-even subalgebra as the two dimensional representation of $\mathfrak{c}_1$ and have charges $\pm 1$ under the $\mathfrak{u}(1)$ factor.  

Next, we identify the unbroken symmetry at a generic point (any point other than the origin) on the singular locus which in this case is $\unbrSYMM=\mathfrak{c}_1\oplus\mathfrak{u}_1$.
Notice that the $\mathfrak{u}_1$ factor is unbroken on the whole singular locus while the $\mathfrak{c}_1$ is unbroken by the choice of VEV. According to our general discussion in the previous section, we can immediately write down the free field realization of the currents associated to the unbroken symmetry to be
\begin{equation}
\label{Jc1}
	(\mathcal{J}_{\mathfrak{c}_1})_{mn}=(\mathcal{J}^{\text{IR}}_{\mathfrak{c}_1})_{mn}+\xi_m\xi_n
\qquad
	\mathcal{J}_{\mathfrak{u}_1}=\mathcal{J}^{\text{IR}}_{\mathfrak{u}_1}\,.
\end{equation}
This symmetry needs to be combined with other current generators to form the full $\mathfrak{c}_2\oplus\mathfrak{u}_1$ flavour symmetry of the UV theory. The  $\mathfrak{c}_2\oplus\mathfrak{u}_1$ decomposes in terms of $\mathfrak{g}^\natural$ as follows
\begin{equation}
	\mathcal{J}_{\mathfrak{c}_{2}} 
	\mapsto
	\mathcal{J}_{\mathfrak{c}_{1}}\oplus
	\mathcal{J}_{\bf{2}}^{\alpha}\oplus
	\mathcal{J}^{\alpha \beta}_{\mathfrak{sl}_{2}}\,,
\end{equation} 
or, in matrix notation,
\begin{equation}
	\mathcal{J}_{\mathfrak{c}_2}=
		\begin{pmatrix}
			
			\mathcal{J}_{\mathfrak{c}_1} & \mathcal{J}_{\mathbf{2}}^+ & \mathcal{J}_{\mathbf{2}}^- \\
			 \mathcal{J}_{\mathbf{2}}^+ & 	\mathcal{J}^{++}_{\mathfrak{sl}_{2}} & \mathcal{J}^{+-}_{\mathfrak{sl}_{2}} \\
			\mathcal{J}_{\mathbf{2}}^- & \mathcal{J}^{-+}_{\mathfrak{sl}_{2}} & \mathcal{J}^{--}_{\mathfrak{sl}_{2}} \\
		\end{pmatrix}\,,
\end{equation}
where $\mathcal{J}^{++}_{\mathfrak{sl}_{2}}=e$ is associated to the moment map that gets a vev.
There is a unique candidate for the free field realization of the currents with non negative weight under $\mathcal{J}^{+-}_{\mathfrak{sl}_{2}}$, namely
\begin{equation}
\label{C1U1easycurrents}
\mathcal{J}^{++}_{\mathfrak{sl}_{2}}
	\,=\,e^{\delta+\varphi}\,,
	\qquad
	\mathcal{J}^{+-}_{\mathfrak{sl}_{2}}
	\,=\,\tfrac{1}{2}\,k\partial \varphi\,,
 \qquad
 \mathcal{J}_{m}^+ = \xi_m \,e^{\frac{\delta+\varphi}{2}}\,,
\end{equation}
with $\mathcal{J}_{\mathfrak{c}_1}$ given in \eqref{Jc1}.
The lowering generator $\mathcal{J}^{--}_{\mathfrak{sl}_{2}}$ is constructed following the general scheme discussed in the previous section and takes the form \eqref{fthetaDELIGNE} for an appropriate choice of $\mathcal{S}^{\natural}$.
Before discussing the explicit form of $\mathcal{S}^{\natural}$, we will construct the non-current generators of the VOA with non negative weight. One candidate is very simple
\begin{equation}
\label{WpinC1U1}
	\mathcal{W}^{+}_{m,\pm}\,=\,
	\mathcal{J}^{\text{IR}}_{m,\pm}\,
 e^{\frac{\delta+\varphi}{2}}\,.
\end{equation}
This is an operator of conformal weight $3/2$ and should be part of a a multiplet of the UV flavor symmetry $\mathfrak{c}_2\oplus\mathfrak{u}_1$. Since it has regular OPEs with the generators $\mathcal{J}_{m}^+$ given in \eqref{C1U1easycurrents}, the combination \eqref{WpinC1U1} is a generalized highest weight state and, from its charges, we can identify unambiguously the relevant representation to which it belongs:
\begin{equation}
	\mathcal{W}^{\pm 1}_{\llbracket MN \rrbracket}
	\mapsto
	\mathcal{W}^{\alpha}_{m,\pm}\oplus 
	\mathcal{W}_{\pm}\,,
\end{equation}
where $M,N=1,\dots,4$ are $\mathfrak{c}_2$ indices and ${\llbracket MN \rrbracket}$ denotes antisymmetric $\Omega$-traceless combination, which in this case gives the five dimensional representation.
The free field realization of $\mathcal{W}_{\pm}$ has also a unique candidate in the free field space, namely
\begin{equation}
	\mathcal{W}_{\pm}\,=\,
	\xi_m^{}\mathcal{J}^{\text{IR}}_{n,\pm} \,\Omega^{mn}\,.
\end{equation}

To conclude the construction we need to build $S^{\natural}$, the lowest weight states $\mathcal{J}^-_m$ and $\mathcal{W}^{-}_{m,\pm}$ will then be obtained acting with the lowering operator $\mathcal{J}_{\mathfrak{sl}_2}^{--}$ on the highest weight state.
As in the example of rank-one and rank-two Deligne theories, the operator $S^{\natural}$ is found by imposing that it has regular OPEs with \eqref{C1U1easycurrents} and that its OPEs with the $\mathfrak{sl}(2)$ highest weight generators  $\mathcal{J}_m^+$, $\mathcal{W}^{+}_{m,\pm}$ and $\mathcal{W}_{\pm}$ does not contain any pole higher then the first. This gives
\begin{equation}
	S^{\natural}=-\frac{1}{8}(\mathcal{J}^{\text{IR}}_{\mathfrak{c}_1})_{mn}(\mathcal{J}^{\text{IR}}_{\mathfrak{c}_1})_{pq}\Omega^{mp}\Omega^{nq}+\frac{3}{4}\xi_m\partial\xi_n\Omega^{mn}-\frac{1}{4}(\mathcal{J}^{\text{IR}}_{\mathfrak{c}_1})_{mn}\xi_p\xi_q\Omega^{mc}\Omega^{pq}\,,
\end{equation}
where the normalization is fixed by \eqref{SnaturalSnaturalOPES}.
Now that we have constructed all the generators $\mathcal{J}_{\mathfrak{c}_2},\mathcal{J}_{\mathfrak{u}(1)}$ and $\mathcal{W}^{\pm}_{\mathbf{5}}$, let us check the levels, close the OPEs and look at the R-filtration.

\paragraph{Central Charges.}
Let us check that the conformal anomaly $c_{2d}$ and flavor central charge $k_{2d}$ are correctly reproduced 
\begin{align}
	c_{\text{UV}}&=c_{\mathfrak{a}_2}+c_{\xi}+c_{\delta,\varphi}\,,
 \qquad   \qquad 
  \qquad k_{UV}=I_{\mathfrak{c}_1\hookrightarrow \mathfrak{a}_2}k_{IR}+k_{\xi}\,,
	\\
	-19&=-8+2 \times (-\tfrac{1}{2})-10\,,
 \qquad  \qquad 
 -2=-1\times\tfrac{3}{2}-\tfrac{1}{2}\;\;\,.\,\;
\end{align}
In fact, these values can be easily derived from the free field construction as follows.
The level of $\mathcal{J}_{\mathfrak{c}_1}$ is fixed in terms of the IR data thanks to the free field realization \eqref{Jc1}. The level of its $\mathfrak{sl}(2)\subset \mathfrak{c}_2$ commutant is in turn fixed and specifies, via \eqref{cdeltaphiSnatural}, the contribution to the central charge $c_{\delta,\varphi}$.

\paragraph{OPEs.}
All the above constructed generators have the following OPEs
\begin{align}
	\mathcal{J}(z)\ \mathcal{J}(0) & ~\sim~ \frac{-1}{z^2} ~,\\
	\mathcal J_{IJ}(z)\ \mathcal J_{KL}(0) & ~\sim~ \frac{-2\ \Omega_{L(I}\Omega_{J)K}}{z^2} ~+~ \frac{2\ \Omega_{(I(K}\ \mathcal J_{J)L)}(0)}{z}~,\displaybreak[0]\\
	\mathcal{J}(z)\ \mathcal{W}^{\pm}_{IJ}(0)& ~\sim~ \frac{\pm\ \mathcal{W}^{\pm}_{IJ}(0)}{z}~, \\
	\mathcal J_{IJ}(z)\ \mathcal{W}^{\pm}_{KL}(0) & ~\sim~ \frac{\ \Omega_{(I[K} \ \mathcal{W}^{\pm}_{J)L]}(0)}{z}~,\\
 \mathcal{W}_{K_1L_1}^+(z)\ \mathcal{W}^-_{K_2L_2}(0) & ~\sim~ \frac{-3\Delta^{\mathbf{5},\mathbf{5}}_{K_1L_1; K_2L_2}}{z^3} + \frac{1}{z^2}\Big(3\, \Delta^{\mathbf{5},\mathbf{5}}_{K_1L_1; K_2L_2}\ \mathcal{J} -\frac{3}{4}\, \Omega_{[K_1[K_2}\ \mathcal J_{L_1]L_2]} \Big)\displaybreak[0] \nn\\
	&+\frac{1}{z}\Big( \frac{3}{2}\, \Delta^{\mathbf{5},\mathbf{5}}_{K_1L_1; K_2L_2}\ \partial \mathcal{J} -\frac{3}{8} \Omega_{[K_1[K_2}\ \partial\mathcal J_{L_1]L_2]} -\frac{3}{2}\, \Delta^{\mathbf{5},\mathbf{5}}_{K_1L_1; K_2L_2}\ (\mathcal{J}\mathcal{J}) \nn \\
	&\qquad \frac{3}{4}\, \Omega_{[K_1[K_2}\ (j\mathcal J_{L_1]L_2]}) -\frac{1}{16} \Delta^{\mathbf{5},\mathbf{5}}_{K_1L_1; K_2L_2}\, \Omega^{PQ}\, \Omega^{RS}\ (\mathcal J_{PR}\mathcal J_{QS}) \nn \\
	&\qquad -\frac{1}{8}\big( (\mathcal J_{[K_1[K_2} \mathcal J_{L_1]L_2]})|_{\Omega\text{-traceless}} \big) \Big)~,
\end{align}
where
\begin{equation}
	\Delta^{\mathbf{5},\mathbf{5}}_{K_1L_1; K_2L_2}\colonequals \Omega_{K_1K_2}\Omega_{L_1L_2}+\Omega_{K_1L_2}\Omega_{K_2L_1} - \frac{1}{2}\Omega_{K_1L_1}\Omega_{K_2L_2}~\,.
\end{equation}

\paragraph{Null States and Superconformal Indices.}
The free field realization has given us an explicit realization of the VOA. One of the advantages of the construction is that it gives a simple quotient of the VOA, which essentially means that null states are zero once expressed in terms of the free fields. What is more, is that it also allows to recover the R-filtration. This helps us to compute different limits of the superconformal index. This method is especially useful since not all theories we know have a class-S realization. 

\paragraph{Schur Index.}
 The vacuum character of the VOA can be computed to be
\begin{equation}
\begin{split}
	\chi_{C_2U_1}(q)&=1+11 q+10 q^{3/2}+72 q^2+90 q^{5/2}+...\\
	&=\text{PE}\left[\frac{11q+10q^{3/2}-5q^2-30q^{5/2}+...}{1-q}\right]\,.
\end{split}
\end{equation}
It can be refined by the flavour fugacities to give
\begin{equation}
\begin{split}
	\chi_{C_2U_1}(q)&=1+(1+\mathbf{10}_0) q+(\mathbf{5}_{1}+\mathbf{5}_{-1})q^{3/2}+(3+\mathbf{14}_0+\mathbf{35}_0+2\cdot \mathbf{10}_0)q^{2}\\&+(2\cdot\mathbf{5}_{+1}+2\cdot\mathbf{5}_{-1}+\mathbf{35}_{+1}+\mathbf{35}_{-1}))q^{5/2}+...\\
	&=\text{PE}\left[\frac{(1+\mathbf{10}_0)q+(\mathbf{5}_{1}+\mathbf{5}_{-1})q^{3/2}-\mathbf{5}_0q^2-(\mathbf{5}_1+ \mathbf{5}_{-1}+ \mathbf{10}_1+\mathbf{10}_{-1})q^{5/2}+...}{1-q}\right]
\end{split}
\end{equation}
This matches with the class-S computation of the Schur Index. The terms at order $q$ are associated to the  AKM currents and $q^{3/2}$ terms to the generators $\mathcal{W}_{\mathbf{5}}^{\pm}$. There are $5$ null states at conformal dimension $2$ of the form
\begin{equation}
\label{Jc2Jc2null}
	\left(\mathcal{J}_{\mathfrak{c}_2}\mathcal{J}_{\mathfrak{c}_2}\right)_{\mathbf{5}}=0\,.
\end{equation}
At conformal weight $\frac{5}{2}$, there are nulls in representation $\mathbf{5}_1$, $\mathbf{5}_{-1}$, $\mathbf{10}_1$ and $\mathbf{10}_{-1}$
\begin{equation}
\label{JWnullsFORC2U1}
\mathcal{J}_{\mathfrak{c}_2}\, \mathcal{W}^{\pm}+
\mathcal{J}_{\mathfrak{u}_1}\,\mathcal{W}^{\pm}
+\partial\mathcal{W}^{\pm}=0\,,
\qquad
\mathcal{J}_{\mathfrak{c}_2}\, \mathcal{W}^{\pm}\big{|}_{\mathbf{10}}=0\,.
\end{equation}
Additional null states at conformal weight $h=3$ can be found in \cite{Beem:2020pry}. 
It is easy to verify that using the generalized free field realization these expressions are either identically zero or proportional to null operators of the IR theory. We will return to a discussion of null states for all rank-one theories with ECB in Section \ref{sec:RfiltrationandNullsgeneral}.

\paragraph{Hall-Littlewood Index.}
The Hall-Littlewood Index can be computed by working in the leading $R$-filtration and is given by
\begin{equation}
	\label{eq:hlredhighlightc2u1}
	\begin{split}
		\mathcal{I}_{HL}&=1+11 t^2+10 t^3+60 t^4+80 t^5+...\\
		&=\text{PE}[11 t^2+10 t^3-6 t^4-30 t^5+...]\,,
	\end{split}
\end{equation}
If we refine by the flavour fugacities we obtain
\begin{equation}
	\begin{split}
		\mathcal{I}_{HL}&=1+(\mathbf{1}+\mathbf{10}_0) t^2+(\mathbf{5}_{1}+\mathbf{5}_{-1})t^{3}+(\mathbf{1}+\mathbf{14}_0+\mathbf{35}_0+ \mathbf{10}_0)t^{4}+\\
  &\,\,\,\,\,\,+(\mathbf{5}_{+1}+\mathbf{5}_{-1}+\mathbf{35}_{+1}+\mathbf{35}_{-1}))t^{5}+...\\
		&=\text{PE}\left[(\mathbf{1}+\mathbf{10}_0)t^2+(\mathbf{5}_{1}+\mathbf{5}_{-1})t^{3}-(\textcolor{red}{\mathbf{1}}+\mathbf{5}_0)t^4-(\mathbf{5}_1+ \mathbf{5}_{-1}+ \mathbf{10}_1+\mathbf{10}_{-1})t^{5}+...\right]
	\end{split}
\end{equation}
The null operators in the VOA given in \eqref{Jc2Jc2null} and \eqref{JWnullsFORC2U1} immediately give relations among the Hall-Littlewood (HB in this case) generators by taking the leading term in the R-filtration limit which corresponds to ignoring the derivative terms in \eqref{JWnullsFORC2U1}.
These is an additional relation in the Higgs branch, corresponding to the term $\textcolor{red}{1}$ in the HL index. This is associated wih the fact that in this example the stress tensor, which has $R=1$, coincides with the Sugawara stress tensor. This gives rise to the additional HB relation
\begin{equation}
\left(\mathsf{J}_{\mathsf{c}_2}\mathsf{J}_{\mathsf{c}_2}+\mathsf{J}^2\right)_{\mathbf{1}}=0\,,
\end{equation}
where $\mathsf{J}$ are the HB avatars of the VOA currents $\mathcal{J}$.

\subsection{\label{sec5c2u1}\texorpdfstring{$A_1U_1$}{A1U1}}
\label{sec:subsectionA1U1}

The example of $A_1U_1$ theory corresponds to the UV-IR pair $(\Delta_{IR},\Delta_{UV})=(\frac{4}{3},4)$ where the infrared Deligne theory is the $\mathfrak{a}_1$ theory. The singular locus is $\mathbb{H}^2/\mathbb{Z}_3$ which we now describe. Let us denote the coordinates of $\mathbb{H}^2=\mathbb{C}^2\times\mathbb{C}^2$ by 
$(X_1,X_2)$ and $(Y^1,Y^2)$. The $\mathbb{Z}_2$ acts on $(X_A,Y^A)$ as follows
\begin{equation}
	(X_A,Y^A)\mapsto (\omega X_A,\omega^{-1}Y^A),\;\;\qquad \omega^3=1\,.
\end{equation}
This action is compatible with the symplectic structure $\left\{X_A,Y^A\right\}=\delta^A_B$. The ring of invariants is generated by
\begin{equation}
	\begin{split}
		b_{ABC}&=X_A X_B X_C,\;\; \qquad\qquad \overline{b}^{ABC}=Y^A Y^B Y^C\,,\\
		M^{B}_A&=X_AY^B-\frac{1}{2}\delta^A_B X_CY^C,\;\;\qquad m=X_C Y^C\,,
	\end{split}
\end{equation}
where $M_A^B$ and $m$ generate $\mathfrak{su}(2)\oplus\mathfrak{u}(1)$ symmetries. 
The subset of $\mathbb{H}^2/\mathbb{Z}_3$ where $\mathsf{e}=b_{222}=(X_2)^3\neq 0$ is identified with 
\begin{equation}
	\frac{ T^*(\mathbb{C}^*)\times \mathbb{C}^{2}}{\mathbb{Z}_{3}}\,.
\end{equation}
In this patch we write 
\begin{equation}
	\label{C2hmodZlgenerators}
		(X_1,X_2)\,=\, (\betaclass,\,\mathsf{e}^{1/3})\,,\qquad
		(Y^1,Y^2)\,=\, (\gammaclass,\mathsf{h}\mathsf{e}^{-1/3})\,,
  \end{equation}
where $\mathsf{e}^{1/3}$ and $\mathsf{h}$ are coordinates of $T^*(\mathbb{C}^*)$
with $\{\mathsf{h},\mathsf{e}^{1/3}\}=\mathsf{e}^{1/3}$ and $\left\{\betaclass,\gammaclass \right\}=-1$. 
The generators of  $\mathbb{H}^2/\mathbb{Z}_3$ can then be written as 
\begin{equation}
	M^B_A=
	\begin{pmatrix}
	-\tfrac{1}{2}\,\mathsf{z} &\,\, \betaclass \mathsf{h}\mathsf{e}^{-1/3}\\
		\gammaclass\mathsf{e}^{1/3} & \,\,\tfrac{1}{2}\,\mathsf{z}
	\end{pmatrix}\,,
	\qquad
m=\betaclass\gammaclass+\mathsf{h}\,,
\end{equation}
\begin{equation}
\label{bandbbarsingularLocus}
	b_{ABC}=\begin{pmatrix}
		\mathsf{e} \\
		\betaclass \mathsf{e}^{2/3}\\
		\betaclass^2 \mathsf{e}^{1/3}\\
		\betaclass^3
	\end{pmatrix},\;\;\;	
 \qquad
\overline{b}^{ABC}=\begin{pmatrix}
	\gammaclass^3 \\
	\gammaclass^2 \mathsf{h}\mathsf{e}^{-1/3}\\
	\gammaclass \mathsf{h}^2\mathsf{e}^{-2/3}\\
	\mathsf{h}^3\mathsf{e}^{-1}
	\end{pmatrix}\,,
\end{equation}
where we introduced the short hand notation
$\mathsf{z}=\mathsf{h}-\betaclass\gammaclass$.
We refer to the generator $\gammaclass\,\mathsf{e}^{1/3}$ as raising operator and to the quantities annihilated by it, like $b_{222}=\mathsf{e}$ and  $\bar{b}^{111}=\gammaclass^3$, as highest weight states.

Having gained some understanding of the singular locus, let us try to fiber the Deligne $\mathfrak{a}_1$ theory over this. The infrared Deligne theory now introduces new ingredients $\mathcal{J}^{\text{IR}}_{\mathfrak{a}_1}$ on which we  need to specify a $\mathbb{Z}_3$ action. 
In this case there is a unique non-trivial choice, it is associated with the branching
\begin{equation}
\mathcal{J}^{\text{IR}}_{\af_1} 
\mapsto 
\mathcal{J}^{\text{IR}}_{\uf_1}\,\oplus \,
\mathcal{J}^{\rm IR}_{-2}\oplus \,
\mathcal{J}^{\rm IR}_{+2}\,,
\end{equation}
so that the generator of $\mathbb{Z}_3$ acts as
\begin{equation}
	\mathcal{J}^{\text{IR}}_{\uf_1}\mapsto \mathcal{J}^{\text{IR}}_{\uf_1}\,,
 \qquad
	\mathcal{J}^{\rm IR}_{\pm 2}
 \mapsto \omega^{\mp 1}\mathcal{J}^{\rm IR}_{\pm 2}\,.
\end{equation}
The flavor symmetry of the UV theory is $\mathfrak{a}_1\oplus \mathfrak{u}_1$. A linear combination of the Cartan generator of the  $\mathfrak{a}_1$ factor and the  $\mathfrak{u}_1$ generator is unbroken by the VEV and a  combination orthogonal to the latter (in the sense of OPEs) is identified with the distinguished $U(1)$ that acts on $e$. The charges of the free field ingredients associated to the coordinates of the singular locus  with respect to $\mathcal{Z}$, the Cartan of $\mathfrak{a}_1$, 
and $\mathcal{J}_{\mathfrak{u}_1}$ are easy to determine to be $(1,\tfrac{1}{3})$ for $\mathsf{e}^{1/3}$ and $(-1,\tfrac{1}{3})$, $(+1,-\tfrac{1}{3})$ for $\betaclass$ and $\gammaclass$ respectively. We also need to assign charges to the IR currents. A possible way to do it is to insist that the highest wight state $b^{111}=\gammaclass^3$ in \eqref{bandbbarsingularLocus} gets modified by IR currents when we move away from the singular locus. We have already encountered this mechanism in the $C_1A_1$ example, in that case $b^{11}$ is part of the $\mathfrak{c}_2$ currents and the corresponding VOA generator takes the form $\xi_2 \xi_2+\mathcal{J}^{\text{IR}}_{22}$. 
In our case, we postulate
\begin{equation}
\label{B111inA1U1}
	b^{111}\,\rightsquigarrow\,
 \overline{\mathcal{B}}^{
	111}(z)
 =\gamma\,\gamma\,\gamma+ \mathcal{J}_{+2}^{\text{IR}}\,\,e^{\frac{1}{3}(\delta+\varphi)}\,,
\end{equation}
where $ \overline{\mathcal{B}}^{ABC}(z)$ is the VOA avatar of $\bar{b}^{ABC}$ given in \eqref{bandbbarsingularLocus}. It is important that the extra term has the correct conformal weight $3/2$ and is invariant under the $\mathbb{Z}_{3}$ action, this condition singles out the choice of exponent.
Now we can read off from \eqref{B111inA1U1} the weight of $\mathcal{J}_{+2}^{\text{IR}}$ under the $\mathfrak{u}_1$s above to be $(3,-1)-(1,-\tfrac{1}{3})=(2,-\tfrac{4}{3})$.  The generator $\mathcal{J}_{-2}^{\text{IR}}$ will have opposite charges.
We conclude that the free field expression for the currents associated to these factors is
\begin{equation}
\label{ZandJcurrentforA1U1}
\mathcal{Z}=h-\beta\gamma+\mathcal{J}_{\mathfrak{u}_1}^{\text{IR}}\,,
\qquad
\mathcal{J}_{\mathfrak{u}_1}=\tfrac{1}{3}\left( h+\beta\gamma-2\mathcal{J}_{\mathfrak{u}_1}^{\text{IR}}
\right)\,,
\end{equation}
where $h(z)=3\,\langle \varphi,\varphi\rangle^{-1} \partial \varphi(z)$.
Imposing that the OPE of $\mathcal{Z}$ with $\mathcal{J}_{\mathfrak{u}_1}$ is regular and the OPE of $\mathcal{Z}$ with itself reproduces\footnote{Recall that \begin{equation}
\label{ZZOPEinfootnote}
 \mathcal{Z}(w)\mathcal{Z}(0)\sim\frac{2 k_{\text{UV}}}{w^2}\,,
 \qquad
\mathcal{J}_{\mathfrak{u}_1}^{\text{IR}}(w)
\mathcal{J}_{\mathfrak{u}_1}^{\text{IR}}(0)\sim\frac{2 k_{\text{IR}}}{w^2}\,,
\qquad
 (\beta \gamma)(w)(\beta \gamma)(0)\sim\frac{-1}{w^2}\,,
\end{equation}} the UV $\mathfrak{a}_1$ level, we obtain the conditions
\begin{equation}
k_{\text{UV}}=3 k_{\text{IR}}-1
\,,
\qquad
\langle \varphi,\varphi\rangle=\tfrac{9}{4k_{\text{IR}}-1}
\,.
\end{equation}
Notice that we did not yet set  $k_{\text{IR}}=-\tfrac{4}{3}$ to emphasize that these properties will be satisfied for any value of $k_{\text{IR}}$. The fact that the correct value of the level $k_{\text{UV}}=-5$ is reproduced is an indication that we are on the right track.
We are ready to complete the construction of $\mathfrak{a}_1$ AKM currents
\begin{equation}
\label{a1inA1U1}
	(\mathcal{J}_{\mathfrak{a}_1})^A_B=
 \begin{pmatrix}
		-\tfrac{1}{2}\mathcal{Z}\, & \,\mathcal{J}^{-}\\ 
  \mathcal{J}_{+}\,& \,+\tfrac{1}{2}\mathcal{Z}
	\end{pmatrix}\,.
\end{equation}
The most general ansatz for $\mathcal{J}_{+}$ and  $\mathcal{J}^{-}$ which is of conformal dimension $1$, $\mathbb{Z}_3$ invariant and consistent with our charge assignement is 
\begin{equation}
\mathcal{J}_{+}=\gamma\,e^{\frac{1}{3}(\delta+\varphi)}\,,
\end{equation}
\begin{equation}
\mathcal{J}^{-}=
\left(\beta\,( c_1\partial \delta + c_2 \partial  \varphi)+c_3\partial \beta
+c_4\,
\beta
\mathcal{J}^{\text{IR}}_{\mathfrak{u}_1}
\right)\,
e^{-\frac{1}{3}(\delta+\varphi)}
+c_5\,
\mathcal{J}^{\text{IR}}_{-2}\,\g\g\,
e^{-\frac{2}{3}(\delta+\varphi)}\,.
\end{equation}
Requiring that the generators in \eqref{a1inA1U1} closes onto the $\mathfrak{a}_1$ current algebra gives the conditions
\begin{equation}
c_1=\tfrac{1}{3}-k_{\text{IR}}\,,
\qquad
c_2=\tfrac{1}{3}\,k_{\text{IR}}\,,
\qquad
c_3=3\,k_{\text{IR}}\,,
\qquad
c_4=1\,,
\end{equation}
where we left the value of $k_{\text{IR}}$ unspecified since the $\mathfrak{a}_1\oplus \mathfrak{u}_1$ current algebra closes for any value of $k_{\text{IR}}$.
The coefficient $c_5$ is in turn fixed by the requirement that the operator $\overline{\mathcal{B}}^{111}$ introduces in \eqref{B111inA1U1} is an AKM primary, i.e. the OPE with $\mathcal{J}^-$ contains only a simple pole. This gives $c_5=-9$.
This completes the construction of $A_1U_1$ AKM currents.
We now use the lowering generator $\mathcal{J}^-$ to build the $\mathcal{B}$ and $\overline{\mathcal{B}}$ modules from their highest weight states $e^{\delta+\phi}$ and \eqref{B111inA1U1} respectively. Notice that these two state have charge $(3,1)$ and $(3,-1)$ respectively with respect to the current $\mathcal{Z}$ and $\mathcal{J}_{\mathfrak{u}_1}$.  The first two descendants of $\mathcal{B}$
take the form
\begin{equation}
\label{somedescendantsofeforA1U1}
\mathcal{J}^-\cdot e\,=\,3 \beta\,e^{\frac{2}{3}(\delta+\varphi)}\,,
\qquad
\mathcal{J}^-\cdot\mathcal{J}^-\cdot e\,=\,6\left( \beta\beta\,e^{\frac{1}{3}(\delta+\varphi)}-9\,\gamma\,\mathcal{J}^{\text{IR}}_{-2}\right)\,.
\end{equation}
Interestingly, requiring the absence of a second order pole when acting one more time with $\mathcal{J}^-$ imposes the condition $k_{\text{IR}}=-\tfrac{4}{3}$ which is the correct value.
This condition further guarantees that the $\mathcal{B}$ and $\overline{\mathcal{B}}$ operators transform in the four dimensional representation of $\mathfrak{a}_1$.

We succeeded in the construction of all the VOA generators which are associated to generators of the chiral ring of the singular locus. As in the $C_2U_1$ example we expect that this does not exhaust  the list of strong generators. The missing generators can be found by closing the OPEs of the generators we have already constructed. The OPEs $\mathcal{B}(z)\mathcal{B}(0)$ and $\overline{\mathcal{B}}(z)\overline{\mathcal{B}}(0)$ close on new generators of conformal dimension two in the singlet of the $\mathfrak{a}_1$ flavor symmetry and with charges $\pm 2$ under the  $\mathfrak{u}_1$ flavor symmetry and are denoted as $\mathcal{W}^{\pm\pm}$. The schematic form of the OPEs is 
\begin{equation}
	\mathcal{B}(z)\mathcal{B}(w)\sim
	\frac{\mathcal{W}^{++}(w)}{(z-w)}\,,
	\qquad
	\bar{\mathcal{B}}(z)\bar{\mathcal{B}}(w)\sim
	\frac{\mathcal{W}^{--}(w)}{(z-w)}\,.
\end{equation}
The form of $\mathcal{W}^{++}$ is very simple and in fact is obtained from the OPE of the two descendants in \eqref{somedescendantsofeforA1U1}. This gives
\begin{equation}
\label{WppVOAinA1U1}
	\mathcal{W}^{++}=\mathcal{J}^{\text{IR}}_{-2}\,\,e^{\frac{2}{3}(\delta+\varphi)}\,.
\end{equation}
The generator $\mathcal{W}^{--}$ is more complicated. Here we present the explicit expression of its leading term in the R-filtration\footnote{To simplify the form of $\mathsf{W}^{--}$ we used the IR Higgs branch relation
$\mathsf{J}^{\text{IR}}_{+2}\mathsf{J}^{\text{IR}}_{-2}+\tfrac{1}{4}\mathsf{J}^{\text{IR}}_{\mathfrak{u}_1} \mathsf{J}^{\text{IR}}_{\mathfrak{u}_1}=0
$.}
\begin{equation}
\label{Wmmclassical}
\mathsf{W}^{--}=
\mathsf{J}^{\text{IR}}_{-2}\,\,\gammaclass^6\,\mathsf{e}^{-\frac{4}{3}} 
+
\mathsf{x}\,\mathsf{J}^{\text{IR}}_{\mathfrak{u}_1}\,\,\gammaclass^3\,\mathsf{e}^{-1} 
-\mathsf{x}^2\,\mathsf{J}^{\text{IR}}_{+2}\,\,\mathsf{e}^{-\frac{2}{3}}\,, 
\end{equation}

where we introduced the combination $\mathsf{x}=\tfrac{1}{6}(\mathsf{J}^{\text{IR}}_{\mathfrak{u}_1}-2(\mathsf{h}+\betaclass\gammaclass))$. Setting to zero the generators of the Higgs branch of the IR theory, which are associated to transverse directions, corresponds to restricting the generators of the UV theory to the singular locus.  
The generator $\mathsf{W}^{--}$ given in \eqref{Wmmclassical}, together with  $\mathsf{W}^{++}$ obtained from \eqref{WppVOAinA1U1}, vanishes on the singular locus as it should.
We finally point out that the stress tensor is also an independent strong generator in this case.

\paragraph{A  $\mathbb{C}^2/\mathbb{Z}_3$ locus and a subVOA.}
One can define a subspace of the Higgs branch by the condition that the symmetry which is unbroken by the VEV, which we call $\unbrSYMM$, is unbroken. This subspace clearly includes the point associated with the VEV. At the level of the VOA we can look at the (or a) subVOA of invariants under $\unbrSYMM$. In the example of $C_2U_1$ presented in the previous section, the unbroken symmetry is $\mathfrak{c}_1\oplus \mathfrak{u}_1$ and the invariant locus is\footnote{The same is true for the rank-one and rank-two Deligne theories reviewed in Section \ref{Delignereview}. } $\mathbb{C}^2/\mathbb{Z}_2$. At the level of the VOA we have an AKM $\mathfrak{sl}_2$ algebra whose free field realization follows the general scheme presented in  Section \ref{Delignereview}. In the example of the $A_1U_1$ theory considered here, the unbroken symmetry is generated by the combination $\mathcal{Z}-3\mathcal{J}_{\mathfrak{u}_1}=3\,\mathcal{J}^{\text{IR}}_{\mathfrak{u}_1}-2\,\beta\gamma$, see
\eqref{ZandJcurrentforA1U1}. The invariant locus of the HB this time is $\mathbb{C}^2/\mathbb{Z}_3$. Let us see how this emerges from our free field realization. 
The VOA (strong) generators that are invariant under this $\mathfrak{u}_1$ are the highest weight state of the $\mathcal{B}$ multiplet, namely $\mathcal{B}_{222}=e^{\delta+\varphi}$, the lowest weight state of the $\overline{\mathcal{B}}$ multiplet, namely $\overline{\mathcal{B}}^{222}$, together with $\mathcal{Z}$, $\mathcal{J}_{\mathfrak{u}_1}$ and $T$.

The leading term in the R-filtration of the $\overline{\mathcal{B}}^{222}$ generator gives
\begin{equation}
\overline{\mathsf{B}}^{222}=
\mathsf{H}\,\mathsf{e}^{-1}
-\mathsf{J}^{\text{IR}}_{+2}\,\betaclass^3\,\mathsf{e}^{-\frac{2}{3}}+9\,
\mathsf{J}^{\text{IR}}_{-2}\,\gammaclass^3 (\mathsf{J}^{\text{IR}}_{\mathfrak{u}_1}-2\mathsf{h}+\betaclass\gammaclass)\,\mathsf{e}^{-\frac{4}{3}}\,,
\end{equation}
where
\begin{equation}
\mathsf{H}=\tfrac{1}{4}(2\mathsf{h}-\mathsf{J}^{\text{IR}}_{\mathfrak{u}_1})^2(2\mathsf{h}+\mathsf{J}^{\text{IR}}_{\mathfrak{u}_1})
-\tfrac{3}{2}\,\mathsf{J}^{\text{IR}}_{\mathfrak{u}_1}\,\betaclass\gammaclass\,
(\mathsf{J}^{\text{IR}}_{\mathfrak{u}_1}-2\mathsf{h}+2\betaclass\gammaclass)\,.
\end{equation}
It is easy to see that on the locus in which the symmetry associated to the  combination $\mathcal{Z}-3\mathcal{J}_{\mathfrak{u}_1}$ is unbroken, the vanishing of the charged Higgs branch generators implies that $\betaclass,\gammaclass$ and the IR HB generators $\mathsf{J}^{\text{IR}}$ all vanish. 
In this limit $\overline{\mathsf{B}}^{222}\mapsto \mathsf{h}^2\mathsf{e^{-1}}$ and $\mathsf{Z}\mapsto \mathsf{h}$. Together with $\mathsf{B}_{222}=\mathsf{e}$ they reproduce the relations of $\mathbb{C}^2/\mathbb{Z}_3$ in the locus where $\mathsf{e}\neq 0$.

One may wonder whether a similar mechanism takes place at the level of the VOA and look for the subVOA which has regular OPE with the current associated with the unbroken symmetry. The expectation is that this VOA might be isomorphic to the Bershadsky-Polyakov algrebra. This is not the case.

\paragraph{Central Charges.}
Let us check if we can reproduce the conformal anomaly $c_{2d}$ and flavor central charge $k_{2d}$.
\begin{align}
	c_{\text{UV}}&=c_{\mathfrak{a}_1}+c_{\beta,\gamma}+c_{\delta,\varphi}\,,
	\\
	-24&=-6+2 \times (-\tfrac{1}{2})-17\,,
\end{align}
The flavor central charge is obtained simply by adding the levels of individual terms in the expression for 
$\mathcal{Z}$ (equation \ref{ZandJcurrentforA1U1}), which reproduces the correct $k_{\text{UV}}$.
\paragraph{OPEs}
Having obtained the free field expression of the generators we can proceed to compute their OPEs and check that they close upon adding the stress tensor $T$ to the list of strong generators. As the OPEs invlolving currents or the stress tensor take a canonical form, here we report only the remaining non-vanishing OPEs \footnote{The notation used here is such that all the indices are totally symmetrized. For example,   $\delta^{A_2}_{A_1}\delta^{B_2}_{B_1}\delta^{C_2}_{C_1}$ is totally symmetrized in $A_1,B_1,C_1$ and also totally symmetrized in $A_2,B_2,C_2$}

\newcommand{\coeffBBAUa}{{15}}
\newcommand{\coeffBBAUb}{{\left(-\frac{15}{2}\right)}}
\newcommand{\coeffBBAUc}{{3}}
\newcommand{\coeffBBAUd}{{\frac{1}{2}}}
\newcommand{\coeffBBAUe}{{\frac{9}{8}}}
\newcommand{\coeffBBAUf}{{\left(-\frac{15}{4}\right)}}
\newcommand{\coeffBBAUg}{{\frac{3}{2}}}
\newcommand{\coeffBBAUh}{{12}}
\newcommand{\coeffBBAUi}{{-\frac{3}{4}}}
\newcommand{\coeffBBAUj}{{\frac{1}{4}}}

\begin{equation}
	\begin{split}
		&\mathcal{B}_{A_1B_1C_1}(z)\overline{\mathcal{B}}^{A_2B_2C_2}(w)\sim \frac{\coeffBBAUa\delta^{A_2}_{A_1}\delta^{B_2}_{B_1}\delta^{C_2}_{C_1}}{(z-w)^3}+\frac{\coeffBBAUc\delta^{A_2}_{A_1}\delta^{B_2}_{B_1}(\mathcal{J}_{\mathfrak{a}_1})^{C_2}_{C_1}+\coeffBBAUb\delta^{A_2}_{A_1}\delta^{B_2}_{B_1}\delta^{C_2}_{C_1}\mathcal{J}_{\mathfrak{u}_1}}{(z-w)^2}\\&+\frac{\delta^{A_2}_{A_1}\delta^{B_2}_{B_1}\delta^{C_2}_{C_1}(\coeffBBAUe T+\coeffBBAUd(\mathcal{J}_{\mathfrak{a}_1}\mathcal{J}_{\mathfrak{a}_1})+\coeffBBAUe\mathcal{J}_{\mathfrak{u}_1}\mathcal{J}_{\mathfrak{u}_1}+\coeffBBAUf\mathcal{J}_{\mathfrak{u}_1}')+\delta^{A_2}_{A_1}\delta^{B_2}_{B_1}((\coeffBBAUg\mathcal{J}_{\mathfrak{a}_1})^{C_2}_{C_1})'+\coeffBBAUi\mathcal{J}_{\mathfrak{u}_1}(\mathcal{J}_{\mathfrak{a}_1})^{C_2}_{C_1})}{(z-w)}\\&+\frac{\coeffBBAUj\delta^{A_2}_{A_1}(\mathcal{J}_{\mathfrak{a}_1})^{B_2}_{B_1}(\mathcal{J}_{\mathfrak{a}_1})^{C_2}_{C_1}}{(z-w)}
	\end{split}
\end{equation}
\newcommand{\coeffBbWxmssone}{{-6}}
\newcommand{\coeffBbWxmsstwo}{{-3}}
\newcommand{\coeffBbWxmssthree}{{\left(\frac{9}{2}\right)}}
\newcommand{\coeffBbWxmssfour}{{\left(-\frac{3}{2}\right)}}

\begin{equation}
	\mathcal{B}_{ABC}(z)\mathcal{W}^{--}(w)\sim \frac{\coeffBbWxmssone\overline{\mathcal{B}}_{ABC}}{(z-w)^2}+\frac{\coeffBbWxmsstwo(\overline{\mathcal{B}}_{ABC})'+\coeffBbWxmssthree\mathcal{J}_{\mathfrak{u}_1}\overline{\mathcal{B}}_{ABC}+\coeffBbWxmssfour\mathcal{J}_{A}^{D}\overline{\mathcal{B}}_{D BC}}{z-w}
\end{equation}

\newcommand{\coeffBbWmssone}{{-6}}
\newcommand{\coeffBbWmsstwo}{{-3}}
\newcommand{\coeffBbWmssthree}{{\left(-\frac{9}{2}\right)}}
\newcommand{\coeffBbWmssfour}{{\left(-\frac{3}{2}\right)}}

\begin{equation}
	\overline{\mathcal{B}}_{ABC}(z)\mathcal{W}^{++}(w)\sim \frac{\coeffBbWmssone\mathcal{B}_{ABC}}{(z-w)^2}+\frac{\coeffBbWmsstwo\mathcal{B}_{ABC}'+\coeffBbWmssthree\mathcal{J}_{\mathfrak{u}_1}\mathcal{B}_{ABC}+\coeffBbWmssfour\mathcal{J}_{A}^D\mathcal{B}_{D BC}}{z-w}
\end{equation}


\newcommand{\coeffWWAUbbone}{{-18}}
\newcommand{\coeffWWAUbbtwo}{{\left(-45\right)}}
\newcommand{\coeffWWAUbbthree}{{\left(-3\right)}}
\newcommand{\coeffWWAUbbfour}{{\left(-\frac{81}{2}\right)}}

\newcommand{\coeffWWAUaaone}{{\left(-18\right)}}
\newcommand{\coeffWWAUaatwo}{{\left(-36\right)}}
\newcommand{\coeffWWAUaathree}{{\left(\frac{3}{2}\right)}}
\newcommand{\coeffWWAUaafour}{{\left(\frac{189}{4}\right)}}
\newcommand{\coeffWWAUaafive}{{\left(27\right)}}
\newcommand{\coeffWWAUaasix}{{\left(-\frac{1}{4}\right)}}
\newcommand{\coeffWWAUaaseven}{{\left(-\frac{81}{8}\right)}}
\newcommand{\coeffWWAUaaeight}{{\left(\frac{3}{4}\right)}}

\begin{equation}
	\begin{split}
		\mathcal{W}^{++}(z)\mathcal{W}^{--}(w) & \sim \frac{90}{(z-w)^4}+\frac{-90\mathcal{J}_{\mathfrak{u}_1}}{(z-w)^3}+\frac{\coeffWWAUbbone\mathcal{T}+\coeffWWAUbbfour\mathcal{J}_{\mathfrak{u}_1}\mathcal{J}_{\mathfrak{u}_1}+\coeffWWAUbbtwo\mathcal{J}_{\mathfrak{u}_1}'+\coeffWWAUbbthree\mathcal{J}_{\mathfrak{a}_1}\mathcal{J}_{\mathfrak{a}_1}}{(z-w)^2}\\ &+\frac{\coeffWWAUaafive \mathcal{T}\mathcal{J}_{\mathfrak{u}_1}+\coeffWWAUaaone\mathcal{T}'+\coeffWWAUaafour\mathcal{J}_{\mathfrak{u}_1}'\mathcal{J}_{\mathfrak{u}_1}+\coeffWWAUaatwo\mathcal{J}_{\mathfrak{u}_1}''+\coeffWWAUaaseven\mathcal{J}_{\mathfrak{u}_1}\mathcal{J}_{\mathfrak{u}_1}\mathcal{J}_{\mathfrak{u}_1}}{z-w}\\&+\frac{\coeffWWAUaathree\mathcal{J}_{\mathfrak{a}_1}'\mathcal{J}_{\mathfrak{a}_1}+\coeffWWAUaasix\mathcal{B}\overline{\mathcal{B}}+\coeffWWAUaaeight\mathcal{J}_{\mathfrak{a}_1}\mathcal{J}_{\mathfrak{a}_1}\mathcal{J}_{\mathfrak{u}_1}}{(z-w)}
	\end{split}
\end{equation}

\paragraph{Schur Index.}
The vacuum character of the VOA can be computed to be 
\begin{equation}
\begin{split}
	\chi_{A_2U_1}(q)&=1+4q+8q^\frac{3}{2}+17q^2+36q^{\frac{5}{2}}+77q^3...\\
	&=	\text{PE}\left[\frac{4 q+8 q^{3/2}+3 q^2-4 q^{5/2}-14q^3...}{1-q}\right]\,.
\end{split}
\end{equation}
 Its refined version is given by
 \begin{equation}
 	\begin{split}
 		\chi_{A_1U_1}(q)&=1+(\mathbf{1}_0+\mathbf{3}_0)q+(\mathbf{4}_{1}+\mathbf{4}_{-1})q^\frac{3}{2}+(4\cdot\mathbf{1}_0+\mathbf{1}_2+\mathbf{1}_{-2}+\mathbf{3}_0+\mathbf{3}_0+\mathbf{5}_0)q^2+\\& +(3\cdot(\mathbf{4}_1+\mathbf{4}_{-1})+\mathbf{6}_1+\mathbf{6}_{-1})q^{\frac{5}{2}}+\\&+(8\cdot\mathbf{1}_0+2\cdot(\mathbf{1}_{2}+\mathbf{1}_{-2})+7\cdot\mathbf{3}_0+\mathbf{3}_2+\mathbf{3}_{-2}+2\cdot\mathbf{5}_0+2\cdot\mathbf{7}_0+\mathbf{7}_{2}+\mathbf{7}_{-2})q^3+...\\
 		&=	\text{PE}\left[
   \frac{\upchi_{\text{gen}}-\upchi_{\text{nulls}}+\dots}{1-q}\right]\,,
 	\end{split}
 \end{equation}
 where
\begin{subequations}
\label{chiA1U1Schur}
\begin{align}
\upchi_{\text{gen}}&=(\mathbf{1}_0+\mathbf{3}_0)q+(\mathbf{4}_{1}+\mathbf{4}_{-1})q^\frac{3}{2}+(\textcolor{blue}{\mathbf{1}_0}+\mathbf{1}_2+\mathbf{1}_{-2})q^2\,,
\\
\upchi_{\text{nulls}}&=(\mathbf{2}_1+ \mathbf{2}_{-1})q^{5/2}+(\mathbf{3}_0+\mathbf{3}_{2}+\mathbf{3}_{-2}+\mathbf{5}_0)q^3\,.
 \end{align}
 \end{subequations}
 The set of strong generators of the VOA is consistent with the index with four AKM currents at order $q$, eight  $\mathcal{B}_{\mathbf{4},\pm}$ generators at $q^{3/2}$ and $\mathcal{W}^{\pm\pm}$ and the stress tensor $T$, which we highlight as $\textcolor{blue}{\mathbf{1}_0}$, at order $q^2$.
 Additionally, there are null states of conformal dimension $5/2$ given by
 \begin{equation}
 \left(\mathcal{J}_{\mathfrak{a}_1}\mathcal{B}\right)_{\mathbf{2}_{+1}}=0\,,
 \qquad \quad
  \left(\mathcal{J}_{\mathfrak{a}_1}\overline{\mathcal{B}}\right)_{\mathbf{2}_{-1}}=0\,,
 \end{equation}

and at conformal weight $3$, there are nulls in representation $\mathbf{3}_0$, $\mathbf{3}_{2}$, $\mathbf{3}_{-2}$ and $\mathbf{5}_0$ given by
\begin{align}
\!\!\!\!\left(
\mathcal{B}\overline{\mathcal{B}}+
\mathcal{J}_{\mathfrak{a}_1}^3+
\!\mathcal{J}_{\mathfrak{a}_1}\partial\mathcal{J}_{\mathfrak{a}_1}+\partial^2\mathcal{J}_{\mathfrak{a}_1}
\!+
\mathcal{J}_{\mathfrak{u}_1}\mathcal{J}_{\mathfrak{a}_1}^2+\mathcal{J}_{\mathfrak{u}_1}^2\mathcal{J}_{\mathfrak{a}_1}+\partial\mathcal{J}_{\mathfrak{u}_1}\mathcal{J}_{\mathfrak{a}_1}+\mathcal{J}_{\mathfrak{u}_1}\partial\mathcal{J}_{\mathfrak{a}_1}+T\mathcal{J}_{\mathfrak{a}_1}
\right)_{\mathbf{3}_0}&=0
\\
\left(\mathcal{B}\,\mathcal{B}+\mathcal{J}_{\mathfrak{a}_1}\mathcal{W}^{++}\right)_{\mathbf{3}_2}&=0\\
\left(\overline{\mathcal{B}}\,\overline{\mathcal{B}}+\mathcal{J}_{\mathfrak{a}_1}\mathcal{W}^{--}\right)_{\mathbf{3}_{-2}}&=0\\
\left(\mathcal{B}\,\overline{\mathcal{B}}+\mathcal{J}_{\mathfrak{u}_1}\mathcal{J}_{\mathfrak{a}_1}^2+\mathcal{J}_{\mathfrak{a}_1}\partial\mathcal{J}_{\mathfrak{a}_1}\right)_{\mathbf{5}_{0}}&=0
\end{align}
There are more nulls at higher conformal weights, but we will not display them here.

\paragraph{Hall-Littlewood Index.}
The Hall-Littlewood Index can be computed by working in the leading $R$-filtration and is given by
\begin{equation}
	\begin{split}
		\mathcal{I}_{HL}&=1+4 t^2+8 t^3+12 t^4+28 t^5+49 t^6+...\\
		&=\text{PE}[4 t^2+8 t^3+2 t^4-4 t^5-15 t^6+...]\,,
	\end{split}
\end{equation}
with its refined version given by
 \begin{equation}
	\begin{split}
			\mathcal{I}_{HL}&=1+(\mathbf{1}_0+\mathbf{3}_0)t^2+(\mathbf{4}_{1}+\mathbf{4}_{-1})t^3+(2\cdot\mathbf{1}_0+\mathbf{1}_2+\mathbf{1}_{-2}+\mathbf{3}_0+\mathbf{5}_0)t^4\\ &+(2\cdot(\mathbf{4}_1+\mathbf{4}_{-1})+\mathbf{6}_1+\mathbf{6}_{-1})t^5\\
		&+((2\cdot\mathbf{1}_0+\mathbf{1}_{2}+\mathbf{1}_{-2})+2\cdot\mathbf{3}_0+\mathbf{3}_2+\mathbf{3}_{-2}+\mathbf{5}_0+(2\cdot\mathbf{7}_0+\mathbf{7}_{2}+\mathbf{7}_{-2}))t^6+...\\
		&=	\text{PE}\left[
  \upchi_{\text{gen}}^{\text{HL}}-\upchi_{\text{nulls}}^{\text{HL}}+\dots  \right]
	\end{split}
\end{equation}
where now
\begin{subequations}
\label{chiA1U1HL}
\begin{align}
\upchi_{\text{gen}}^{\text{HL}}&= (\mathbf{1}_0+\mathbf{3}_0)t^2+(\mathbf{4}_{1}+\mathbf{4}_{-1})t^3+(\mathbf{1}_2+\mathbf{1}_{-2})t^4\,,
\\
\upchi_{\text{nulls}}^{\text{HL}}&= (\mathbf{2}_1+ \mathbf{2}_{-1}) t^5+(\textcolor{red}{\mathbf{1}_0}+\mathbf{3}_0+\mathbf{3}_{2}+\mathbf{3}_{-2}+\mathbf{5}_0)t^6\,.
 \end{align}
 \end{subequations}
 Comparing this expression with its Schur counterpart \eqref{chiA1U1Schur} we see that the VOA has an additional generator identified with the stress tensor, see the $\textcolor{blue}{\mathbf{1}_0}$ in \eqref{chiA1U1Schur}, while the HB has an additional relation not associated with a null state in the VOA denoted by $\textcolor{red}{\mathbf{1}_0}$. As in the previous example this HB relation is accociated to a composite operator in the VOA, which we call $W_3$, with lower degree in the R-filtration compared to the one of its constituents. This operator should be regarded as the analogue of the stress tensor for the $C_2U_1$ theory previously discussed.
 The operator  $W_3$ transforms in the $\mathbf{1}_0$ representation and has conformal weight $3$, there are $8$ operators in the VOA with these quantum numbers, half of which are quasi-primary with three being Virasoro primary\footnote{The quasi-primary which is not Virasoro primay is $T\,\mathcal{J}_{\mathfrak{u}_1}-\tfrac{1}{2}\partial^2 \mathcal{J}_{\mathfrak{u}_1}$.}. A single linear combination of these three Virasoro primaries has degree $R=2<3$ and it is schematically given by
 \begin{equation}
\label{eq:ora1u1}
W_3=\left(\mathcal{B}\overline{\mathcal{B}}+\mathcal{J}_{\mathfrak{a}_1}^2\mathcal{J}_{\mathfrak{u}_1}^{}
+\mathcal{J}_{\mathfrak{a}_1}\partial\mathcal{J}_{\mathfrak{a}_1}
+
\mathcal{J}_{\mathfrak{u}_1}\partial\mathcal{J}_{\mathfrak{u}_1}
+\partial^2\mathcal{J}_{\mathfrak{u}_1}+T\mathcal{J}_{\mathfrak{u}_1}+\partial T
\right)_{\mathbf{1}_0}\,.
\end{equation}
Notice that only seven of the eight terms appear in this linear combination with the term $\mathcal{J}_{\mathfrak{u}_1}^3$ appearing with coefficient zero.
The relative coefficients between $\mathcal{B}\overline{\mathcal{B}}$ and $\mathcal{J}_{\mathfrak{a}_1}^2\mathcal{J}_{\mathfrak{u}_1}^{}$ is fixed in such a way that there is a drop of the degree in $R$ while the remaining coefficients are fixed by the requirement that $W_3$ is a Virasoro primary.
Let us take a closer look at the explicit expression of $W_3$. In the free field realization, the drop in the $R$ filtration of the composite operator $W_3$ is a consequence of the existence of the following null state\footnote{This is a special feature of the $\mathfrak{a}_1$ Deligne theory, in the other cases, expect for $\mathfrak{a}_0$, there are nulls already at conformal weight two.} of the IR VOA 
\begin{equation}
\mathcal{O}^{\text{IR}}_{\text{null}}=T_{\text{IR}}\,
\mathcal{J}^{\text{IR}}_{\mathfrak{u}_1}+
\partial\mathcal{J}^{\text{IR}}_{+2}\,\mathcal{J}^{\text{IR}}_{-2}
-\mathcal{J}^{\text{IR}}_{+2}\,\partial\mathcal{J}^{\text{IR}}_{-2}
-\tfrac{1}{3} \partial^2\mathcal{J}^{\text{IR}}_{\mathfrak{u}_1}\,.
\end{equation}
Once we set this to zero, as it should be done,  the leading term of $W_3$ in the R-filtration takes the form 
\begin{equation}
\label{W3inA1U1Leadingfiltration}
\begin{split}
&(20\,
T_{\text{IR}}+13\,(T_{\beta\gamma}-T_{\delta\varphi}))\,(3v_-+2j_{\beta\gamma})
-\mathcal{J}^{\text{IR}}_{\mathfrak{u}_1} 
(14\, T_{\delta\varphi}+52\, T_{\beta\gamma}
+22\, v_+ j_{\beta\gamma}
-33\,\partial j_{\beta\gamma}
)\\
&11\,(\mathcal{J}^{\text{IR}}_{\mathfrak{u}_1})^2\,v_+
+198(v_+\gamma-3\,\partial\gamma)\gamma^2\,\mathcal{J}^{\text{IR}}_{-2}\,e^{-\frac{1}{3}}
+19\,(\mathcal{J}^{\text{IR}}_{+2}\partial \mathcal{J}^{\text{IR}}_{-2}-
\mathcal{J}^{\text{IR}}_{-2}\partial \mathcal{J}^{\text{IR}}_{+2})
-\tfrac{33}{2}\,
\mathcal{J}^{\text{IR}}_{\mathfrak{u}_1} \partial \mathcal{J}^{\text{IR}}_{\mathfrak{u}_1} 
\,,
\end{split}
\end{equation}
where
\begin{equation}
v_+=\partial(\delta+\varphi)\,,
\quad
v_-=-\tfrac{1}{\langle \varphi,\varphi \rangle}\partial(\delta-\varphi)\,,
\quad
j_{\beta\gamma}=\beta\gamma\,,
\quad 
T_{\beta\gamma}=\tfrac{1}{2} \left(\beta  \partial \gamma -\gamma  \partial\beta \right)\,,
\quad 
T_{\delta\varphi}=\frac{v_- v_+}{2}\,.
\end{equation}
The composite operator $W_3$ should be regarded as the generalization of the stress tensor in the $C_2U_1$ example.

\section{Free field realization for all rank-one theories with ECB}
\label{FFallrank1}

In this section we present the free field realization of all rank-one theories with an ECB.
As in the examples discussed in the previous section, we will first describe the singular locus of the HB which is $\mathbb{C}^{2\quatdim}/\mathbb{Z}_{\ell}$ and explain in which sense the Deligne rank-one theories are fibered over it. This requires specifying an appropriate $\mathbb{Z}_{\ell}$ action on the IR Deligne currents.
Next we turn to the construction of the VOA generators. 
Again, some of the VOA generators are very easy to construct in terms of free fields, the remaining ones follow from the OPEs once a relatively simple ``lowering generator'' is built. 
Finally, we show that the OPEs close and present two important applications of the free field construction: determination of null states and the R-filtration. As we will recall, both are related to the relations satisfied by the HB generators. 

\subsection{The free field realization}
\label{sec:ffrgeneral}
\paragraph{The singular locus.}
Let us start by looking at the space $W=\mathbb{C}^{2\quatdim}/\mathbb{Z}_{\ell}$.
We denote by $(X_A,Y^A)$ with $A=1,\dots,\quatdim$, the coordinates of $\mathbb{C}^{2\quatdim}$ on which $\mathbb{Z}_{\ell}$ acts as
\begin{equation}
(X_A,Y^A)\mapsto (\omega_{\ell}^{}\,X_A,\omega_{\ell}^{-1}\,Y^A)\,,
\qquad
\omega_{\ell}=e^{2\pi i /\ell}\,.
\end{equation}
The symplectic structure is given by $\{Y^A,X_B\}:=\delta^A_B$ and is compatible with the $\mathbb{Z}_{\ell}$ action.
The ring of invariants is generated by 
\begin{align}
\label{C2hmodZlgenerators}
\begin{split}
b_{A_1\dots A_{\ell}}\,=&\, X_{A_1}\cdots X_{A_{\ell}}\,,\qquad\qquad
\bar{b}^{A_1\dots A_{\ell}}\,=\,Y^{A_1}\cdots\, Y^{A_{\ell}}\,,\\
M_A^B\,=&\,X_{A}Y^B-\tfrac{1}{\quatdim}\,\delta^A_B\,\,X_C Y^C\,,\qquad\,\,\,\,\,
m\,=\,X_C Y^C\,,
\end{split}
\end{align}
where $M^A_B$ and $m$ generate the symmetry $\mathfrak{su}(n)\oplus \mathfrak{u}(1)$ which is enhanced to $\mathfrak{usp}(2n)=\mathfrak{c_n}$ for $\ell=2$. In this case $W$ is the closure of the minimal nilpotent orbit of $\mathfrak{c}_\quatdim$.

As recalled in Section \ref{Delignereview}, the generalized free field realization should be thought as an inverse Higgsing performed at the level of the VOA. In this case, the relevant Higgsing corresponds to a point on the singular locus $W$ for which  the only non vanishing generator is the component  $b_{\quatdim\quatdim\dots \quatdim}\neq 0$ of $b$.
Associated to this Higgsing there is a Zariski open subset of $W$ defined by the condition  $b_{\quatdim\quatdim\dots \quatdim}=:\mathsf{e}\neq 0$ which can be identified with
\begin{equation}
\frac{ T^*(\mathbb{C}^*)\times \mathbb{C}^{2(\quatdim-1)}}{\mathbb{Z}_{\ell}}\,.
\end{equation}
Accordingly, we write
\begin{align}
\label{C2hmodZlgeneratorsXY}
\begin{split}
(X_1,\dots,X_{\quatdim-1},X_\quatdim)\,=\,& (\betaclass_1,\dots,\betaclass_{\quatdim-1},\,\mathsf{e}^{1/\ell})\,,\\
(Y^1,\dots,Y^{\quatdim-1},Y^\quatdim)\,=\,& (\gammaclass^1,\dots,\gammaclass^{\quatdim-1},\mathsf{h}\mathsf{e}^{-1/\ell})\,,
\end{split}
\end{align}
where $\mathsf{e}^{1/\ell},\mathsf{h}$ are coordinates of $T^*(\mathbb{C}^*)$
with $\{\mathsf{h},\mathsf{e}^{1/\ell}\}=\mathsf{e}^{1/\ell}$ and
$\{\gammaclass^a,\betaclass_b\} =\delta^a_b$, see \eqref{Zellactiononfreefields}.
For later convenience we collect the form of $M_A^B$ and $m$ in this patch
\begin{equation}
\label{MABandminpatch}
\begin{aligned}
M^B_A&=
\begin{pmatrix}
\hat{M}_a^b-\tfrac{1}{\quatdim}\,\mathsf{z}\,\delta^b_a &\,\, \betaclass_a \mathsf{h}\mathsf{e}^{-1/\ell}\\
\gammaclass^b\mathsf{e}^{1/\ell} & \,\tfrac{\quatdim-1}{\quatdim}\,\mathsf{z}
\end{pmatrix}\,,
\\
m&=\,\,\mathsf{h}+\betaclass_c\gammaclass^c\,,
\end{aligned}
\qquad\quad
\begin{aligned}
\hat{M}_a^b&=\betaclass_a\gammaclass^b-\tfrac{1}{\quatdim-1}\delta^b_a\,\betaclass_c\gammaclass^c\,\\
\mathsf{z}&=\mathsf{h}-\tfrac{1}{\quatdim-1}\betaclass_c\gammaclass^c\,,
\end{aligned}
\end{equation}
where $a,b,c=1,\dots,\quatdim-1$. We denote by $\mathfrak{u}(1)_m$ and $\mathfrak{u}(1)_{\mathsf{z}}$ the abelian algebras corresponding to $m$ and $\mathsf{z}$  and by $\mathfrak{u}(1)_{\text{unb}}$ the combination that acts trivially on $\mathsf{e}$ whose generator is  $\mathsf{z}-m=-\tfrac{\quatdim}{\quatdim-1} \betaclass_c \gammaclass^c$.
We will call generalized highest weight (ghw) states the states that are annihilated by the raising generator  represented by  $\gammaclass^b\mathsf{e}^{1/\ell}$. For the $b$ and $\bar{b}$ generators the corresponding generalized highest weight states take the form
\begin{equation}
\label{bandbarbgeneralizedHWS}
b^{\text{ghw}}=\mathsf{e}\,,
\qquad
(\bar{b}^{\text{ghw}})^{a_1\dots a_{\ell}}=\gammaclass^{a_{1}}\dots \gammaclass^{a_{\ell}}\,.
\end{equation}
As we will see, the first one does not change as we move away from the singular locus $W$ onto the general point of the HB, while the second is modified.
\paragraph{The $\mathbb{Z}_{\ell}$ action on IR currents.}
Next, let us present the relevant $\mathbb{Z}_{\ell}$ automorphisms of Deligne currents.
An important uniform feature of the choice of $\mathbb{Z}_{\ell}$ automorphisms is that the   $\mathbb{Z}_{\ell}$ invariant part of the IR symmetry takes the form 
\begin{equation}
\label{gIR0general}
(\mathfrak{g}_{\text{IR}})_0 
=
\begin{cases}
\mathfrak{c}_{\quatdim-1} \oplus \mathfrak{f}\,, & \ell=2\,,\\
\mathfrak{a}_{\quatdim-2} \oplus \mathfrak{f}\,, & \ell\neq 2\,,
\end{cases}
\end{equation}
where $\quatdim$,  the quaternionic dimension of the singular locus $W$, and the factor $\mathfrak{f}$ depend on the theory: $\mathfrak{f}=\mathfrak{a}_1$ for the $C_3 A_1$ theory, $\mathfrak{f}=\mathfrak{u}_1$ for the $C_2 U_1$ and $A_1U_1$ theories and trivial in the other cases.
The symmetry \eqref{gIR0general} should be compared with the UV flavor symmetry which follows the same uniform pattern
\begin{equation}
\label{gUValltheories}
\mathfrak{g}_{\text{UV}}
=
\begin{cases}
\,\,\,\mathfrak{c}_{\quatdim} \,\,\,\oplus \mathfrak{f}\,, & \ell=2\,,\\
\mathfrak{a}_{\quatdim-1} \oplus \mathfrak{f}\,, & \ell\neq 2\,.
\end{cases}
\end{equation}
The list of the relevant $\mathbb{Z}_{\ell}$ automorphisms is the following.
For the $\mathbb{Z}_2$ cases we have the branching rules
\begin{subequations}
\label{Z2actiononCurrents}
\begin{align}
\mathcal{J}^{\text{IR}}_{\mathfrak{e}_6}&
\rightarrow 
\mathcal{J}^{\text{IR}}_{\mathfrak{c}_4}
\oplus
\mathcal{J}^{\text{IR}}_{\mathbf{42}}=
\mathcal{J}^{\text{IR}}_{mn}
\oplus
\mathcal{J}^{\text{IR}}_{\llbracket mnpq\rrbracket}\,,
\\
\mathcal{J}^{\text{IR}}_{\mathfrak{d}_4}&
\rightarrow 
\mathcal{J}^{\text{IR}}_{\mathfrak{c}_2\oplus\mathfrak{a}_1}
\oplus
\mathcal{J}^{\text{IR}}_{(\mathbf{5},\mathbf{3})}=
\left(
\mathcal{J}^{\text{IR}}_{mn}
\oplus
\mathcal{J}^{\text{IR}}_{(IJ)}
\right)
\oplus
\mathcal{J}^{\text{IR}}_{\llbracket mn\rrbracket,(IJ)}\,,
\\
\mathcal{J}^{\text{IR}}_{\mathfrak{a}_2}&
\rightarrow 
\mathcal{J}^{\text{IR}}_{\mathfrak{c}_1\oplus\mathfrak{u}_1}
\oplus
\mathcal{J}^{\text{IR}}_{\mathbf{2}_+\oplus \mathbf{2}_-}=
\left(
\mathcal{J}^{\text{IR}}_{mn}
\oplus
\mathcal{J}^{\text{IR}}
\right)
\oplus
\left(
\mathcal{J}^{\text{IR}}_{ m,+}
\oplus 
\mathcal{J}^{\text{IR}}_{ m,-}
\right)\,,
\end{align}
\end{subequations}
where the first term is $\mathbb{Z}_2$ even and the second is $\mathbb{Z}_2$ odd.
For the $\mathbb{Z}_3$ cases the relevant branchings are  
\begin{subequations}
\label{Z3actiononCurrents}
\begin{align}
\mathcal{J}^{\text{IR}}_{\mathfrak{d}_4}&
\rightarrow 
\mathcal{J}^{\text{IR}}_{\mathfrak{a}_3}
\oplus
\mathcal{J}^{\text{IR}}_{\mathbf{10}}
\oplus
\mathcal{J}^{\text{IR}}_{\bar{\mathbf{10}}}=
(\mathcal{J}^{\text{IR}})^{a}_b
\oplus
(\mathcal{J}^{\text{IR}})_{(abc)}
\oplus
(\mathcal{J}^{\text{IR}})^{(abc)}\,,
\\
\label{Z3forA1U1}
\mathcal{J}^{\text{IR}}_{\mathfrak{a}_1}&
\rightarrow 
\mathcal{J}^{\text{IR}}_{\mathfrak{u}_1}
\oplus
\mathcal{J}^{\text{IR}}_{-2}
\oplus
\mathcal{J}^{\text{IR}}_{+2}\,,
\end{align}
\end{subequations}
where the first, second and third terms after the arrow have eigenvalues $1,\omega_3^{},\omega_3^{-1}$ under the  $\mathbb{Z}_3$ action respectively.
Finally, in the $\mathbb{Z}_4$ case  only a $\mathbb{Z}_2$ acts non trivially with even and odd part respectively given by
\begin{equation}
\label{Z4action}
\mathcal{J}^{\text{IR}}_{\mathfrak{a}_2}\rightarrow 
\mathcal{J}^{\text{IR}}_{\mathfrak{a}_1}
\oplus
\mathcal{J}^{\text{IR}}_{\mathbf{5}}=
(\mathcal{J}^{\text{IR}})^{a}_b
\oplus
(\mathcal{J}^{\text{IR}})_{(abc d)}
\end{equation}
Notice that we used $m,n,..=1,\dots, 2r$ for indices of $\mathfrak{c}_r=\mathfrak{sp}(2r)$ and $a,b,...=1,\dots,r+1$ for  $\mathfrak{a}_{r}=\mathfrak{su}(r+1)$.
The notation $\llbracket m n \dots \rrbracket$ indicates antisymmetrization and removal of $\Omega$ traces while $(ab..)$ indicates that the indices are totally symmetrized.

\paragraph{Fibering the Deligne rank-one theories over the singular locus.}
As already discussed in detail, the starting point for the genealized free field construction is the identification of the patch of the Higgs branch where the generator that is getting a VEV, which we call $\mathsf{e}$, is non vanishing.
As in \eqref{Ueidentification} this open subset takes the form
\begin{equation}
\mathcal{U}_{\mathsf{e}}\simeq
\frac{ T^*(\mathbb{C}^*)\times \mathbb{C}^{2(\quatdim-1)} \times \mathcal{M}_H[\mathcal{T}_{IR}]}{\mathbb{Z}_{\ell}}\,,
\end{equation}
where the $\mathbb{Z}_{\ell}$ action on the $T^*(\mathbb{C}^*)$ and $\mathbb{C}^{2(\quatdim-1)}$ is the same as given in \eqref{Zellactiononfreefields} while the action of the IR currents, which are the generators of the $\mathcal{M}_H[\mathcal{T}_{IR}]$ chiral ring, has been introduced in equations \eqref{Z2actiononCurrents}, \eqref{Z3actiononCurrents} and \eqref{Z4action}.
Accordingly we will realize the VOA as a subVOA 
\begin{equation}
\label{VsubVOAallrank1}
\mathbb{V}[\mathcal{T}]
\subset  \left(\Pi_{\frac{1}{\ell}} \otimes \mathbb{V}_{\xi}\otimes V_{-\frac{h^\vee}{6}-1}(\mathfrak{g}_{\text{IR}})\right)^{\mathbb{Z}_{\ell}}\,,
\end{equation}
where $\mathcal{T}\in \{C_5,C_3A_1,C_1U_1,A_3,A_1U_1,A_2\}$ and for each of the six cases we have already discussed the choice of $\mathfrak{g}_{\text{IR}}$, $\mathbb{Z}_{\ell}$ action and number of symplectic bosons $\xi$.

The next step is to identify the part of the UV symmetry, denoted by $\mathfrak{g}^{\natural}_{\text{UV}}$, which is not broken by the choice of VEV  (mathematically, it is the semi-simple factor of the subalgebra of $\mathfrak{g}_{\text{UV}}$ that acts trivially on $\mathsf{e}$) and the distinguished $\mathfrak{u}(1)\subset \mathfrak{g}_{\text{UV}}$ that commutes with\footnote{If $\mathfrak{g}^{\natural}_{\text{UV}}$ contains abelian factors we should add the requirement that the current associated with the distinguished $\mathfrak{u}(1)$ has regular OPE with all the $\mathfrak{g}^{\natural}_{\text{UV}}$ currents.}  $\mathfrak{g}^{\natural}_{\text{UV}}$. These symmetries are important since the associated quantum numbers of both the (generalized) free field ingredients and UV VOA are under control.
The unbroken symmetry can have two factors:
\begin{itemize}
\item[(i)] symmetries which remain unbroken everywhere on the singular locus $W=\mathbb{C}^{2\quatdim}/\mathbb{Z}_{\ell}$,
\item[(ii)] symmetries that do act on $W$, but are unbroken by the VEV $\mathsf{e}$.
\end{itemize}
Concerning the second factor, it is easy to identify the symmetries of $W$ that remain unbroken by the VEV, they are $\mathfrak{c}_{\quatdim-1}$ in the $\ell=2$ cases and $\mathfrak{a}_{\quatdim-2}\oplus \mathfrak{u}(1)_{\text{unb}}$ in the $\ell=3,4$ cases,
where the form of $\mathfrak{u}(1)_{\text{unb}}$ is spelled out below \eqref{MABandminpatch}.
Naively, one would conclude that these are the unbroken symmetries of type (ii).
While this is correct for the $\ell=2$ cases, and this is related to the fact that the symmetries of $W$ in these instances have no abelian factors, it is not always true for $\ell=3,4$. In these cases the contribution to the unbroken symmetries (ii) can be smaller.
This happens for the $A_3$ and $A_2$ theories (for which $d=4$ and $3$ respectively) as it can be anticipated  since the symmetry of $W$ is bigger then the one of  $\mathcal{T}_{\text{UV}}$ by an extra $\mathfrak{u}(1)$ factor generated by $m$, see \eqref{C2hmodZlgenerators}. In these two cases the generator $m$ is not the restriction of a function\footnote{Apart from these two exceptions, the remaining generators of $\mathbb{C}[W]$, $b,\bar{b}$ and $M$ are associated to  the restriction of functions on $\mathcal{M}_H[\mathcal{T}_{UV}]$ to the singular locus $W$.} on  $\mathcal{M}_H[\mathcal{T}_{UV}]$. Since the generator of $\mathfrak{u}(1)_{\text{unb}}$ contains a factor of $m$, it is not part of the symmetries of $\mathcal{T}_{\text{UV}}$ and in particular cannot be unbroken\footnote{Not all the symmetries of the singular locus have to come from the restriction of a symmetry of the Higgs Branch to the singular locus. More generally, functions on the singular locus are not necessarily the restriction of functions on the whole HB to the singular locus.}. We conclude that for the $A_3$ and $A_2$ theories the unbroken symmetry is $\mathfrak{a}_{\quatdim-2}$.
The only remaining case is $\mathcal{T}=A_1U_1$ for which $d=2$. We know how the Cartan of $A_1$ acts on $\mathsf{e}$ since it is the highest weight state of a four dimensional representation and it will have some charge under $U_1$.  Now we can form two linear combinations of the Cartan generator of the $A_1$ factor and the generator of the $U_1$ factor. One combination will act trivially on $\mathsf{e}$ and the other (defined up to the addition of the one with trivial action) will act non trivially. We conclude that the combination of type (i) and (ii) unbroken symmetries for the $A_1U_1$ theory is a single $\mathfrak{u}(1)$. 

To conclude the analysis we need to discuss unbroken symmetries of type (i). It is easy to see by looking at the UV flavor symmetries  \eqref{gUValltheories} that the factor $\mathfrak{f}$ for the $\ell=2$ theories, namely $C_3A_1$ and $C_2U_1$, cannot act on the singular locus $W$ so it is part of the unbroken symmetries of type (i). The only remaining case with a non-trivial $\mathfrak{f}$ factor is the $A_1U_1$ theory that we already discussed.
The important conclusion is that the unbroken symmetry coincides, at the Lie algebra level, with the $\mathbb{Z}_{\ell}$ invariant part of the IR currents given in \eqref{gIR0general}, in equation
\begin{equation}
\label{gIR0generalisgunbroken}
\unbrSYMM=(\mathfrak{g}_{\text{IR}})_0 
\,.
\end{equation}
Notice that to derive this result we used the knowledge of the flavor symmetry of the UV theory, in a bottom-up construction we could have taken the fact that the symmetry unbroken by the VEV coincides with the $\mathbb{Z}_{\ell}$ invariant part of the IR flavor symmetry as the starting point to bootstrap the UV theory.

As reviewed in Section \ref{Delignereview}, the free field realization of the currents associated with the unbroken symmetries follows the general pattern given in equation \eqref{JnaturalasSUMgeneral}. In our case it gives
\begin{equation}
\label{gnaturalUVminusfcurrents}
\begin{cases}
\mathcal{J}_{\mathfrak{c}_{\quatdim-1}} =\mathcal{J}^{\text{IR}}_{\mathfrak{c}_{\quatdim-1}}+ 
\xi\xi\,,\,\,\,\,& \ell=2\,,\\
\mathcal{J}_{\mathfrak{a}_{\quatdim-2}} =\mathcal{J}^{\text{IR}}_{\mathfrak{a}_{\quatdim-2}}+\beta \gamma\,,\,\,\,\, & \ell\neq 2\,,
\end{cases}
\end{equation}
where $\beta \gamma$ here is a shorthand for $\beta_a\gamma^b-\tfrac{1}{\quatdim-1}\delta_a^b\gamma_c\beta^c$
and
\begin{equation}
\label{gnaturalfcurrents}
\begin{cases}
\mathcal{J}_{\mathfrak{f}}^{}=\mathcal{J}_{\mathfrak{f}}^{\text{IR}}\,,\,\,\,\,& \text{ $\mathcal{T}=C_3A_1$ or $\mathcal{T}=C_2U_1$}\,,\\
\mathcal{J}_{\mathfrak{f}=\mathfrak{u}_1}^{}=
\tfrac{1}{3}\left( h+\beta\gamma-2\mathcal{J}_{\mathfrak{u}_1}^{\text{IR}}
\right)\,,\,\,\,\,&  \,\,\,\text{$\mathcal{T}= A_1U_1$}\,,
\end{cases}
\end{equation}
while for the remaining theories the factor $\mathfrak{f}$ is absent.
For the $A_1U_1$ theory the form  of $\mathcal{J}_{\mathfrak{f}}$ was derived in Section \ref{sec:subsectionA1U1}, see \eqref{ZandJcurrentforA1U1}. Notice that (only) in this case $\mathcal{J}_{\mathfrak{f}=\mathfrak{u}_1}$ acts non trivially on $e(z)$.
As a confirmation that we are on the right track we can compute the levels of the (non-abelian) unbroken symmetries and verify that it reproduces the correct values, see Table \ref{levelsTable}. It is interesting to notice that the value of the levels is an output of our construction.

\begin{table}[t]
\begin{center}
\begin{adjustbox}{center}
\begin{tabular}{| c | c| c | c | c|c | c |c | c | c | }

 \hline
 
 Theory & $\ell$ & $\quatdim$ & $(\mathfrak{g}_{IR})_0 \subset \mathfrak{g}_{IR}$ & 
 $\mathcal{R}_{\xi}$ & $k_{IR}$ & $k_{\xi}$ & 
 $I_{G\hookrightarrow H}$&$k_{UV}$\\ 
 \hline
 \hline
 $C_5$ & $2$ & 5 & $\mathfrak{c}_4\subset \mathfrak{e}_6$ &
$\mathbf{8}$  & $-3$& $-\frac{1}{2}$ & $1$ &$-\frac{7}{2}$\\  
 $C_3A_1$  & $2$ & 3 & $\mathfrak{c}_2\oplus\mathfrak{a}_1\subset\mathfrak{d}_4$  &
 $(\mathbf{4},\mathbf{1})$  & $(-2,-2)$ & $(-\frac{1}{2},0)$ & $(1,2)$ & $(-\frac{5}{2},-4)$\\
 $C_2U_1$ & $2$ &2& $\mathfrak{c}_1\oplus \mathfrak{u}_1\subset\mathfrak{a}_2$ &
 $(\mathbf{2},\mathbf{1})$ &  $-\frac{3}{2}$& $-\frac{1}{2}$ & $1$ & $-2$\\
 \hline
 $A_3$ & $3$ & 4 &$\mathfrak{a}_2\subset\mathfrak{d}_4$  &
 $\mathbf{3}\oplus\overline{\mathbf{3}}$ &$-2$ &$-1$ &$3$ & $-7$\\  
 $A_1U_1$
 &$3$ & 2 & $\mathfrak{u}_1\subset\mathfrak{a}_1$ &
 $\mathbf{1}_+\oplus\mathbf{1}_-$  &$-\frac{4}{3}$ & & &$-5$\\
 \hline
 $A_2$ & $4$ & $3$ & $\mathfrak{a}_1\subset\mathfrak{a}_2$
 &
$\mathbf{2}\oplus\overline{\mathbf{2}}$ &
 $-\frac{3}{2}$ & $-1$ & $4$ & $-7$ \\  
 \hline
\end{tabular}
\end{adjustbox}
\end{center}
\caption{This table gives data on how to obtain AKM level $k_{UV}$ from $k_{IR}$ using the formula $k_{UV}=I_{G\hookrightarrow H}k_{IR}+k_{\xi}$ which follows from \eqref{gnaturalUVminusfcurrents} and \eqref{gnaturalfcurrents} for the non-abelian symmetries.
More precisely, this formula applies to the non-abelian part of the $(\mathfrak{g}_{IR})_0$ symmetry.
The level of the $A_1$ factor for the $A_1U_1$ theory is determined from the explicit free field realization.
}
\label{levelsTable}
\end{table}

Having identified the unbroken symmetry we will now identify the distinguished  $U(1)$ that commutes, more precisely that has regular OPEs with,  the unbroken symmetry.
For the $\ell=2$ cases, there is a full $\mathfrak{c}_1=\mathfrak{sl}(2)$ that commmutes with $\mathfrak{c}_{\quatdim-1}\subset \mathfrak{c}_{\quatdim}$ and its Cartan generator is associated with the distinguished  $U(1)$. For the $A_2$ and $A_3$ theories it is the affinization of the generator $\mathsf{z}$ introduced in \eqref{MABandminpatch} and, finally,  for the $A_1U_1$ theory it is a combination of the affinization of $\mathsf{z}$ and $\mathcal{J}_{\mathfrak{u}_1}^{\text{IR}}$.
After fixing  the normalization of this distinguished $\mathfrak{u}(1)$ from the condition that the associated currents $j(z)$ satisfies
\begin{equation}
\label{jeOPEs}
j(z)e(w)\sim \frac{\ell\, e(w)}{(z-w)}\,,
\qquad
e(z)=e^{\delta+\varphi}\,,
\end{equation}
we thus have
\begin{equation}
\label{jzformula}
j(z)=
\begin{cases}
h(z)\,, \qquad &\text{for $\ell=2$,}\\
\mathcal{Z}(z)=h-\tfrac{1}{d-1}\beta_c\gamma^c\,, \qquad\qquad &\text{for the $A_3$ ($d=4$) and $A_2$ ($d=3$) theories,}\\
\mathcal{Z}(z)=h-\beta\gamma+\mathcal{J}_{\mathfrak{u}_1}^{\text{IR}} 
\,,
 \qquad\qquad &\text{for the $A_1U_1$ theory,}
\end{cases}
\end{equation}
where $h(z)=\tfrac{\ell}{\langle \varphi, \varphi \rangle}\,\partial \varphi(z)$ and $\mathcal{Z}(z)$ should be regarded the affinization of $\mathsf{h}$ and $\mathsf{z}$ respectively (with the latter receiving further corrections determined in Section \ref{sec:subsectionA1U1} for the $A_1U_1$ theory).
At this point the value of $\langle \varphi, \varphi \rangle$ has not yet been fixed. 
This can be done by recalling that we know how $j(z)$ is embedded in the $\mathfrak{c}_{\quatdim}$ or $\mathfrak{a}_{\quatdim-1}$ factor of the flavor symmetry. 
In the case $\ell=2$ this is the Cartan of $\mathfrak{sl}(2)$ at level $k_{\mathfrak{c}}$ and one easily finds that $\langle \varphi, \varphi \rangle=2/k_{\mathfrak{c}}$, so that  $h(z)=k_{\mathfrak{c}}\,\partial \varphi$.
In the cases $\ell \neq 2$, we have that
$\mathcal{Z}=\tfrac{\quatdim}{\quatdim-1}\,\mathcal{J}^{\quatdim}_{\quatdim}$ (no sum over $\quatdim$), so that\footnote{Recall the general form of the OPEs given in \eqref{SUNOPEs}.}
\begin{equation}
\label{ZZOPEsgeneral}
\mathcal{Z}(z)\mathcal{Z}(w)\sim 
\frac{
\left(\tfrac{\quatdim}{\quatdim-1}\right)
\,k_{\mathfrak{a}}
}{(z-w)^2}\,,
\end{equation}
where $k_{\mathfrak{a}}$ is the $\mathfrak{a}_{\quatdim-1}$ level. Computing this OPE using the expressions given in \eqref{jzformula} gives the condition
\begin{equation}
\langle \varphi, \varphi \rangle=
\begin{cases}
\frac{(d-1)\ell^2}{1+d\,k_{\mathfrak{a}}}\,,
\qquad\qquad \text{for $A_2$ and $A_3$}\,,
\\
\frac{3^2}{ 1+2(k_{\mathfrak{a}}-k_{\text{IR}})}\,,
\qquad\,
\text{for $A_1U_1$}\,,
\end{cases}
\end{equation}
see also \eqref{ZZOPEinfootnote} for the $A_1U_1$ theory. We could cover also the case $\ell=2$ by these considerations by recalling that in this case  $\mathcal{Z}=h-\tfrac{1}{d-1}(\beta_c\gamma^c+\mathcal{J}^{\text{IR}}_{\mathfrak{u}_1\subset \mathfrak{c}_{\quatdim-1}})$.
In this case \eqref{ZZOPEsgeneral} give us back the condition  $\langle \varphi, \varphi \rangle=2/k_{\mathfrak{c}}$ after recalling that $k_{\mathfrak{a}}= 2k_{\mathfrak{c}}$ from its embedding.
We can also compute the central charge $c$ that follows from the free field construction as $c=c_{\text{IR}}+c_{\beta \gamma}+c_{\delta,\varphi}$ where $c_{\delta,\varphi}$ is given in \eqref{cdeltaphiSectionreview}. For convenience we also report the $\mathfrak{a}_{\quatdim-2}\oplus \mathfrak{u}(1)\subset \mathfrak{a}_{\quatdim-1}$ quantum numbers of $\mathsf{e}$, $\mathsf{h}$, $\betaclass$, $\gammaclass$ in Table \ref{tablebranchingsuh}.
\begin{table}
\centering
    \begin{tabular}{|c|c|c|c|c|}
\hline
&$\mathsf{h}$  & $\mathsf{e}^{1/\ell}$ &    $\beta$   &   $\gamma$     \\
\hline
$\mathfrak{a}_{\quatdim-2}\oplus \mathfrak{u}(1)$  &  $(1,0)$  & $(1,-(\quatdim-1))$ &
$(\quatdim-1,1)$& $(\bar{\quatdim-1},-1)$\\
    \hline
    Conformal weight  &  $1$  & $1$ &    $\frac{1}{2}$&$\frac{1}{2}$\\
    \hline
\end{tabular}
\caption{
The normalization of the $\mathfrak{u}(1)$ is chosen so that the fundamental representation of $\mathfrak{a}_{\quatdim-1}$ decomposes as $\quatdim \mapsto (\quatdim-1,1)\oplus (1,-(\quatdim-1))$ and it corresponds to $\beta_c\gamma^c-(\quatdim-1)\mathsf{h}=-d\,\mathsf{u}$.}
\label{tablebranchingsuh}
\end{table}

\vspace{0.3cm}

Having under good control the unbroken symmetry together with its free field realization and the distinguished  $U(1)$ generated by $j(z)$, we are ready to construct the remaining generators of the VOA.
As usual, some of them are very easy to build since there is a unique candidate in the free field space with the appropriate quantum numbers. In our case, they are the following combinations associated with HB generators that do not vanish on the singular locus 
\begin{equation}
\label{SimpleGeneratorssurvingSING}
e(z)=e^{\delta+\varphi}\,,
\qquad
(\mathcal{J}^-_a\cdot e)(z)=\ell\,
\beta_a\,e^{\frac{\ell-1}{\ell}(\delta+\varphi)}\,,
\qquad
\mathcal{J}_+^a(z)=
\gamma^a\,e^{\frac{1}{\ell}(\delta+\varphi)}\,,
\end{equation}
and the following which is associated to an HB generator that do vanish on the singular locus
\begin{equation}
\label{Wghw}
\mathcal{W}^{\text{ghw}}=
[\mathcal{J}^{\text{IR}}\,]_{\omega_{\ell}}\,e^{\frac{\ell-1}{\ell}(\delta+\varphi)}\,,
\end{equation}
where $[\mathcal{J}^{\text{IR}}]_{\omega_{\ell}}$ denotes the set of currents that pick a factor $\omega_{\ell}$ under the $\mathbb{Z}_{\ell}$ action\footnote{Notice that in the $\ell=4$ this eigenspace is empty, see \eqref{Z4action}, so that there is no $\mathcal{W}$ type generator.}.
Several comments are in order. Firstly all these objects are $\mathbb{Z}_{\ell}$ invariant. Notice that we did not include other $\mathbb{Z}_{\ell}$ invariant combinations like $\beta_a\,e^{-\frac{1}{\ell}(\delta+\varphi)}$ and $\gamma^a\,e^{-\frac{\ell-1}{\ell}(\delta+\varphi)}$, but these are clearly to be discarded since they have non-positive conformal weight $0$ and $\tfrac{1}{2}(2-\ell)$ respectively.
The generator $e(z)$ in \eqref{SimpleGeneratorssurvingSING} is the generalized highest weight state of a multiplet of operators which we call $\mathcal{B}$, with conformal dimension $\tfrac{1}{2}\ell$, that is the VOA avatar of the $b$ generator given in \eqref{bandbarbgeneralizedHWS}. Similarly, we will denote by $\overline{\mathcal{B}}$, to be constructed momentarily, the VOA avatar of $\bar{b}$.
The operator $\mathcal{W}^{\text{ghw}}$, when present, has conformal weight $\tfrac{1}{2}(\ell+1)$.
The transformation properties of  $\mathcal{B}$,   $\overline{\mathcal{B}}$  and $\mathcal{W}$ under the UV flavor symmetry are uniquely fixed by the fact that we know their quantum numbers under the unbroken symmetry and the distinguished\footnote{For the $\mathcal{B}$ and  $\overline{\mathcal{B}}$  operator the same conclusion can be reached by recalling that $\mathcal{B}$ and  $\overline{\mathcal{B}}$ must transform in the same representation as $b$ and $\bar{b}$.} $U(1)$.
The relevant representations together with their branching ratios  are collected in Table \ref{tab:branchings}.

The next task is to construct the generalized highest height state of $\mathcal{\bar{B}}$ and the current associated to the lowering generator $\mathcal{J}^-_a$. We know from the analysis of the singular locus, that the associated HB generators restricted to the singular locus $W$ take the simple form $\gammaclass^{a_1}\dots \gammaclass^{a_{\ell}}$ and 
$\betaclass_a\, \mathsf{h}\,\mathsf{e}^{-1/\ell}$, see \eqref{C2hmodZlgenerators}, \eqref{C2hmodZlgeneratorsXY} and \eqref{MABandminpatch}. 
The quantum number assignment is very restrictive and we find that the only candidate for the generalized highest height state of $\mathcal{\bar{B}}$ is
\begin{equation}
\label{BbarghsinVOA}
(\bar{\mathcal{B}}^{\text{ghw}})^{a_1\dots a_{\ell}}(z)=
\gamma^{a_1}(z)\dots\gamma^{a_{\ell}}(z)+x\,(\mathcal{J}^{\text{IR}})^{a_1\dots a_{\ell}}\,
e^{\frac{\ell-2}{\ell}(\delta+\varphi)}\,,
\end{equation}
where the component of the IR currents appearing here is singled out by $\mathbb{Z}_{\ell}$ invariance do that it has eigenvalues $\omega_{\ell}^2=(1,\omega_3^{-1},-1)$ under the $\mathbb{Z}_{\ell}$ action for the cases $\ell=2,3,4$ respectively. It is a non-trivial fact, which is necessary for the consistency of the construction, that the two terms summed in \eqref{BbarghsinVOA} have the same transformation properties under the unbroken symmetry.
Notice that in the second term in \eqref{BbarghsinVOA} we introduced a coefficient $x$. Its value is arbitrary as long as it is non-zero and it could be set two one without loss of generality\footnote{This freedom already appeared  and has been discussed in the free field realization presented in \cite{Bonetti:2018fqz}.}. 

The last generator we need to construct to have a complete set of (non-strong) generators is the lowering operator current $\mathcal{J}^-_{a}(z)$. To determine it, we build the most general operator with the correct quantum numbers in the free field space, namely
\begin{equation}
\label{Jminusgeneral}
\begin{split}
\mathcal{J}^-_{a}(z)=
\left(\beta_a\,( \alpha_1\partial \delta + \alpha_2 \partial  \varphi)+\alpha_3\partial \beta_a
+\alpha_4\beta_a \beta_c\gamma^c
+\alpha_5 (\beta
\mathcal{J}^{\text{IR}}_0)_a
\right)\,
&e^{-\frac{1}{\ell}(\delta+\varphi)}\,+\,\\
+\,\,\alpha_6\,
(\mathcal{J}^{\text{IR}})_{a b_1\dots b_{\ell-1}}\gamma^{b_1}\dots \gamma^{b_{\ell-1}}\,
e^{-\frac{\ell-1}{\ell}(\delta+\varphi)}\,,
&
\end{split}
\end{equation}
and fix the coefficients $\alpha_1,\dots,\alpha_6$ by the conditions that the generators $\langle \mathcal{J}^-_{a}, \mathcal{J}_+^{a}, \hat{\mathcal{J}}^{a}_b,\mathcal{Z}\rangle$ close into the $\mathfrak{a}_{d-1}$ algebra and that the AKM primaries proposed in
\eqref{Wghw} and  \eqref{BbarghsinVOA} are indeed primaries.
Notice that the IR currents that appear in the term multiplying\footnote{In the $\ell=2$ case the term $\alpha_5 (\beta
\mathcal{J}^{\text{IR}}_0)_a$ should be understood as the sum of two terms: $\alpha_5^{(1)}\beta_b
(\mathcal{J}^{\text{IR}}_{\mathfrak{a}_{d-2}\subset \mathfrak{c}_{d-1}})^b_a+\alpha_5^{(2)}\beta_a
(\mathcal{J}^{\text{IR}}_{\mathfrak{u}_{1}\subset \mathfrak{c}_{d-1}})$.}
$\alpha_5$ are  $\mathbb{Z}_{\ell}$ invariant (as the index $0$ indicates) while the IR currents appearing in the term  multiplying $\alpha_6$ have eigenvalues $\omega_{\ell}^{-2}=(1,\omega_3,-1)$ under the $\mathbb{Z}_{\ell}$ action for the cases $\ell=2,3,4$ respectively. The first condition we impose is
\begin{equation}
\label{JplusJminusOPE}
    \mathcal{J}_+^{a}(z)\,\mathcal{J}_b^{-}(w)\sim 
  \frac{k_{\mathfrak{a}_{\quatdim-1}}\,\delta^a_b}{(z-w)^2}
  +\frac{\delta^a_b\,\mathcal{Z}(w)-\hat{\mathcal{J}}^a_b(w)}{(z-w)}\,.
\end{equation}
This condition implies
\begin{equation}
\alpha_1=\tfrac{k_{\mathfrak{a}_{\quatdim-1}}}{\ell}\,,
\quad
\alpha_2=\tfrac{k_{\mathfrak{a}_{\quatdim-1}}}{\ell}+\tfrac{\ell}{\langle \varphi, \varphi \rangle}\,,
\quad
\alpha_3=1+k_{\mathfrak{a}_{\quatdim-1}}\,,
\quad
\alpha_4=0\,,
\quad
\alpha_5=-1\,,
\end{equation}
and no restriction on the value of $\alpha_6$. The value of $\langle \varphi, \varphi \rangle$ has already been fixed from the $\mathcal{Z}$-$\mathcal{Z}$ OPE but we left it free here since \eqref{JplusJminusOPE} holds for any value of $\ell,\quatdim, k$ and $\langle \varphi,\varphi \rangle$. 
The fact that the coefficient $\alpha_4=0$ is not surprising and it could have been anticipated by taking the leading term in the R-filtration. When $\quatdim>2$, the only exception being the $A_1U_1$ theory already presented in Section \ref{sec:subsectionA1U1}, we know from the form of the unbroken generators in \eqref{gnaturalUVminusfcurrents} that $k_{\mathfrak{a}_{\quatdim-1}}=k_{\mathfrak{a}_{\quatdim-2}}^{\text{IR}}-1$. Now all parameters, except for $\alpha_6$, are fixed in terms of $k_{\mathfrak{a}_{\quatdim-2}}^{\text{IR}}$ but we still need to impose that the  $\mathcal{J}^{-}$-$\mathcal{J}^{-}$ OPE is regular.

Finally, the condition that  $\bar{\mathcal{B}}^{\text{ghw}}$ given in  \eqref{BbarghsinVOA} is an AKM primary, namely that its OPE with $\mathcal{J}^{-}$ has no double (or higher) order poles gives the condition\footnote{
We have fixed the normalization of the IR currents in such a way that
\begin{equation}
(\mathcal{J}^{\text{IR}})_{a_1\dots a_{\ell}}(z)
\,
(\mathcal{J}^{\text{IR}})^{a_1\dots a_{\ell}}(w)
\,\sim\,
\frac{\# \,k_{\text{IR}}\,
\delta^{b_1}_{(a_1}\dots \delta^{b_{\ell}}_{a_{\ell})}
}{(z-w)^2}+\dots 
\end{equation}
with the total symmetrization being unit normalized.
}
\begin{equation}
\# \,k_{\text{IR}}\,x\,\alpha_6+ (k_{\mathfrak{a}_{\quatdim-1}}+1)=0\,,
\end{equation}
which fixes the value of $\alpha_6$. We have constructed all VOA generators that are associated to HB generators.
We will know close the OPEs and establish if some extra strong generator\footnote{In cases where the stress tensor is Sugawara (i.e.~$C_2U_1$,$C_3A_1$) one can verify that the strong generators $\mathcal{W}$ are AKM and Virasoro primaries satisfying the relation between the conformal dimension $\Delta$ and the Quadratic Casimir of the highest weight representation $\Lambda$ which we call $c_\Lambda$ : $\Delta=\frac{c_\Lambda}{2(k+h^\vee)}$. Here $k$ is the AKM level and $h^\vee$ is the dual coxeter number.}
 of the VOA should be added. Our findings are collected in Table \ref{tab:stonggen} . 

\paragraph{Remarks on the case $\ell=2$.}
In the discussion above we presented the $\ell=2$ case in a uniform way with its $\ell\neq 2$ relatives. In this case, the operator that is getting a VEV is a nilpotent current and the commutant of the unbroken symmetry is an $\mathfrak{sl}_2$ triplet. To construct the free field realization it is sufficient to build the $\mathfrak{sl}_2$ primaries and the $\mathfrak{sl}_2$ lowering generator $f(z)$ following the general scheme given in equation \eqref{fthetaDELIGNE}. The operator $S^{\natural}$ takes the following form\footnote
{We normalize things in such a way that 
$(\mathcal{J}^{\text{IR}})^2=(\mathcal{J}^{\text{IR}}_{\mathfrak{c}_{\quatdim-1}})^2+(\mathcal{J}^{\text{IR}}_{\mathfrak{f}})^2+(\mathcal{J}^{\text{IR}}_{\text{odd}})^2$
and recall that $T_{\mathfrak{g}}^{\text{Sug}}=\tfrac{1}{2(k+h^{\vee})}\,(\mathcal{J}_{\mathfrak{g}})^2$.
}
\begin{equation}
S^{\natural}\,=\,
c_1 \,(\mathcal{J}^{\text{IR}}_{\mathfrak{c}_{\quatdim-1}})^2
+
c_2 \,(\mathcal{J}^{\text{IR}}_{\mathfrak{f}})^2
+
c_3 \,(\mathcal{J}^{\text{IR}}_{
\text{odd}
})^2
+c_4\,T_{\xi}
+c_5\,\xi\xi\,\mathcal{J}^{\text{IR}}_{\mathfrak{c}_{\quatdim-1}}\,.
\end{equation}
To fix the coefficients we first demand that $S^{\natural}$ has regular OPEs with the flavor currents given in 
\eqref{gnaturalUVminusfcurrents} and \eqref{gnaturalfcurrents}, this gives 
\begin{equation}
\#\,c_1+\#c_3=c_5\,,
\qquad
c_4=-k_{\text{IR}}\,c_5\,,
\qquad
\#\,c_1+\#c_2+\#c_3=0\,,
\end{equation}
Additionally we require that the OPE of $f(z)$ with the primary $\xi\,e^{\frac{1}{2}(\delta+\varphi)}$ and $\mathcal{J}^{\text{IR}}_{\text{odd}}\,e^{\frac{1}{2}(\delta+\varphi)}$ has at most a simple pole. This implies $c_4=-\tfrac{1}{2}(2k+1)=\#\,c_1+\#c_2+\#c_3$.
The coefficients $c_3$ turns out to be zero in a non trivial way: it is porportional to $C_2(\mathcal{R})+k_{\text{IR}}(h^{\vee}_{\mathfrak{c}_{\quatdim-1}}+k_{\text{IR}})$ which (recalling that $C_2(\mathbf{42})=6$, $C_2(\mathbf{5})=2$, $C_2(\mathbf{2})=\tfrac{3}{4}$ and $h^{\vee}_{\mathfrak{c}_{\quatdim-1}}=\quatdim$ with $\quatdim=5,3,2$) is zero for the $\ell=2$ theories.
The coefficients are all fixed but we still have to check that $c^{\natural}$ takes the correct value and that $\xi \mathcal{J}^{\text{IR}}_{\text{odd}}$ projected into the appropriate representation has regular OPE with $S^{\natural}$.
This is, non trivially, the case for the VOAs constructed here.  
The rigidity of the construction is a feature that makes it particularly promising for a bottom-up clasification.

\begin{table}[t]
\begin{center}
\begin{adjustbox}{center}
\begin{tabular}{| c | c| c | c |c   | }

\hline
 
 Theory & Affine Currents  & $\mathcal{B},\bar{\mathcal{B}}$ & $\mathcal{W}$ & 
Extra
\\ 
 \hline
 \hline
 $C_5$ & $\mathcal{J}_{\mathfrak{c}_5}=\mathcal{J}_{MN}$ &  & $\mathcal{W}_{\mathbf{132}}=\mathcal{W}_{\llbracket MNPQR \rrbracket}$ & $T$
 \\  
 $C_3A_1$ & $\mathcal{J}_{\mathfrak{c}_3}=\mathcal{J}_{MN}$,  $\mathcal{J}_{\mathfrak{a}_1}=\mathcal{J}_{IJ}$  &  & $\mathcal{W}_{(\mathbf{14}',\mathbf{3})}=\mathcal{W}^{(IJ)}_{\llbracket MNP \rrbracket}$& $-$
 \\
 $C_2U_1$ & 
 $\mathcal{J}_{\mathfrak{c}_2}=\mathcal{J}_{MN}$,  $\mathcal{J}_{\mathfrak{u}_1}=\mathcal{J}$ 
 & &  $\mathcal{W}_{(\mathbf{5},\pm 1)}=\mathcal{W}_{\llbracket MN \rrbracket}^{\pm}$ &  $-$
 \\
 \hline
 $A_3$ &$\mathcal{J}_{\mathfrak{a}_3}=\mathcal{J}_{A}^B$  & $\mathcal{B}_{\bar{\mathbf{20}}''}=\mathcal{B}_{(ABC)}$, 
 $\bar{\mathcal{B}}_{\mathbf{20}''}=\bar{\mathcal{B}}^{(ABC)}$
 &$\mathcal{W}_{\mathbf{50}}=\mathcal{W}^{(ABC)}_{(EFG)}$ &$T$
 \\  
 $A_1U_1$& $\mathcal{J}_{\mathfrak{a}_1}=\mathcal{J}_{A}^B$, $\mathcal{J}_{\mathfrak{u}_1}=\mathcal{J}$   & $\mathcal{B}_{(\mathbf{4},+1)}=\mathcal{B}_{(ABC)}$, $\bar{\mathcal{B}}_{(\mathbf{4},-1)}=\bar{\mathcal{B}}^{(ABC)}$  &
 $\mathcal{W}_{(\mathbf{1},\pm 2)}=\mathcal{W}^{\pm\pm}$
 &
 $T$
 \\
 \hline
 $A_2$ &$\mathcal{J}_{\mathfrak{a}_2}=\mathcal{J}_{A}^B$  & 
 $\mathcal{B}_{\mathbf{15}'}=\mathcal{B}_{(ABCD)}$, $\bar{\mathcal{B}}_{\bar{\mathbf{15}}'}=\bar{\mathcal{B}}^{(ABCD)}$  
 &$-$ & $T$, $W_3$
 \\  
 \hline
\end{tabular}
\end{adjustbox}
\end{center}
\caption{This table contains a summary of the strong generators of the VOAs discussed in this section. 
$\mathcal{J}$ are affine currents, $\mathcal{B}$ and $\bar{\mathcal{B}}$ are VOA generators of conformal weight $h=\ell/2$ associated to HB generators that do not vanish on the singular locus (in the $\ell=2$ case we include them as part of the currents).
The generators $\mathcal{W}$ are associated to HB generators that vanish on the singular locus. They have conformal weight $3/2$ and $2$ for the cases $\ell=2$ and $\ell=3$ respectively and are absent in the $\ell=4$ case.
The pattern of representations of the non-current HB generators becomes more uniform is we use Dynkin labels.
In the $\ell=2$ case the relevant representations for the  $\mathcal{W}$ generators are
$\mathbf{132}=[00001]$,
$\mathbf{14}'=[001]$,
$\mathbf{5}=[01]$.
For $A_3$,  $\bar{\mathbf{20}}''=[300]$, $\mathbf{20}''=[003]$,  $\mathbf{50}=[030]$ while for $A_2$ we have $\mathbf{15}'=[4,0]$, $\bar{\mathbf{15}}'=[0,4]$.
The entry ``Extra'' shows strong generators of the VOA that are not associated with Higgs branch generators.
}
\label{tab:stonggen}
\end{table}

\vspace{0.5cm}

\begin{table}[htpb]
	\begin{center}
		\begin{adjustbox}{center}
			\begin{tabular}{| c | c| c| c| c| }
				\hline
				Theory & \makecell{$\mathfrak{g}_{\text{IR}}$\\ Deligne theory} & $\textcolor{blue}{(\mathfrak{g}_{\text{IR}})_0}$ & \makecell{IR Branching\\ $\mathfrak{g}_{\text{IR}}\rightarrow \textcolor{blue}{(\mathfrak{g}_{\text{IR}})_0}$} & UV Branching \\
				\hline
				$C_5$ & $\mathfrak{e}_6$ &$\mathfrak{c}_4$&		
    $\mathcal{J}^{\text{IR}}_{\mathfrak{e}_6} 
					\mapsto 
					\mathcal{J}^{\text{IR}}_{\mathfrak{c}_4}\,\oplus \,
					\mathcal{J}^{\text{IR}}_{\textcolor{purple}{\bf{42}}}\,,$

				& 	\makecell{
					$\mathfrak{c}_5\mapsto\textcolor{blue}{\mathfrak{c}_4}\oplus\textcolor{ForestGreen}{\mathfrak{a}_1}$
    \\
					$\mathcal{J}_{\mathfrak{c}_{5}} 
					\mapsto
					\mathcal{J}_{\mathfrak{c}_{4}}
     \oplus
					\mathcal{J}_{\mathfrak{a}_1}\oplus
					\mathcal{J}_{(\mathbf{8},\mathbf{2})}\,,$ 
					\\
					$\mathcal{W}:\mathbf{132}
					\mapsto
					(\textcolor{purple}{\mathbf{42}},\mathbf{2})\oplus 
					(\mathbf{48},\mathbf{1})\,
					$}\\
				\hline
				$C_3A_1$& $\mathfrak{d}_4$ &$\mathfrak{c}_2\oplus \mathfrak{a}_1$ &
    $\mathcal{J}^{\text{IR}}_{\mathfrak{d}_4} 
					\mapsto 
					\mathcal{J}^{\text{IR}}_{\mathfrak{c}_2}\,\oplus \,
					\mathcal{J}^{\text{IR}}_{\mathfrak{a}_1}\,\oplus \,
					\mathcal{J}^{\text{IR}}_{(\textcolor{purple}{\mathbf{5}},\textcolor{purple}{\mathbf{3}})}$
				& 
    \makecell{$\mathfrak{c}_3\oplus\mathfrak{a}_1\mapsto\textcolor{blue}{\mathfrak{c}_2}\oplus\textcolor{ForestGreen}{\mathfrak{a}_1}\oplus \textcolor{blue}{\mathfrak{a}_1}$\\
					$\mathcal{J}_{\mathfrak{c}_{3}} 
					\mapsto
					\mathcal{J}_{\mathfrak{c}_{2}}\oplus
					\mathcal{J}_{\mathfrak{a}_1}\oplus
					\mathcal{J}_{(\mathbf{4},\mathbf{2},\mathbf{1})}$  
					\\
					$	\mathcal{J}_{\mathfrak{a}_{1}} 
					\mapsto
					\mathcal{J}_{\mathfrak{a}_{1}}\,  $
					\\
					$\mathcal{W}:{(\mathbf{14'},\mathbf{3})}
					\mapsto
					{(\textcolor{purple}{\mathbf{5}},\mathbf{2},\textcolor{purple}{\mathbf{3}})}\oplus 
					{(\mathbf{4},\mathbf{1},\mathbf{3})}$}\\
				\hline
				$C_2U_1$ & $\mathfrak{a}_2$&$\mathfrak{c}_1\oplus \mathfrak{u}_1$ &	
    $\mathcal{J}^{\text{IR}}_{\mathfrak{a}_2} 
					\mapsto 
					\mathcal{J}^{\text{IR}}_{\mathfrak{c}_1}\,\oplus \,
					\mathcal{J}^{\text{IR}}_{\mathfrak{u}_1}\,\oplus \,
					\mathcal{J}^{\text{IR}}_{(\textcolor{purple}{\mathbf{2}},\textcolor{purple}{\pm 1})}\,,$
				& \makecell{		
					$\mathfrak{c}_2\oplus\mathfrak{u}_1\mapsto\textcolor{blue}{\mathfrak{c}_1}\oplus\textcolor{ForestGreen}{\mathfrak{a}_1}\oplus\textcolor{blue}{\mathfrak{u}_1}$\\$
					\mathcal{J}_{\mathfrak{c}_{2}} 
					\mapsto
					\mathcal{J}_{\mathfrak{c}_{1}}\oplus
					\mathcal{J}_{\mathfrak{a}_1}\oplus
					\mathcal{J}_{(\mathbf{2},\mathbf{2})_0}\,,  $
					\\
					$	\mathcal{J}_{\mathfrak{u}_{1}} 
					\mapsto
					\mathcal{J}_{\mathfrak{u}_{1}}\,,  $
					\\
					$\mathcal{W}:{\mathbf{5}_{\pm 1}}
					\mapsto
					(\textcolor{purple}{\mathbf{2}},\mathbf{2})_{\textcolor{purple}{\pm 1}}\oplus 
					(\mathbf{1},\mathbf{1})_{\pm 1}\,.$}\\
				\hline
				\hline
				$A_3$ & $\mathfrak{d}_4$&$\mathfrak{a}_2$&
   $\mathcal{J}^{\text{IR}}_{\mathfrak{d}_4} 
					\mapsto 
					\mathcal{J}^{\text{IR}}_{\mathfrak{a}_2}\,\oplus \,
					\mathcal{J}^{\text{IR}}_{\textcolor{purple}{\mathbf{10}}}\,\oplus \,
					\mathcal{J}^{\text{IR}}_{\textcolor{purple}{\overline{\mathbf{10}}}}\,,$	
				& \makecell{
					$\mathfrak{a}_3\mapsto\textcolor{blue}{\mathfrak{a}_2}\oplus
     \textcolor{ForestGreen}{\mathfrak{u}_1}$\\	$\mathcal{J}_{\af_3} 
					\mapsto 
					\mathcal{J}_{\af_2}\,\oplus 
					\mathcal{J}_{\mathfrak{u}_1}\,\oplus{\mathcal{J}_{\bf{3}}^{\pm\frac43}}$\\
					$\mathcal{B}: {\bf{20}''}
					\mapsto
					{\bf1}_3\oplus 
					\overline{\bf3}_{\frac53}\oplus 
					{\bf6}_{\frac13}\oplus 
					\overline{\bf{10}}_{-1}$\\
					$\overline{\mathcal{B}}:{\overline{\bf{20}''}}
					\mapsto
					\textcolor{purple}{\bf{10}}_1\oplus 
					\overline{\bf6}_{-\frac13}\oplus 
					{\bf3}_{-\frac53}\oplus{\bf1}_{-3}$\\
					$\mathcal{W}:{\bf{50}}
					\mapsto		
					\textcolor{purple}{\overline{\bf{10}}}_2\oplus 
					\overline{\bf{15}}_{\frac23}\oplus	{\bf{15}}_{-\frac23}\oplus \bf{10}_{-2}$}\\
				\hline
				$A_1U_1$& $\mathfrak{a}_1$ &$\mathfrak{u}_1$ &		
    $\mathcal{J}^{\text{IR}}_{\af_1} 
					\mapsto 
					\mathcal{J}^{\text{IR}}_{\uf_1}\,\oplus \,
					\mathcal{J}^{\text{IR}}_{\textcolor{purple}{-2}}\,\oplus \,
					\mathcal{J}^{\text{IR}}_{\textcolor{purple}{+2}}$		& \makecell{		
				$\mathfrak{a}_1\oplus\mathfrak{u}_1\mapsto	\textcolor{blue}{\mathfrak{u}_1}\oplus\textcolor{ForestGreen}{\mathfrak{u}_1}$\\	$\mathcal{J}_{\mathfrak{a}_{1}}
					\mapsto
					\mathcal{J}_{\uf_1}\oplus
					\mathcal{J}_{\pm2}$
					\\
					$\mathcal{J}_{\uf_1}
					\mapsto \mathcal{J}_{\uf_1}$
					\\
					$\mathcal{B}:{\bf 4}
					\mapsto
							({+},{3})\oplus 
					({+},{1})\oplus ({+},{-1})\oplus 
				({+},-3)
    $ 
					\\
					$\overline{\mathcal{B}}:{\bf 4}
					\mapsto
					({-},\textcolor{purple}{3})\oplus 
					({-},{1})\oplus({-},{-1})\oplus 
					({-},{-3})$ 
					\\
					$\mathcal{W}:{\bf 1}_{+2}
					\mapsto
					(\textcolor{purple}{+2},{0})\,,
      \,\,\,\,
     {\bf 1}_{-2}
					\mapsto
					(- 2,{0})
     $
     }
     \\
				
				\hline
				\hline
				$A_2$& $\mathfrak{a}_2$ &$\mathfrak{a}_1$ &		
    $\mathcal{J}^{\text{IR}}_{\af_2} 
					\mapsto 
					\mathcal{J}^{\text{IR}}_{\af_1}\,\oplus \,
					\mathcal{J}^{\text{IR}}_{\textcolor{purple}{\bf{5}}}$
				
				& \makecell{	
				$\mathfrak{a}_2\mapsto
    \textcolor{blue}{\mathfrak{a}_1}\oplus	\textcolor{ForestGreen}{\mathfrak{u}_1}
    $
    \\	$\mathcal{J}_{\mathfrak{a}_{2}} 
					\mapsto
					\mathcal{J}_{\af_1}\oplus
     \mathcal{J}_{\af_1}\oplus
					\mathcal{J}_{\bf{2}}^{\pm\frac32}$  \\
					$\mathcal{B}:{\overline{\bf 15'}}
					\mapsto		
					{\bf1}_4\oplus 
					{\bf2}_{\frac52}\oplus{\bf3}_1\oplus 
					{\bf4}_{-\frac12}\oplus\bf5_{-2}  $ \\	
					$\overline{\mathcal{B}}:\bf 15'
					\mapsto			
					{\textcolor{purple}{\bf{5}}}_2\oplus {\bf{4}}_{\frac12}\oplus{\bf{3}}_{-1}\oplus{\bf{2}}_{-\frac52}	\oplus{\bf{1}}_{-4} $  }\\
				\hline
			\end{tabular}
		\end{adjustbox}
	\end{center}
	\caption{ Branchings for IR VOA AKM currents and UV VOA generators. We use the color \textcolor{blue}{blue} to indicate the $\mathbb{Z}_{\ell}$ invariant part of the IR flavor symmetry which coincides with part of the UV symmetry unbroken by the VEV: $\unbrSYMM=\textcolor{blue}{(\mathfrak{g}_{\text{IR}})_0}$. In \textcolor{ForestGreen}{green} we indicate the commutant of $\textcolor{blue}{(\mathfrak{g}_{\text{IR}})_0}$ in the UV symmetry.
Finally, we displayed in \textcolor{purple}{purple} the components of $\overline{\mathcal{B}}$ and $\mathcal{W}$ that reduce, after Higgsing, to the   IR currents that are not $\mathbb{Z}_{\ell}$ invariant, see equations \eqref{Wghw} and \eqref{BbarghsinVOA}. 
 The $A_1U_1$ is a little special from the point of view of purple coloring, this is related to the way $\mathcal{J}^{\text{IR}}_{\uf_1}$ enters in the $\textcolor{blue}{\mathfrak{u}_1}\oplus\textcolor{ForestGreen}{\mathfrak{u}_1}$ currents, see equation \eqref{ZandJcurrentforA1U1}.
 Lets also recall that for $\ell=2$ the generator $e(z)$ is part of the 
 $\mathcal{J}_{\textcolor{ForestGreen}{\mathfrak{a}_1}}$ 
 currents while for the other cases is the generalized highest weight state of the $\mathcal{B}$ operator.
 }
 \label{tab:branchings}
\end{table}

\subsection{The OPEs}

We will now collect all the OPEs in schematic form. 
We will not report OPEs among currents and between currents and the other generators since they take the standard form. 
We report once for all the OPEs for $\mathfrak{c}_{\quatdim}$ currents
\begin{equation}
\mathcal{J}_{MN}(z)\mathcal{J}_{PQ}(w)\sim
\frac{k\,\Omega_{P(M}\Omega_{N)Q}}{(z-w)^2}
+\frac{2\,(\Omega_{M(P}\mathcal{J}_{Q)N}(w)+\Omega_{N(Q}\mathcal{J}_{P)M}(w))}{(z-w)}\,,
\end{equation}
where we use conventions $v_{(AB)}=\tfrac{1}{2}(v_{AB}+v_{BA})$
and for $\mathfrak{a}_{\quatdim-1}$ currents
\begin{equation}
\label{SUNOPEs}
\mathcal{J}^A_B(z)\mathcal{J}^C_D(w)\sim 
\frac{k(\delta^A_D\delta^C_B-\tfrac{1}{d}\delta^A_B\delta^C_D)}{(z-w)^2}+\frac{\delta^A_D\mathcal{J}^C_B-\delta^C_B\mathcal{J}^A_D}{z-w}\,.
\end{equation}
We remark that in all cases but the $A_1U_1$ theory the closure of the OPEs requires that  IR nulls vanish. For this reason, in the practical verification of the OPEs, we employed the free field realization for the IR VOA as well.
In Table \ref{tab:IRbranching} we report the branching of the IR VOA in the basis in which the $\mathbb{Z}_{\ell}$ action is diagonal.
\begin{table}
	\begin{center}
		\begin{adjustbox}{center}
	\begin{tabular}{|c|c|}
		\hline
		(UV Theory, IR Theory) & IR Branching  \\
		\hline
		($C_5$,$E_6$) & 	
		\makecell{$\mathfrak{c}_4\to\mathfrak{c}_3\oplus\mathfrak{a}_1$\\$\mathbf{36} \to \bf \textcolor{blue}{(1,3)}\oplus\textcolor{teal}{(21,1)}\oplus \textcolor{red}{(6,2)}$\\
		$\mathbf{42}\to\bf \textcolor{red}{(14',2)}\oplus \textcolor{teal}{(14,1)}$}
		 \\
		 \hline
		 ($C_3 A_1$,$D_4$) & 	
		 \makecell{$\mathfrak{c}_2\oplus\mathfrak{a}_1\mapsto\mathfrak{c}_1\oplus\mathfrak{a}_1\oplus\mathfrak{a}_1$\\ 
		 	$\mathbf{(10,3)}\to\textcolor{red}{\mathbf{(3,1,1)}}\textcolor{blue}{\oplus\mathbf{(2,2,1)}}\oplus\textcolor{teal}{\mathbf{(1,3,1)}}$\\
		 			$\mathbf{(1,3)}\to \textcolor{teal}{\mathbf{(1,1,3)}}$\\
		 			$\mathbf{(5,3)}\to \textcolor{blue}{\mathbf{(2,2,3)}}\oplus\textcolor{teal}{\mathbf{(1,1,3)}}$}
		\\
		 \hline
		 \textcolor{purple}{($C_2 U_1$,$A_2$)} & 	
		 \makecell{$\mathfrak{c}_1\oplus\mathfrak{u}_1\mapsto\mathfrak{c}_1\oplus\mathfrak{u}_1$\\ 
		 	$\textcolor{red}{\mathbf{3}_{0}}\to\textcolor{red}{\mathbf{3}_{0}}$\\
		 	$\textcolor{teal}{\mathbf{1}_0}\to\textcolor{teal}{\mathbf{1}_0}$\\
		 	$\textcolor{blue}{\mathbf{2}_{\pm1}}\to\textcolor{blue}{\mathbf{2}_{\pm1}}$}
		 \\
		 \hline
		 ($A_3$,$D_4$) & 	
		 \makecell{$\mathfrak{a}_2\to\mathfrak{a}_1\oplus\mathfrak{u}_1$\\$
		 		\mathbf{8}\to\textcolor{red}{\mathbf{1_0}}\oplus\textcolor{blue}{\mathbf{2_3}}\oplus\textcolor{blue}{\mathbf{2_{-3}}}\oplus \textcolor{teal}{\mathbf{3_0}}$\\
		 	$\mathbf{10}\to\textcolor{red}{\mathbf{1_{-6}}}\oplus\textcolor{blue}{\mathbf{2_{-3}}}\oplus\textcolor{teal}{\mathbf{3_{0}}}\oplus\textcolor{blue}{ \mathbf{4_{3}}}$\\
		 	$\overline{\mathbf{10}}\to\textcolor{red}{\mathbf{1_{6}}}\oplus\textcolor{blue}{\mathbf{2_{3}}}\oplus\textcolor{teal}{\mathbf{3_{0}}}\oplus \textcolor{blue}{\mathbf{4_{-3}}}$}
		 \\
		 \hline
		  \textcolor{purple}{($A_1 U_1$,$A_1$)} & 	
		 \makecell{$\mathfrak{u}_1\oplus\mathfrak{u}_1\mapsto\mathfrak{u}_1\oplus\mathfrak{u}_1$\\ 
		 	$\textcolor{red}{\mathbf{1}_{0}}\to\textcolor{red}{\mathbf{1}_{0}}$\\
		 	$\textcolor{teal}{\mathbf{1}_{+2}}\to\textcolor{teal}{\mathbf{1}_{+2}}$\\
		 	$\textcolor{blue}{\mathbf{1}_{-2}}\to\textcolor{blue}{\mathbf{1}_{-2}}$}
		 \\
		 \hline
		  ($A_2$,$A_2$) & 	
		 	\makecell{$\mathfrak{a}_2\to\mathfrak{a}_1$\\$\mathbf{8}\to\textcolor{red}{\mathbf{3}}\oplus\textcolor{blue}{\mathbf{5}}$}\\
		 \hline
	\end{tabular}
\end{adjustbox}	
 \end{center}
	\caption{Branchings of the IR VOA. The branching in the second column shows a convenient basis. The black representations are the ones which comprise the IR adjoint representation. The IR VOA can be decomposed as $\mathfrak{g}=$\textcolor{teal}{$\mathfrak{g}^\natural$}$\oplus$\textcolor{red}{$\mathfrak{sl_2}$}$\oplus$\textcolor{blue}{$(\mathfrak{R},2)$}. The colors correspond to where does the representation lie in this decomposition. The purple theories are a little special since here we don't require further branching for the IR theory for the free fields.}
	\label{tab:IRbranching}	
\end{table}

\subsubsection{\texorpdfstring{$\ell=2$}{l=2}}

\paragraph{The case of $C_5$.}
The HB generators are the $\mathfrak{c}_5$ currents $\mathcal{J}$ and additional operators $\mathcal{W}$ of conformal dimension $3/2$ transforming in the $\mathbf{132}=[00001]$ representation, see Table \ref{tab:stonggen}. The set of strong generators of the VOA consists of generators associated to these HB generators together with the stress tensor. The $\mathcal{J}$-$\mathcal{J}$ OPEs and $\mathcal{J}$-$\mathcal{W}$ OPE take the standard form and we are left with specifying the $\mathcal{W}$-$\mathcal{W}$ OPE. The structure of the latter is dictated  by the $\mathfrak{c}_5$ symmetry recalling that
\begin{equation}
(\mathbf{132}\otimes \mathbf{132})_{\text{S}}=\mathbf{4719}\oplus \mathbf{4004}\oplus \mathbf{55}\,,
\qquad
(\mathbf{132}\otimes \mathbf{132})_{\text{A}}=\mathbf{7865}\oplus \mathbf{780}\oplus \mathbf{1}\,,
\end{equation}
and that $\mathcal{J}^2$ transforms in the representation 
\begin{equation}
(\mathbf{55}\otimes \mathbf{55})_{\text{S}}=\mathbf{715}\oplus \mathbf{780}\oplus \mathbf{44}\oplus \mathbf{1}\,.
\end{equation}
The OPEs take then the schematic form
\begin{equation}
\label{WWOPEC5theory}
\mathcal{W}(z)\mathcal{W}(w)\sim
\frac{\mathbb{I}}{(z-w)^3}+
\frac{\mathcal{J}(w)}{(z-w)^2}+
\frac{\mathcal{J}^2(w)\big{|}_{\mathbf{780}}+\mathcal{J}^2(w)\big{|}_{\mathbf{1}}+T(w)+\mathcal{J}'(w)}{(z-w)}\,.
\end{equation}
Let us make a few remarks on how \eqref{WWOPEC5theory} is obtained from our free field expressions.
To do this we apply the UV  branchings for $\mathcal{J}$ and $\mathcal{W}$ given in Table \ref{tab:branchings} together with $\mathbf{780}\rightarrow (\mathbf{3},\mathbf{36})+\dots$.
The generalized highest weight state of $\mathbf{780}$, for example, takes the form
\begin{equation}
\left(
\mathcal{J}^2(w)\big{|}_{\mathbf{780}}
\right)^{\text{ghw}}=\mathcal{J}_{\mathfrak{c}_4}\,e-(\xi e^{\frac{1}{2}})(\xi e^{\frac{1}{2}})
=
\mathcal{J}_{\mathfrak{c}_4}^{\text{IR}}\,e\,,
\end{equation}
and is easy to reproduce form the OPEs. The OPEs \eqref{WWOPEC5theory} closes only when IR null states are taken into account.

\paragraph{The case of $C_3A_1$.}
This case is very similar to the previous one and we only need to specify the $\mathcal{W}$-$\mathcal{W}$ OPEs where $\mathcal{W}$ is an AKM primary transforming in the representation $(\mathbf{14}',\mathbf{3})=[001;2]$, see Table \ref{tab:stonggen}, and with conformal dimension $3/2$. In this case the stress tensor is not an additional generator but is the Sugawara stress tensor. The structure of the OPEs is constrained by the rules    
\begin{subequations}
\label{tensorProdforC3A1}
    \begin{align}
    ((\mathbf{14}',\mathbf{3})\otimes (\mathbf{14}',\mathbf{3}))_{\text{S}}=&
(\mathbf{84}\oplus \mathbf{21},\mathbf{1}\oplus \mathbf{5})
\oplus
(\mathbf{90}\oplus \mathbf{1}, \mathbf{3})\,,
  \\
  ((\mathbf{14}',\mathbf{3})\otimes (\mathbf{14}',\mathbf{3}))_{\text{A}}=&
  (\mathbf{84}\oplus \mathbf{21},\mathbf{3})
\oplus
(\mathbf{90}\oplus \mathbf{1},\mathbf{1}\oplus \mathbf{5})\,,
\end{align}
\end{subequations}
and the contribution form $\mathcal{J}^2$ by the tensor product
\begin{equation}
\begin{split}
&( ((\mathbf{21},\mathbf{1})\oplus (\mathbf{1},\mathbf{3}))
  \otimes 
 ((\mathbf{21},\mathbf{1})\oplus (\mathbf{1},\mathbf{3}))
)_{\text{S}}
  \\
  = & (\mathbf{126}'\oplus \mathbf{90}\oplus \mathbf{14}\oplus \mathbf{1},\mathbf{1})
  \oplus (\mathbf{1},\mathbf{1}\oplus \mathbf{5})
  \oplus (\mathbf{21},\mathbf{3}) \,.
\end{split}
\end{equation}
The relevant representations of $\mathcal{J}^2$ that can appear are the one overlapping with \eqref{tensorProdforC3A1}
and in fact they overlap only with the antisymmetric tensor product.
They are the representations $(\mathbf{90},\mathbf{1})$, $(\mathbf{1},\mathbf{5})$, $(\mathbf{21},\mathbf{3})$ and twice the singlet $(\mathbf{1},\mathbf{1})$. Their explicit expression is given by
\begin{equation}
\label{J2part1C3A1}
\left(
\mathcal{J}^2(w)\big{|}_{(\mathbf{90},\mathbf{1})}
\right)^{\text{ghw}}=\mathcal{J}_{\mathfrak{c}_2}\,e-(\xi e^{\frac{1}{2}})(\xi e^{\frac{1}{2}})
=
\mathcal{J}_{\mathfrak{c}_2}^{\text{IR}}\,e\,,
\end{equation}
\begin{equation}
\label{J2part2C3A1}
(\mathcal{J}_{\mathfrak{f}=\mathfrak{a}_1}^2)\big{|}_{\mathbf{5}}\,,
\qquad
\mathcal{J}_{\mathfrak{c}_3}\,\mathcal{J}_{\mathfrak{f}=\mathfrak{a}_1}\,,
\qquad
(\mathcal{J}_{\mathfrak{c}_3}^2)\big{|}_{\mathbf{1}}\,,
\qquad
(\mathcal{J}_{\mathfrak{f}=\mathfrak{a}_1}^2)\big{|}_{\mathbf{1}}\,,
\end{equation}
where we used the branchings $\mathbf{90}\mapsto (\mathbf{3}, \mathbf{10})+\dots$.
The OPEs take then the schematic form
\begin{equation}
\label{WWOPEC3A1theory}
\mathcal{W}(z)\mathcal{W}(w)\sim
\frac{\mathbb{I}}{(z-w)^3}+
\frac{\mathcal{J}(w)}{(z-w)^2}+
\frac{\mathcal{J}^2(w)+\mathcal{J}'(w)}{(z-w)}\,,
\end{equation}
where the $\mathcal{J}^2$ term includes the five contributions from \eqref{J2part1C3A1} and \eqref{J2part2C3A1}.

\paragraph{The case of $C_2U_1$.}
The OPEs for this theory have been presented in detail in Section \ref{sec:C2U1indetails} and will not be repeated here.

\subsubsection{\texorpdfstring{$\ell=3$}{l=3}}

\paragraph{The case of $A_3$.}
The HB generators are the $\mathfrak{a}_3$ currents $\mathcal{J}(z)$ and $\mathcal{B}(z), \bar{\mathcal{B}}(z), \mathcal{W}(z)$ in the $\bar{\mathbf{20}}'', \mathbf{20}''$ and $\mathbf{50}$
with conformal weight  $3/2,3/2$ and $2$ respectively. The stress tensor $T(z)$ is an additional strong generator. The OPEs take the schematic form
\begin{equation}
\mathcal{B}(z)\bar{\mathcal{B}}(w)\sim
\frac{\mathbb{I}}{(z-w)^3}+
\frac{\mathcal{J}(w)}{(z-w)^2}+
\frac{\mathcal{J}^2(w)+\mathcal{J}'(w)+T(w)}{(z-w)}\,,
\end{equation}
\begin{equation}
\mathcal{B}(z)\mathcal{B}(w)\sim
\frac{\mathcal{W}(w)}{(z-w)}\,,
\qquad
\bar{\mathcal{B}}(z)\bar{\mathcal{B}}(w)\sim
\frac{\mathcal{W}(w)}{(z-w)}\,,
\end{equation}
\begin{equation}
\label{WBOPEinA3}
\mathcal{W}(z)\mathcal{B}(w)\sim
\frac{\bar{\mathcal{B}}(w)}{(z-w)^2}+
\frac{\mathcal{J}\bar{\mathcal{B}}(w)+\bar{\mathcal{B}}'(w)}{(z-w)}\,,
\quad
\mathcal{W}(z)\bar{\mathcal{B}}(w)\sim
\frac{\mathcal{B}(w)}{(z-w)^2}+
\frac{\mathcal{J}\mathcal{B}(w)+\mathcal{B}'(w)}{(z-w)}\,,
\end{equation}
\begin{equation}
\mathcal{W}(z)\mathcal{W}(w)\sim
\frac{\mathbb{I}}{(z-w)^4}+
\frac{\mathcal{J}(w)}{(z-w)^3}+
\frac{\mathcal{J}^2(w)+\mathcal{J}'(w)+T(w)}{(z-w)^2}+
\frac{(\mathcal{B}\bar{\mathcal{B}})(w)+\mathcal{J}^3(w)+\dots}{(z-w)}\,.
\end{equation}
The form of the OPE is very constrained by the $A_3$ symmetry. In particular no $\mathcal{J}\mathcal{W}$ term is allowed in the right hand side of the last OPE.
We notice that there can be no contamination from null operators on the right hand side of these OPEs.
The first potentially dangerous term is in the $\mathcal{W}\mathcal{B}$ OPE, where the combination $\mathcal{J}\bar{\mathcal{B}}$ appears. By $A_3$ symmetry it can appear here only once projected in the representations $\mathbf{140}=[112]$ and  $\mathbf{20}''=[003]$, but the nulls of the schematic form $\mathcal{J}\bar{\mathcal{B}}$ are in the  $\mathbf{20}=[011]$ so they cannot appear in this OPE. The case of the  $\mathcal{W}\bar{\mathcal{B}}$ is obtained by conjugation.
We are left to analyze the simple pole in the $\mathcal{W}\mathcal{W}$ OPE.
In this case the simple pole contains operators in the $\mathbf{300}'=[303]$, $\mathbf{175}=[121]$
and $\mathbf{15}=[101]$ and there is no null operator of conformal weight three transforming in these representations.

\paragraph{The case of $A_1U_1$.}

The OPEs for this theory have been presented in detail in Section \ref{sec:subsectionA1U1} and will not be repeated here.

\subsubsection{\texorpdfstring{$\ell=4$}{l=4}}

\paragraph{The case of $A_2$.}
The HB generators are the $\mathfrak{a}_2$ currents $\mathcal{J}(z)$, the generators $\mathcal{B}(z), \bar{\mathcal{B}}(z)$ in the $\mathbf{15}'$ and $\bar{\mathbf{15}}'$
with conformal weight  $2,2$ respectively. The stress tensor $T(z)$ is an additional strong generator together with an extra operator of conformal weight $3$, denoted as $W_3$, which is a singlet under the flavor symmetry.
The OPEs take the form
\begin{equation}
\mathcal{B}(z)\bar{\mathcal{B}}(w)\sim
\frac{\mathbb{I}}{(z-w)^4}+
\frac{\mathcal{J}(w)}{(z-w)^3}+
\frac{\mathcal{J}^2(w)+\mathcal{J}'(w)+T(w)}{(z-w)^2}+
\frac{\mathcal{O}_3}{(z-w)}\,,
\end{equation}
\begin{equation}
\mathcal{B}(z)\mathcal{B}(w)\sim
\frac{\bar{\mathcal{B}}(w)}{(z-w)^2}+
\frac{(\mathcal{J}\bar{\mathcal{B}})(w)+\bar{\mathcal{B}}'(w)}{(z-w)}\,,
\quad
\bar{\mathcal{B}}(z)\bar{\mathcal{B}}(w)\sim
\frac{\mathcal{B}(w)}{(z-w)^2}+
\frac{(\mathcal{J}\mathcal{B})(w)+\mathcal{B}'(w)}{(z-w)}\,,
\end{equation}
Also in this case, there can be no contamination from nulls in the OPEs.
We found that the operator appearing in the simple pole of the $\mathcal{B}\bar{\mathcal{B}}$ OPEs cannot be written as a composite of the remaining generators. We thus introduce a new generator, that we call $W_3$,
which is a Virasoro primary of conformal weight $3$ and singlet under the flavor symmetry such that
\begin{equation}
\mathcal{O}_3\big{|}_{\mathbf{1}}=W_3+\#\,\mathcal{J}^3 + \text{Virasoro descendants}
\end{equation}
We fix the value of $\#$, and as a consequence the form of $W_3$, by the requirement that its leading term in the R-filtration has $R<3$. Notice that with this choice $W_3$ is not an AKM primary.
The explicit expression of the strong generator $W_3$ in terms of free fields in not particularly illuminating. Its structure in the leading R-filtration is similar to the one of the composite operator $W_3$ in the $A_1U_1$ theory given explicitly in \eqref{W3inA1U1Leadingfiltration}.

To complete the analysis we have to compute the OPEs of $W_3$ with the remaining generators and its self-OPE and show that they close on the proposed list of strong generators.
The OPE with the $\mathfrak{a}_2$ currents gives
\begin{equation}
\mathcal{J}^a_b(z)W_3(w)\sim\frac{\partial\mathcal{J}^a_b+\mathcal{J}^a_x\mathcal{J}^x_b+\delta^a_b\mathcal{J}^2}{(z-w)^2}\,.
\end{equation}
It should be noticed that the simple pole is absent since $W_3$ is a singlet, while the double pole is present indicating that $W_3$ is not an AKM primary. Its OPEs with the $\mathcal{B}$, $\overline{\mathcal{B}}$ have the  form (indices are totally symmetrised)
\begin{equation}
	W_3(z)\mathcal{B}_{abcd}(w)\sim\frac{\mathcal{B}_{abcd}}{(z-w)^3}+\frac{\mathcal{B}_{abcd}'+\mathcal{J}_a^x\mathcal{B}_{xbcd}}{(z-w)^2}+\frac{\mathcal{B}_{abcd}''+{\mathcal{J}_a^x}'\mathcal{B}_{xbcd}+\mathcal{J}_a^x\mathcal{B}_{xbcd}'+T\mathcal{B}_{abcd}}{(z-w)}
\end{equation}
and similarly for $\overline{\mathcal{B}}$. Finally, its self OPE is given by 
\begin{equation}
	\begin{split}    W_3(z)W_3(w)&\sim\frac{1}{(z-w)^{6}}+\frac{T+\mathcal{J}^2}{(z-w)^4}+\frac{T'+\mathcal{J}'\mathcal{J}}{(z-w)^3}+\frac{T^2+T''+\mathcal{J}''\mathcal{J}+\mathcal{J}'\mathcal{J}'+T\mathcal{J}^2}{(z-w)^2}\\
		& +\frac{T'''+T T'+\mathcal{J}''\mathcal{J}'+\mathcal{J}\mathcal{J}'''+T\mathcal{J}^2+T\mathcal{J}'\mathcal{J}}{z-w}\,.
	\end{split}
\end{equation}

\subsection{R-filtration, nulls and HB relations}
\label{sec:RfiltrationandNullsgeneral}

We will now elaborate on two important (conjectural) aspects of the free field realization: (i) all null states are identically zero,
(ii) the four dimensional R-filtration, and the Higgs Branch chiral ring, can be easily reconstructed.
For the subset of rank-one theories with ECB that have a known class-S realization, determined in \cite{Giacomelli:2020jel}, see Table \ref{ClassSrealization}, we can compare our findings with the index\footnote{We thank Wolfger Peelaers for providing these indices back in 2019.}.
\begin{table}
\begin{center}
\begin{tabular}{|c|c|c|}
\hline
Theory \cite{Argyres:2016xmc}  & S-fold theory \cite{Giacomelli:2020jel} &  Class S realization\\
\hline
      $C_5$ & $\mathcal{S}_{E_6,2}^{(1)}$ & $\mathcal{S}_{E_6,2}^{(1)}\otimes HM^{\otimes 3}\leftrightarrow ([3,2^2,1],[3,2^2,1],[2^2,1^4])_{\mathfrak{d}_4}$\\
    $C_3A_1$ &  $\mathcal{S}_{D_4,2}^{(1)}$ &  $\mathcal{S}_{D_4,2}^{(1)}\otimes HM^{\otimes 1}\leftrightarrow ([2,1^2],\underline{[2^2,1]},\underline{[2^2,1]})_{\mathfrak{a}_3}$\\
     $C_2U_1$ &  $\mathcal{S}_{A_2,2}^{(1)}$ &
     $\mathcal{S}_{A_2,2}^{(1)}\leftrightarrow ([2,1],\underline{[1,1]},\underline{[1,1]})_{\mathfrak{a}_2}$
     \\[1ex]
\hline
     $A_3\rtimes\mathbb{Z}_2$ &
     $\mathcal{S}_{D_4,3}^{(1)}$ &
     $\mathcal{S}_{D_4,3}^{(1)}\leftrightarrow ([5,3],[A_1]_\omega,[A_1]_{\omega^2})_{\mathfrak{d}_4}$\\
\hline
\end{tabular}
\caption{\label{ClassSrealization}Known class $\mathcal{S}$ realization of some of the rank-one theories with ECB}
\end{center}
\end{table}
In the HL limit, our results also confirm the Hilbert series proposal based on magnetic quivers given in \cite{Bourget_2020}.
Having discussed these issues in the examples of the $C_2U_1$ and $A_1U_1$ theories in Section \ref{OverviewofStrategyandC2U1}, here 
we will first take a closer look at the null states for the $C_5$ theory as an illustrative example and emphasize how the latter are proportional to the null states of the associated IR theory. We will then discuss Higgs Branch relations that are associated with a drop in the R-filtration degree rather than to null states in the VOA.

\paragraph{Nulls in the $C_5$ theory.}
The first null state involves only the $\mathfrak{c}_5$ currents and takes the form $\mathcal{J}^2|_{\mathbf{44}}=0$.
Its generalized highest weight (recalling that $\mathbf{44}\rightarrow (\mathbf{2},\mathbf{8})+\dots$) has the schematic form
\begin{equation}
\label{JminusC5null}
\mathcal{J}^-_m\,\mathcal{J}^{++}_{\mathfrak{sl}_2}+ 
(\mathcal{J}_{\mathfrak{c}_4})_{mn}\,\Omega^{np}\mathcal{J}^+_p+ 
\mathcal{J}^+_m\,\mathcal{J}^{+-}_{\mathfrak{sl}_2}+ 
\partial\mathcal{J}^-_m=0\,,
\end{equation}
and is easily checked to vanish in the free field realization.
Alternatively, one could solve equation \eqref{JminusC5null}  to find $\mathcal{J}^-_m$ once the remaing generators are given.
At conformal dimension $3$ we have null states of the  form $\mathcal{J}\mathcal{W}$ projected in the representation\footnote{Recall that the currents $\mathcal{J}$ and the generators $\mathcal{W}$ transform in the $\mathbf{55}$ and $\mathbf{132}$ irreps of $\mathfrak{c}_5$ respectively  so that their product decomposes as $\mathbf{55}\otimes \mathbf{132}=\mathbf{5720}\oplus \mathbf{1408}\oplus \mathbf{132}$.} $\mathbf{1408}$.
Once again, it is sufficient to check the relation with maximal weight under the Cartan of $\mathfrak{sl}_2$. Recalling that $\bf{1408}\rightarrow (\bf{3},\bf{48})+\dots$, the relevant component is
\begin{subequations}
\begin{align}
\label{firstpiecenullC5}
	\mathcal{J}^{++}_{\mathfrak{sl}_{2}}\,
	\mathcal{W}_{[mnp]}\,&=\,
	\mathcal{W}^{+}_{[mnpq]}\,\Omega^{qr}\,\mathcal{J}_{r}^{+}\,,
 \\
 	e\,
	(\mathcal{J}^{\text{IR}}_{mnpq}\,\Omega^{qr}\,\xi_{r})\,&=\,
	(\mathcal{J}^{\text{IR}}_{mnpq}\,e^{\tfrac{1}{2}})\,\Omega^{qr}\,(\xi_{r}\,e^{\tfrac{1}{2}})\,\,.
 \end{align}
\end{subequations}
In the second line we added the free field realization which makes the equality \eqref{firstpiecenullC5} obvious.
There are additional relations at order $q^3$. 
The presence of these relations can be anticipated by looking at the Schur index but with some caution. These relations have the same conformal weight as the generator $\mathcal{J}$ times the relation of smallest conformal weight, namely $\mathcal{J}^2|_{\mathbf{44}}$. This implies that $(1-q)$ times the Plethystic logarithm of the Schur index contains at $q^3$ new nulls ``polluted'' by the contributions mentioned above\footnote{
An example of this phenomenon can be seen for the minimal nilpotent orbit of $\mathfrak{a}_2$ (but it is not present in the case of the minimal nilpotent orbit of $\mathfrak{a}_1$). In this case the Hilbert series is
\begin{equation}
\text{PE}\left[
t^2\,\chi_{[1,1]}-t^4\,(\chi_{[1,1]}+\chi_{[0,0]})+2 t^6\,\chi_{[1,1]}+\dots 
\right]\,,
\end{equation}
The term $2 t^6\,\chi_{[1,1]}$ is not associated to new generators but comes, roughly, from the fact that the expression $J^2|_{[1,1]}$ and $J^2|_{[0,0]}$ associated to the null automatically satisfied certain relations once multiplied by the generators. These are added back by the term $2 t^6\,\chi_{[1,1]}$. }.
The first null state at this order is of the form $(\mathcal{J}\mathcal{J}\mathcal{J}+\mathcal{W}\mathcal{W})\big{|}_{\bf{4004}}$.
As before, it is sufficient to check the relation with maximal weight under the Cartan of $\mathfrak{sl}_2$.
Since 
$\bf{4004}\rightarrow (\bf{3},\bf{308})+\dots$, the relevant component is
\begin{equation}
\label{4004null}
	\mathcal{J}^{++}_{\mathfrak{sl}_{2}}\,
	(\mathcal{J}_{\mathfrak{c}_4}^2)\Big{|}_{\bf{308}}\,=\,
	(\mathcal{W}^{+}_{\underline{\bf{42}}}\,\mathcal{W}^{+}_{\underline{\bf{42}}})\Big{|}_{\bf{308}}\,.
\end{equation}
It is not hard to see that the difference of the operators in \eqref{4004null} is proportinoal to a null state in the IR VOA. More precisely, the Joseph relation of the $\mathfrak{e}_6$ Deligne theory corresponds to the $\mathbf{650}$ of $\mathfrak{e}_6$, which under $\mathfrak{e}_6\rightarrow \mathfrak{c}_4$  decomposes as $\mathbf{650}\rightarrow \mathbf{308}\oplus \mathbf{315}\oplus \mathbf{27}$.
Finally there is a null state of the schematic form 
\begin{equation}
\label{55nullC5}
\left(\mathcal{J}\mathcal{J}\mathcal{J}+\mathcal{W}\mathcal{W}
+T\mathcal{J} +\mathcal{J}\partial \mathcal{J}+\partial^2 \mathcal{J}\right)\big{|}_{\mathbf{55}}\,.
\end{equation}
The component with maximal weight under $\mathfrak{sl}_2$ vanishes thanks to the fact that the contribution to $T$ from the IR VOA is precisely the IR Sugawara stress tensor. This ensures that the operator \eqref{55nullC5} is zero.

\paragraph{Higgs branch relations not associated with VOA nulls.}
We will now briefly discuss Higgs branch relations that are not associated with VOA nulls but rather to a drop of the R-filtration. This is a common phenomenon already observed in e.g.~\cite{Beem:2019tfp,Beem_2020rank2inst,Beem:2020pry}.
For the $\ell=2$ theories of type $C_3A_1$ and $C_2U_1$ there is a flavor singlet Higgs branch relation with $R=2$. From the VOA point of view, the stress tensor, which has $R=1$ from four-dimensional considerations, coincides with the Sugawara stress tensor. This implies that the combination of affine currents that equates the stress tensor presents a drop of its $R$ degree from two to one implying the existence of an Higgs branch relation.
The case of $\ell=3,4$ is more interesting. In each of  these three cases there is a flavor singlet Higgs branch relation of degree $R=\ell$ of the schematic form
\begin{equation}
\mathcal{J}^{\ell}+\mathcal{B}\bar{\mathcal{B}}\,.
\end{equation}
At the level of VOA this null originates from a composite operator which has a drop in $R$-degree. 
For the  $A_1U_1$ theory this operator was presented in equation \eqref{eq:ora1u1} and below it.
\subsection{Theories with enhanced supersymmetry \texorpdfstring{$\mathcal{N}=3,4$}{(N=3,4)}}
\label{sec:moresusy}

In this section we will discuss the remaining rank-one theories: theories with $\mathcal{N}\geq 3$. The enhanced supersymmetry has implications both on the structure of the moduli space of vacua and on the associated VOA, see \cite{Bonetti:2018fqz}. The associated VOAs in the rank-one case have been  bootstrapped in \cite{Nishinaka:2016hbw} and a  free field realization was given in \cite{Bonetti:2018fqz} in terms of a $\beta\gamma bc$ system.
For $\mathcal{N}\geq 3$ theories the full moduli space of vacua coincides with the ECB and is given by
\begin{equation}
\mathcal{M}_{\text{ECB}}=\frac{\mathbb{C}\times \mathbb{H}}{\mathbb{Z}_{\ell}}\,.
\end{equation}
From considerations on the low energy effective theory on the Coulomb branch only the values $\ell=2,3,4,6$ are allowed, but the VOA, and its free field realization, exist for any value of $\ell$. In the case $\ell=2$ the supersymmetry is further enhanced to $\mathcal{N}=4$.
The action of $\mathbb{Z}_{\ell}$ is as in the rest of examples considered in this work. 
The first step to set up the free field realization is to identify an open patch  the HB $\mathbb{H}/\mathbb{Z}_{\ell}$ with $T^*(\mathbb{C}^*)$. Over this space we will fiber $r=1$ symplectic fermions $\eta_1,\eta_2$, see\footnote{We take $\omega_{12}=2$.} \eqref{symplecticFermions}, where the action of $\mathbb{Z}_{\ell}$ is given by $(\eta_1,\eta_2)\mapsto (\omega_{\ell}^{}\,\eta_1,\omega_{\ell}^{-1}\eta_2)$. According to the general rules we will realize the associate VOA as a subVOA
\begin{equation}
\mathcal{V}^{(\ell)}_{\mathcal{N}\geq 3}
\subset
(
\Pi_{\frac{1}{\ell}}\otimes \mathbb{V}_{\eta}
)^{\mathbb{Z}_{\ell}}\,,
\end{equation}
where the factor $\Pi_{\frac{1}{\ell}}$ is associated to the chiral bosons $(\delta,\varphi)$, see equation \eqref{Pi1elldef}.
These VOAs  possess a $U(1)$ outer automorphism for $\ell\neq 2$ which is 
  enhanced to a $SL(2)$ outer automorphism\footnote{While $r$ symplectic fermions have a $SP(2r)$ group of outer automorphism, the $\Gamma=\mathbb{Z}_{\ell}$ quotient breaks it in part.} for $\ell=2$. From the four dimensional point of view this $U(1)$, or the Cartan of  $SL(2)$, is interpreted as the $U(1)_r$ R-symmetry.

As a consequence of $\mathcal{N}=3$ superconformal symmetry in four dimensions these VOAs posses an $\mathcal{N}=2$ super-Virasoro subalgebra, generated by
$\langle \mathcal{J},
\mathcal{G},
\widetilde{\mathcal{G}},
\mathcal{T}\rangle$, where $\mathcal{J}$ is a $U(1)$ current, $\mathcal{T}$ the stress tensor and $\mathcal{G}$ and
$\widetilde{\mathcal{G}}$ are fermionic AKM primary generators of dimensions $3/2$.
Let us present some of the OPEs
\begin{equation}
\label{N2someOPES}
 \begin{aligned}
    \mathcal{J}(z)\mathcal{J}(w)&\sim\frac{2k}{z-w}\,,\\
     \mathcal{G}(z)\widetilde{\mathcal{G}}(w)&\sim\frac{2k}{(z-w)^3}+\frac{\mathcal{J}}{(z-w)^2}+\frac{ \mathcal{T}+\frac{1}{2}\partial\mathcal{J}}{z-w}\,,
        \end{aligned}
    \qquad \qquad
    \begin{aligned}
      \mathcal{J}(z)\mathcal{G}(w) & \sim\frac{+\,\mathcal{G}(w)}{z-w}\,,\\
    \mathcal{J}(z)\widetilde{\mathcal{G}}(w) & \sim\frac{-\,\widetilde{\mathcal{G}}(w)}{z-w}\,,
     \end{aligned}
\end{equation}
 see \cite{Bonetti:2018fqz} for the full list. Central charge and level are related as $c=6k$.
The remaining generators are organized into two short chiral/antichiral $\mathfrak{osp}(2|2)$ multiplets whose superconformal primaries are $\mathcal{W}$ and $\widetilde{\mathcal{W}}$ with conformal dimension $\ell/2$ and their susy descendants of dimension $(\ell+1)/2$ constructed as 
\begin{equation}
\label{WmultipletsNis3}
    \mathcal{W}\xrightarrow[]{\widetilde{\mathcal{G}}} \mathcal{L}\,,
\qquad
    \widetilde{\mathcal{W}}\xrightarrow[]{\mathcal{G}}\widetilde{\mathcal{L}}\,,
\end{equation}
For $\ell=2$ these enhanced the $\mathcal{N}=2$ super-Virasoro subalgebra to the small $\mathcal{N}=4$ super-Virasoro algebra, with $\mathcal{W}$ and $\widetilde{\mathcal{W}}$ providing the extra current to form a $\mathfrak{sl}_2$.
The HB generator $\mathsf{e}$ to which we give a VEV is the avatar of the VOA generator $\mathcal{W}$. So we set
\begin{equation}
\mathcal{W}(z)=e^{\delta+\varphi}\,.
\end{equation}
It follows that the current is $\mathcal{J}=\ell\,\langle \varphi,\varphi \rangle^{-1} \partial \varphi$ with $\langle \varphi,\varphi \rangle=\ell^2/(2k)$.
The stress tensor takes the canonical form, see \eqref{TUVgeneral}, namely $T=T_{\delta,\varphi}+T_{\eta}$ with $T_{\eta}=-\tfrac{1}{2}\eta_1\eta_2$. 
Next we build the fermionic generator of the super-Virasoro algebra by first making the most general ansatz compatible with $\mathcal{J}$ quantum numbers, conformal weight and $\mathbb{Z}_{\ell}$ invariance
\begin{equation}
\label{GandtildeGsupercharges}
\mathcal{G}=\eta_1\, e^{\frac{\delta+\varphi}{\ell}}\,,
\qquad
\widetilde{\mathcal{G}}=
\left(c_1\,\eta_2\,\partial\delta+c_2\,\partial \eta_2\right)e^{-\frac{\delta+\varphi}{\ell}}\,.
\end{equation}
Requiring that the $\mathcal{G}$-$\widetilde{\mathcal{G}}$ OPE given in  \eqref{N2someOPES} is satisfied implies that $c_1=-\frac{k}{\ell}$ and $c_2=\frac{2k+1}{4}$. The $\widetilde{\mathcal{G}}$-$\widetilde{\mathcal{G}}$ OPE is then automatically regular as long as $\langle\varphi,\varphi\rangle=\ell^2/(2k)$ without imposing any further condition on the values of $k$ and $\ell$.
Given $\widetilde{\mathcal{G}}$ we can construct the SUSY descendants of $\mathcal{W}$, see \eqref{WmultipletsNis3} which takes the simple form
\begin{equation}
    \mathcal{L}=\eta_2\,e^{\frac{\ell-1}{\ell}(\delta+\varphi)}\,.
\end{equation}
Notice that, for $\ell>2$, the symplectic fermions $\eta_1$ and $\eta_2$ enter asymmetrically in the construction, compare the generator $\mathcal{G}$ in \eqref{GandtildeGsupercharges} with $\mathcal{L}$.
The final generator that we need is $\widetilde{\mathcal{W}}$. Also in this case, we write the most general ansatz compatible with its  quantum numbers\footnote{In the case $\ell=2$, $\widetilde{\mathcal{W}}$ is part of the $\mathfrak{sl}(2)$ currents of the small $\mathcal{N}=4$ super-Virasoro algebra and its free-field realization coincides with  $f(z)$  given in \eqref{fthetaDELIGNE} upon taking $S^{\natural}\sim T_{\eta}$.}
\begin{equation}
\label{Wcoefficients}
\widetilde{\mathcal{W}}=
((\partial\delta)^{\ell}+\#_1\,T_{\eta}\,(\partial\delta)^{\ell-2}+\#_2\,\partial^2\delta(\partial\delta)^{\ell-2}+\dots)e^{-(\delta+\varphi)}\,,
\end{equation}
and fix the coefficients by the requirement that $\widetilde{\mathcal{W}}$ is a  $\mathcal{N}=2$ super-Virasoro chiral primary of dimension $\tfrac{\ell}{2}$. This includes the relations
\begin{equation}
   \widetilde{\mathcal{G}}(z)
   \widetilde{\mathcal{W}}(w)\sim 0\,,
   \qquad
  \mathcal{G}(z)\widetilde{\mathcal{W}}(w)\sim \frac{\widetilde{\mathcal{L}}(w)}{z-w}\,.
\end{equation}
Imposing these conditions not only fixes the coefficients in \eqref{Wcoefficients}, but also produces a finite list of allowed values of $k$ for a given $\ell$. The interesting values are given by\footnote{Recall that this fixes also the value of the central charge since for the $\mathcal{N}=2$ super-Virasoro $c=6k$.}
$k=-\tfrac{(2\ell-1)}{2}$ while the remaining one are associate to discrete quotients\footnote{For the first few values of $\ell$ the allowed level $k$ are given by
\begin{equation}
\begin{split}
&\ell=2
,\;k=-(1/2),  -(3/2)\\
&\ell=3
,\;k=-(1/2),  -(5/2)\\
&\ell=4
,\;k=-(1/2),  -(3/2),  -(7/2)\\
&\ell=5
,\;k=-(1/2),  -(9/2)\\
&\ell=6
,\;k=-(1/2),  -(3/2),  -(5/2),  -(11/2) 
\end{split}
\end{equation}
The last value correspond to the desired one, namely $k=-\tfrac{(2\ell-1)}{2}$. The fist $k=-1/2$ to the $\mathbb{Z}_{\ell}$ quotient of the $\beta$-$\gamma$ VOA. The intermediate values are other quotient, for examples for $\ell=6$ they correspond to the $\mathbb{Z}_2$ quotient of the $\ell=3$ VOA and $\mathbb{Z}_3$ quotient of the $\ell=2$ VOA.}.
Focusing on the case of interest $k=-\tfrac{(2\ell-1)}{2}$,  we have checked that the OPE among the generators we constructed close so that they provide a complete list of strong generators.

\acknowledgments

It is a pleasure to thank Philip Argyres for helpful conversations, and especially Wolfger Peelaers who collaborated with us in the early stages of this work. The work of CB is supported in part by ERC Consolidator Grant 864828 ``Algebraic Foundations of Supersymmetric Quantum Field Theory (SCFTAlg)'' and by the Simons Collaboration for the Nonperturbative Bootstrap under grant 494786 from the Simons Foundation. CB is also supported in part by the STFC through grant number ST/T000864/1. MM is supported in part by the STFC through grant number ST/X000753/1. The work of AD and LR is supported in part by NSF grant PHY-2210533 and by the Simons Foundation grant  681267 (Simons Investigator Award).
CM has received funding from the European Union’s Horizon 2020 research and innovation programme under the Marie Sklodowska Curie grant agreement No 754496.
The work of CM is also supported in part by Istituto Nazionale di Fisica Nucleare (INFN) through the “Gauge and String Theory” (GAST) research project.

\appendix
\section{Anomaly matching }
\label{appB}

One of the simple, but powerful, implications of the generalized free field construction is that we can predict the central charge and AKM levels of the VOA from the IR data. In this appendix we discuss how these relations emerge  by anomaly matching on the moduli space of vacua, see e.g.~\cite{Shimizu:2017kzs}.
We will match anomalies associated to symmetries that are preserved in the Higgsing procedure all the way from the UV to the IR.

First, let us recall that the conformal anomalies $a,c$ and level\footnote{Here we are a bit schematic since each simple factor of the flavor symmetry is a associated to a different level $k$.} $k$ are given by
\begin{equation}
\Tr\big( r^3\big)=48(a-c)\,,
\qquad
\Tr\big( r R^2\big)=2( 2a-c)\,,
\qquad
\Tr\big( r\,F^{a}F^{b}\big)=-\tfrac{k}{2}\delta^{ab}\,,
\end{equation}
where $r$ is the generator of $U(1)_r$ normalized so that supercharges have charge one, $R$ is the Cartan generator, normalized to be $\pm\tfrac{1}{2}$ on the two-dimensional representation, of $SU(2)_R$ and $F^a$ are generators of the flavor symmetry.
On the Higgs branch $U(1)_r$ is, by definition, unbroken so that the $\Tr r^3$ anomaly can be immediately matched. The $SU(2)_R$ on the other hand is broken but we can still extract information by recalling that, as emphasized in Section \ref{sec:ffrgeneral},  the choice of VEV preserves a combination of $R$ and a generator of the flavor symmetry given by\footnote{For a special class of Higgsing, the $\ell=2$ cases in the construction presented here,  there is a whole $SU(2)$ that can be constructed in this way and is interpreted and the IR R-symmetry. }
\begin{equation}
\label{overlineRdef}
	\overline{R}=R-\tfrac{1}{2}\,\mathsf{j}\,,
\end{equation}
where $\mathsf{j}$ is the avatar of the VOA generator $j(z)$ given in \eqref{jzformula}.
Recall from \eqref{jeOPEs} that $\mathsf{j}\cdot \mathsf{e}=\ell\,\mathsf{e}$ and, from Table \ref{tab:Rdegree}, that $R[\mathsf{e}]=\tfrac{\ell}{2}$,
so that giving a VEV to $\mathsf{e}$ preserves $\overline{R}$.
Additionally, the  semi-simple part of the unbroken flavor symmetry, whose algebra has been denoted by $\unbrSYMM$ can also be matched. To summarize, we can match the anomalies 
\begin{equation}
\label{Summaryoftheanomaliestocompute}
\Tr\big( r^3\big)\,,
\qquad
\Tr\big( r \overline{R}^2\big)\,,
\qquad
\Tr\big( r\,\hat{F}^{\alpha}\hat{F}^{\beta}\big)\,,
\end{equation}
where $\hat{F}$ denote the generators of $\unbrSYMM$. Lets compute and match these three quantities in the UV and in the IR.

Concerning the $\Tr r^3$ anomaly, recalling that each full hyper contributes a factor $-2$ to the anomaly (associated to the fermions in the hypermupltiplet which have $r$ charge $-1$) and each vector multiplet contributes  with a factor $+1$, we have
\begin{equation}
\label{matchingrcube}
\Tr\big( r^3\big)\big{|}_{\text{UV}}=24(a-c)_{\text{UV}}\,,
\qquad
\Tr\big( r^3\big)\big{|}_{\text{IR}}=24(a-c)_{\text{IR}}- (n_{\beta\gamma}+1)+n_{\text{v}}\,,
\end{equation}
where $n_{\beta\gamma}+1$ and  $n_v$ are the total number of hypers and vectors in the IR. 
Notice that the $\beta\gamma$ and $(\mathsf{e}^{1/\ell},\mathsf{h}\mathsf{e}^{-1/\ell})$ hypers give the same contribution to $\Tr r^3$.
Let us turn to the second anomaly in \eqref{Summaryoftheanomaliestocompute}. In the UV it is easy to compute using the explicit form of $\overline{R}$
\begin{equation}
\label{trrR2UV}
\Tr\big( r \overline{R}^2\big)\big{|}_{\text{UV}}=
\Tr\big( r R^2\big)\big{|}_{\text{UV}}+
\tfrac{1}{4}\Tr\big( r\, \mathsf{j}^2\big)\big{|}_{\text{UV}}=
2( 2a-c)_{\text{UV}}-\tfrac{1}{2}\,I_{\mathfrak{u}(1)\hookrightarrow\mathfrak{g}_{\text{UV}}} k_{\text{UV}}\,,
\end{equation}
$\mathsf{j}$ is embedded in the UV symmetry with an embedding index $I_{\mathfrak{u}(1)\hookrightarrow\mathfrak{g}_{UV}}$. The embedding index can be evaluated based on discussions in Section \ref{sec:ffrgeneral} and equals $1,\frac{2}{3}$ and $\frac{3}{4}$ for $\ell=2,3$ and $\ell=4$ respectively.
We now turn to the evaluation of $\Tr r  \overline{R}^2$ in the IR.
In this case the $\beta\gamma$  and  $(\mathsf{e}^{1/\ell},\mathsf{h}\mathsf{e}^{-1/\ell})$ hypers give different contributions as they have different $\overline{R}$ assignment.
The latter have weight $(0,1)$ as it should since the generator $\mathsf{e}$ getting a VEV is uncharged under $\overline{R}$. The $\overline{R}$ assignment of $\beta\gamma$ is collected in Table \ref{tab:Rassignbetagamma} and follows from Table \ref{tab:Rdegree} and \eqref{jeOPEs}.
\begin{table}[t]
\begin{center}
\begin{tabular}{|c|c|c|c|}
	\hline
	Theory & $R[(\beta,\gamma)]$ &
	 $\mathsf{j}[(\beta,\gamma)]$ & $\overline{R}[(\beta,\gamma)]$ \\
	 \hline
	 $C$-series & $(1/2,1/2)$ & $(0,0)$ & $(1/2,1/2)$ \\
	 \hline
	 \hline
	$A_3$ & $(1/2,1/2)$ & $(1/3,-1/3)$ & $(1/2,-1/2)$ \\
	$A_1U_1$ & $(1/2,1/2)$ & $(1/2,-1/2)$ &$(0,1)$\\
	\hline
	\hline
	$A_2$ & $(1/2,1/2)$ & $(1/2,-1/2)$ & $(1/4,3/4)$\\
	\hline
\end{tabular}
\caption{Charge assignments for the $\beta\gamma$ pairs under $R$, $\mathsf{j}$ and $\overline{R}$}
\label{tab:Rassignbetagamma}
\end{center}
\end{table}
The contribution of a full hypermultiplets with $\overline{R}$ charges $(\lambda,1-\lambda)_{r=0}$ to the anomaly in question comes from the fermions which have charges $(\lambda-\frac{1}{2})_{-1}$ and $(\frac{1}{2}-\lambda)_{-1}$ respectively and is given by
\begin{equation}
	\Tr\big(r\overline{R}^2\big)=-2\left(\lambda-\tfrac{1}{2}\right)^2\,.
\end{equation}
This implies that $(\mathsf{e}^{1/\ell},\mathsf{h}\mathsf{e}^{-1/\ell})$, which corresponds to $\lambda=0$, contribute $-\tfrac{1}{2}$ and each $\beta\gamma$ pair contributes with a factor $\alpha(\lambda)$ collected in Table \ref{tab:cdeltaphiallrank1}.
\begin{table}[t]
	\begin{center}
		\begin{adjustbox}{center}
			\begin{tabular}{|c|c|c|c|c|}
				\hline
				Theory & $\overline{R}[(\beta,\gamma)]$& $n_{\beta\gamma}$ &$n_{\beta\gamma}c^{(\lambda)}_{\beta\gamma}$ &$c_{\delta\varphi}$ \\
				\hline
				$E_8$& $(\frac{1}{2},\frac{1}{2})$ & $29$& $-29$&$-34 $ \\	 $E_7$& $(\frac{1}{2},\frac{1}{2})$ &$17$& $-17$  & $-22 $ \\	 $E_6$& 
				$(\frac{1}{2},\frac{1}{2})$ & $11$& $-11$ &$-16 $ \\
				$D_4$& $(\frac{1}{2},\frac{1}{2})$&$5$& $-5$ &$-10 $ \\
				$A_2$& $(\frac{1}{2},\frac{1}{2})$&$2$& $-2$  &$-7 $
				\\
				$A_1$& $(\frac{1}{2},\frac{1}{2})$&$1$& $-1$  &$-6 $
				\\
				\hline
				$C_5$& $(\frac{1}{2},\frac{1}{2})$&$5$& $-5$ & $-19$
				\\
				$C_3A_1$& $(\frac{1}{2},\frac{1}{2})$&$3$& $-3$  & $-13$
				\\
				$C_2U_1$& $(\frac{1}{2},\frac{1}{2})$&$2$& $-2$ & $-10$\\
				\hline
				$A_3$& $(\frac{2}{3},\frac{1}{3})$&$3$ & $-2$ & $-25 $\\
				
				$A_1U_1$& $(\frac{2}{3},\frac{1}{3})$&$1$ & $-\frac{2}{3}$ & $-17$ \\
				
				\hline
				$A_2$& $(\frac{3}{4},\frac{1}{4})$&$2$ & $-\frac{1}{2}$  & $-28$\\
				\hline
			\end{tabular}
		\end{adjustbox}
\caption{$c_{\delta\varphi}$ for the rank-one theories.}
\label{tab:cdeltaphiallrank1}
	\end{center}
\end{table}
As vector  multiplets in our cases, when present, are not charged  under $\mathsf{j}$, they  give a contribution $\tfrac{1}{2}n_{\text{vector}}$ to this anomaly.
Putting the pieces together we get\footnote{For the $A_1U_1$ theory this formula needs a small modification since in this case $\mathsf{j}$ (the avatar of  $j(z)=h-\beta\gamma+\mathcal{J}_{\mathfrak{u}_1}^{\text{IR}}$) acts non-trivially on the interacting part of the IR theory.  Due to the additional $\mathcal{J}_{\mathfrak{u}_1}^{\text{IR}}$ piece, in the infrared we also have contribution from  $\tfrac{1}{4}\Tr r\left(\mathcal{J}_{\mathfrak{u}_1}^{\text{IR}}\right)^2=-\frac{k_{\mathfrak{u}_1}}{2}$. The effect of this term is to simply shift $k^{2d}\to k^{2d}-k_{\mathfrak{u}_1}$ in the final formula for the central charges.}
\begin{equation}
\label{trrR2IR}
\Tr\big( r \overline{R}^2\big)\big{|}_{\text{IR}}=
2( 2a-c)_{\text{IR}}-\tfrac{1}{2}+n_{\beta\gamma}\,\alpha(\lambda)+ \tfrac{1}{2}n_{\text{v}}\,.
\end{equation}
By equating \eqref{trrR2IR} to \eqref{trrR2UV} and the two expressions in \eqref{matchingrcube}
we immediately obtain\footnote{Notice that this is possible thanks to the fact that the $a$ central charge always appears in the combination $a_{\text{UV}}-a_{\text{IR}}$.}
\begin{equation}
 -12\,c_{\text{UV}}=   
  -12\,c_{\text{IR}}+2-3I_{\mathfrak{u}(1)\hookrightarrow\mathfrak{g}_{\text{UV}}} k_{\text{UV}}
  -n_{\beta\gamma}(6 \alpha(\lambda)+1)-2 n_{\text{v}}\,,
\end{equation}
which, translated to $2d$ central charges gives
\begin{equation}
\label{cUVandCIR}
 c^{\text{2d}}_{\text{UV}}=   
 c^{\text{2d}}_{\text{IR}}+2+6 I_{\mathfrak{u}(1)\hookrightarrow\mathfrak{g}_{UV}} \,k^{\text{2d}}_{\text{UV}}
  +n_{\beta\gamma}\,c_{\beta\gamma}^{(\lambda)}-2 n_{\text{v}}\,,
\end{equation}
where we introduced 
$c_{\beta\gamma}^{(\lambda)}=-(6 \alpha(\lambda)+1)$ which is the central charge of a $\beta \gamma$ pair with conformal weights $(\lambda,1-\lambda)$.
Let us compare this expression to the expression coming from the free field construction \eqref{TUVgeneral}
\begin{equation}
\label{cfreefields}
 c^{\text{2d}}_{\text{UV}}= c_{\delta,\varphi} -n_{\beta\gamma}-2 n_{\text{v}}+ c^{\text{2d}}_{\text{IR}}\,.
\end{equation}
Comparing \eqref{cUVandCIR} to \eqref{cfreefields} we obtain
\begin{equation}
\label{cdeltaphiAPP}
c_{\delta,\varphi}=2+6I_{\mathfrak{u}(1)\hookrightarrow\mathfrak{g}_{UV}} \,k^{\text{2d}}_{\text{UV}}+n_{\beta\gamma}\,(c_{\beta\gamma}^{(\lambda)}+1)\,.
\end{equation}
In particular if the $(\beta,\gamma)$ have canonical $\overline{R}$ assignment $(\tfrac{1}{2},\tfrac{1}{2})$ we get $c_{\beta\gamma}^{(\lambda)}=-1$ and recover the formula from 
\cite{Beem:2019tfp} of which  \eqref{cdeltaphiAPP} is a generalization.

Finally we match the third anomaly in \eqref{Summaryoftheanomaliestocompute} recalling that the IR ingredients transform under the unbroken symmetry as dictated by \eqref{JnaturalasSUMgeneral}. This gives
\begin{equation}
	k_{\text{UV}}=I_{\mathfrak{g}_{\text{IR}}\hookrightarrow\unbrSYMM} k_{\text{IR}}+k_{\xi}\,,
\end{equation}
reproducing Table \ref{levelsTable}.

\bibliographystyle{JHEP}
\bibliography{ref.bib}

\end{document}